\documentclass[aip, jmp, bmf, sd, rsi, amsmath,amssymb, preprint, reprint, author-year, author-numerical, Conference Proceedings]{revtex4-1}

\usepackage{graphicx}
\usepackage{dcolumn}
\usepackage{bm}
\usepackage[utf8]{inputenc}
\usepackage[T1]{fontenc}
\usepackage{mathptmx}
\usepackage{etoolbox}
\usepackage{physics}
\usepackage{blkarray}
\usepackage{orcidlink}
\usepackage{hyperref}

\graphicspath{ {./} } 

\newcommand{\be}{\begin{equation}}
\newcommand{\ee}{\end{equation}}

\newcommand{\bq}{\mathbf{q}}
\newcommand{\bA}{\mathbf{A}}
\newcommand{\bJ}{\mathbf{J}}

\makeatletter
\def\@email#1#2{%
 \endgroup
 \patchcmd{\titleblock@produce}
  {\frontmatter@RRAPformat}
  {\frontmatter@RRAPformat{\produce@RRAP{*#1\href{mailto:#2}{#2}}}\frontmatter@RRAPformat}
  {}{}
}%
\makeatother

\begin{document}

\preprint{AIP/123-QED}

\title[]{Breaking coexistence: zealotry vs. nonlinear social impact}

\author{Christopher R. Kitching\,\orcidlink{0000-0002-1451-9757}}
\email{christopher.kitching@manchester.ac.uk}
\thanks{Corresponding author: \href{mailto:christopher.kitching@manchester.ac.uk}{christopher.kitching@manchester.ac.uk}}
\affiliation{Department of Physics and Astronomy, School of Natural Sciences, The University of Manchester, Manchester M13 9PL, UK}

\author{Luc\'ia S. Ramirez\,\orcidlink{0000-0003-3529-0581}}
\email{lucia.s.ramirez@gmail.com}
\affiliation{Departament de F\'isica de la Mat\`eria Condensada, Universitat de Barcelona, Spain}
\affiliation{Instituto de F\'isica Interdisciplinar y Sistemas Complejos, IFISC (CSIC-UIB), Campus Universitat Illes Balears, E-07122 Palma de Mallorca, Spain}

\author{Maxi San Miguel\,\orcidlink{0000-0001-6237-425X}}
\email{maxi@ifisc.uib-csic.es}
\affiliation{Instituto de F\'isica Interdisciplinar y Sistemas Complejos, IFISC (CSIC-UIB), Campus Universitat Illes Balears, E-07122 Palma de Mallorca, Spain}

\author{Tobias Galla\,\orcidlink{0000-0003-3402-2163}}
\email{tobias.galla@ifisc.uib-csic.es}
\affiliation{Instituto de F\'isica Interdisciplinar y Sistemas Complejos, IFISC (CSIC-UIB), Campus Universitat Illes Balears, E-07122 Palma de Mallorca, Spain}

\date{\today}

\begin{abstract}
We study how zealotry and nonlinear social impact affect consensus formation in the nonlinear voter model, evolutionary games, and the partisan voter model. In all three models, consensus is an absorbing state in finite populations, while coexistence is a possible outcome of the deterministic dynamics. We show that sufficiently strong zealotry, i.e. the presence of agents who never change state, can drive infinite populations to consensus in all three models. However, while evolutionary games and the partisan voter model permit zealotry-induced consensus for all values of their model parameters, the nonlinear voter model does not. Central to this difference is the shape of the social impact function, which quantifies how the influence of a group scales with size, and is therefore a measure of majority and minority effects. We derive general conditions relating the slope of this function at small group sizes to the local stability of consensus. Sublinear impact favours minorities and can override zealotry to prevent consensus, whereas superlinear impact promotes majorities and therefore facilitates consensus. We extend the analysis to finite populations, exploring the time-to-consensus, and the shape of quasi-stationary distributions. 
\end{abstract}

\maketitle

\begin{quotation}
We consider several individual-based models of opinion dynamics and simple two-player games. These models share the property of either allowing indefinite coexistence of opinions (or strategies) or converging to a consensus state. Zealotry, that is, the presence of agents who never change their state, is known to promote consensus, whereas certain types of nonlinear interactions tend to favour coexistence. In this paper, we study the balance between these two opposing effects, focusing on nonlinear social impact: how the influence of a group of agents depends on the group's size. We find that sublinear social impact can override the effects of zealotry and prevent consensus formation. This result is particularly relevant, as sublinear impact has been observed in experiments with human participants.
\end{quotation}

\section{Introduction}
A system is considered `complex' when it consists of many interacting entities which give rise to system-wide emergent behaviour that cannot be understood by examining its constituents in isolation. Examples span physics, biology, and other sciences \cite{anderson1972more, may1972will, herbert1962architecture, mitchell2009complexity, strogatz2001exploring, barabasi_albert}, and statistical physics has played a key role in understanding these phenomena \cite{bak1995complexity, castellano2009statistical, mezard2009information, sornette2006critical}.

Complex dynamics arise in opinion formation  in social (or socio-technical) systems and population evolution in game theory \cite{nowak2006evolutionary, hofbauer2003evolutionary, castellano2009statistical}. Individual-based stochastic models commonly used in these two areas to describe finite populations are closely related to one another. For example, the celebrated voter model \cite{clifford1973model, liggett1994coexistence,redner2019reality} can be seen as the limit of neutral selection of certain evolutionary processes \cite{fotouhi2019evolution, allen2012mutation, shakarian2012review}. The phenomena studied in both fields also mirror each other. In opinion dynamics, one focus is on the emergence of consensus, i.e. situations in which all agents share the same opinion. In evolutionary models, the analogue is a `monomorphic' state, where all individuals are of the same type or species, or in which all involved players use the same strategy. Hence, the distinction between consensus and diversity in opinion dynamics parallels fixation and coexistence in evolutionary systems.

Zealots, i.e. agents who do not change state, are known to affect outcomes in both domains \cite{galam1997, galam2007,mobilia2015nonlinear, masuda2012evolution, nakajima2015evolutionary, khalil2018zealots}. In the standard voter model without zealots, the two opinion states are symmetric, and consensus on either is equally likely. A single zealot breaks this symmetry, ensuring consensus in finite populations at the state represented by the zealot. Zealots can also influence infinite populations \cite{mobilia_2003}, and similar effects are observed in network co-ordination games with extremists \cite{kearns2009behavioural}, and in the two-opinion naming game with committed minorities \cite{xie2011social}. Generally, one-sided zealotry breaks symmetry, while mixed zealotry promotes coexistence \cite{mobilia2007role, yildiz2013binary, acemouglu2012opinion}. For a further review of zealotry in opinion dynamics, see Sec.~2.1 of Ref.~\cite{cao2024discrete}.

In this paper, we ask if and when zealotry can bring about consensus in different models of opinion dynamics and evolutionary games. We consider three settings: the nonlinear voter model \cite{castellano2009nonlinear,ramirez2023ordering, schweitzer2009nonlinear, min2017fragmentation}, two-player two-strategy evolutionary games \cite{nowak2006evolutionary,traulsen2009stochastic}, and the so-called partisan voter model \cite{masuda2010heterogeneous, masuda2011can,llabres2023partisan}. In all three models, agents can be in one of two discrete states, and in the absence of zealots consensus states are absorbing; that is, once such a state is reached, no further dynamics can occur. Our focus is on conditions under which zealotry and nonlinear interactions shift a system from coexistence to consensus by breaking the symmetry between states.

We approach this through deterministic rate equations for infinite populations and also by investigating finite-size effects: time-to-consensus (or fixation) and the shape of quasi-stationary distribution that describes the system at long times but before absorption. For each of these we establish the similarities and differences across the three models.

A key finding is that the local stability of consensus states in the deterministic rate equations is, among other things, determined by `social impact' \cite{latane}. This describes the influence a group of agents has on other agents. The strength of this influence can be a nonlinear function of the size of the group of influencers. We quantify this using a `social impact function', showing that sublinear impact destabilises consensus, while superlinear impact can stabilise it.

Our paper is organised as follows: In Sec.~\ref{sec:gen} we describe the general framework of populations with susceptible agents and zealots. Section~\ref{sec:defs_rate_eqs} defines the models: the (linear and nonlinear) voter models, simple evolutionary games, and the partisan voter model, along with their corresponding rate equations. In Sec.~\ref{sec: fixed point analysis}, we determine the fixed points of these rate equations in the presence of zealots and analyse their stability. This provides a comprehensive study of the effects of zealotry in finite populations. We establish common behaviour across the three models, in particular we show that show that zealotry above a model-dependent critical magnitude can enforce consensus in many circumstances. In Sec.~\ref{sec: social impact} we explore the interplay of nonlinearity and zealotry in more detail, and we discuss the importance of the social impact function. In particular we derive conditions on this function for the local stability of consensus states, and investigate how sublinear and superlinear impact affect this stability. In Sec.~\ref{sec:finite}, we turn to finite populations, characterising the mean time-to-consensus or fixation in opinion dynamics and evolution, respectively, as a function of the level of zealotry. We also show how zealotry affects the shape of the quasi-stationary distribution before absorption for different dynamics. Finally, in Sec.~\ref{sec:concl} we summarise our findings, draw conclusions, and outline directions for future work.

\section{General model setup}\label{sec:gen}
All models in this paper describe a population of $N$ agents with `all-to-all' interaction. Every agent has a state $\pm 1$, representing the agent's current opinion or strategy. These states are equivalent to one another; the model is symmetric with respect to an interchange of the two states. Each agent is also either susceptible or a zealot. Susceptible agents can change state, whereas the state of a zealot is fixed in time. We note that zealots whose states are truly fixed are referred to as `inflexible' zealots \cite{mobilia2007role, mobilia2015nonlinear}. This is opposed to `flexible' zealots, considered for example in Refs.~\cite{mobilia_2003, mobilia2005voting}, which are agents that have an internal preference but may still change opinion. We write the number of zealots in states $\pm 1$ as $Z_{\pm}$, and the corresponding fraction relative to the total as $z_{\pm}=Z_{\pm}/N$. The total proportion of zealots in the population is $z \equiv z_+ + z_- \in[0, 1)$. We exclude the case in which the entire population is made of zealots, $z=1$, because no dynamics can then occur.

The total number of susceptible agents is $S\equiv N-Z_{+}-Z_{-}$, and we write $s=S/N=1-z$ for the proportion of susceptible individuals in the population. The number of susceptible agents in the state $+1$ is denoted by $n$, and we write $x=n/N\in[0,1-z]$.  The susceptible agents change their state according to specific rules, which differ across the different models, and which will be described below. 

We will look at two different scenarios. The first is a situation in which there are zealots in both states, $\pm 1$. These do not have to be present in equal proportions. To describe any imbalance we introduce the parameter $\delta\in[0,1]$ by setting
\be
z_{\pm}=\frac{1}{2}(1\pm\delta)z.
\ee
The case $\delta=0$ corresponds to equal number of zealots on either side, $z_\pm=z/2$. In the second scenario there is only one type of zealot. Without loss of generality, we assume this to be the state $+1$, so that we have $z_{-}=0$ and $z_+ = z$, i.e. $\delta=1$.

\section{Model definitions and rate equations in infinite populations}\label{sec:defs_rate_eqs}
We will consider several different model dynamics: linear and nonlinear voter models \cite{holley1975ergodic,suchecki,ramirez2023ordering, schweitzer2009nonlinear, min2017fragmentation}, evolutionary game dynamics \cite{taylor, traulsen2005coevolutionary, kitching2024qdeformed, traulsen2009stochastic, traulsen2007pairwise, nowak2006evolutionary} and the partisan voter model \cite{masuda2010heterogeneous, masuda2011can, llabres2023partisan}. Except for the linear voter model, the three models have consensus and coexistence solutions in the absence of zealots. The stability of these fixed points depends on model parameters.

In this section we will define these models and formulate the corresponding rate equations, valid in the limit of infinite populations. These definitions can be found in the existing literature, but we include these details here to make the paper more self-contained. Readers who are familiar with these models can proceed directly to Sec.~\ref{sec: fixed point analysis}.

\subsection{General setup and main objectives}
With minor adaptations in the partisan voter model, all model dynamics in finite populations in this paper define one-step processes on a set of integers $n=0,\dots, S$ (where we recall that $n$ is the number of susceptible individuals in state $+1$). Each event in the population converts an individual of type $+1$ into one of type $-1$ or vice versa, and thus the state of the population changes from $n$ to $n\pm 1$. We write $T^\pm_n$ for these rates if the system is currently in state $n$. The rates $T^\pm_n$ are generally of order $N$, indicating that the number of events in the population per unit time is proportional to the size of the population. We can thus write $T^\pm_n=N T^\pm(x)$ where $x=n/N$ as defined above.

The rate equations for infinite populations are then given by \cite{traulsen2005coevolutionary, mobilia2007role}
\be
    \frac{\dd x}{\dd t} = T^+(x)-T^-(x). \label{eq main: general rate eqn}
\ee
We can find the fixed points of these equations for the different models, and determine their linear stability. We do this for different scenarios of zealotry. 

Our primary interest is to study if and when consensus is a local attractor, or, more strictly, when zealotry can enforce consensus. All models we study involve a copying mechanism. That is to say, any susceptible agent can adopt the state of its interaction partners. These partners can be susceptible agents or zealots. It is then clear that persisting consensus can only be achieved if all zealots are of one single type (we call this `one-sided' zealotry). If there are zealots in both states $\pm 1$ it is possible for the group of susceptible agents to reach consensus temporarily. However, this state will not be absorbing, as the susceptible agents can copy the opposite state from the group of zealots representing that opposing state.

In the standard voter model a non-zero fraction of one-sided zealots drives the system to consensus, even in infinite populations \cite{mobilia2007role}. A natural question is then if related models show similar behaviour, and if there is always a critical level of zealotry that will enforce consensus. We will study this in the nonlinear voter model, in evolutionary games, and in the partisan voter model. After briefly describing the standard voter model, we will now define these models in turn in Secs.~\ref{sec: nonlinear vm}, \ref{sec: games} and \ref{sec: partisan}, respectively.

\subsection{The standard voter model} \label{sec: standard voter model}
The standard voter model, originally introduced in Ref.~\cite{holley1975ergodic}, is one of the simplest models of interacting agents. A discrete-time dynamics in a population with all-to-all interaction and with zealots can be defined as follows:
\begin{enumerate}
    \item At each time step select a random agent for update.
    \item If this is a zealot do nothing, and go to step 5.
    \item Select a second agent randomly from the population of zealots and susceptibles.
    \item The agent chosen in step 1 copies the state of the agent chosen in step 3.
    \item Increment time by $\frac{1}{N}$ and go to step 1.
\end{enumerate}
This can be extended to continuous time, and simulations can then be carried out using the Gillespie method \cite{gillespie1976general}. The rates $T^{\pm}(x)$ for this process are
\begin{align}
\begin{split} \label{eq main: standard vm rates}
    T^{+}(x) &= (s-x)(x+z_{+}), \\
    T^{-}(x) &= x(s-x+z_{-}).
\end{split}
\end{align}
The first term in $T^{+}(x)$ is the probability that the agent chosen in step 1 is susceptible and in state $-1$. The second factor is then the probability that the agent chosen in step 3 is in state $+1$. This latter agent can be susceptible or a zealot. In $T^-(x)$ the roles of the two states $\pm 1$ are reversed.

Using Eqs.~(\ref{eq main: standard vm rates}) in Eq.~(\ref{eq main: general rate eqn}) gives the rate equation
\be
    \frac{\dd x}{\dd t} = (s-x)(x+z_+) - x(s-x+z_-).
\ee
We note that this reduces to $\dot x=0$ in the absence of zealots ($s=1, z_\pm=0$); any point $x\in[0,1]$ is then a fixed point and marginally stable. One-sided zealotry in the linear voter model leads to a preferred stable consensus state. If there are zealots of both types, the only fixed point is the stable coexistence point at $x^*=sz_+/z$.

\subsection{The nonlinear voter model} \label{sec: nonlinear vm}
The nonlinear voter model \cite{ramirez2023ordering, schweitzer2009nonlinear, min2017fragmentation} is a variation of the standard voter model. The rates for this model are a direct extension of Eqs.~(\ref{eq main: standard vm rates}),
\begin{align}
\begin{split} \label{eq main: nonlinear vm rates}
    T^{+}(x) &= (s-x)(x+z_{+})^{q}, \\
    T^{-}(x) &= x(s-x+z_{-})^{q}.
\end{split}
\end{align}
The parameter $q$ takes positive real values, i.e. $q>0$. Analyses of the nonlinear voter model have shown that the dynamics are very different in the cases $q<1$, $q=1$ and $q>1$ \cite{ramirez2023ordering, castello}. Simulations of this model can be performed using transition rates $T_n^\pm= NT^\pm(x=n/N)$ [with $T^\pm(x)$ as in Eqs.~(\ref{eq main: nonlinear vm rates})], and again with the Gillespie algorithm \cite{gillespie1976general}.

The nonlinear voter model is closely related to `nonlinear $q$-voter models' \cite{castellano2009nonlinear}. In this earlier work, $q$ was assumed to be a positive integer. In discrete time the dynamics can then be understood as follows: At each time step, pick an agent, then sample $q$ of its neighbours with replacement. If these $q$ agents are all in the opposite state to that of the focal agent, the focal agent changes state. We note that there are subtle differences between the nonlinear voter model in Eq.~(\ref{eq main: nonlinear vm rates}) and the `nonlinear $q$-voter model' \cite{castellano2009nonlinear}. In the latter there is a separate probability for the focal agent to change state if the $q$ neighbours do not all agree. 

Nonlinear voter models with zealots have been studied previously \cite{mobilia2015nonlinear}; as part of our work, we generalise this to real valued $q$. We highlight that there is empirical evidence from social impact theory \cite{latane} and work on language dynamics \cite{abrams2003modelling} motivating the choice of real-valued $q$. We will return to this in Sec.~\ref{sec: social impact} when we discuss social impact in more detail. We further note recent work in Refs.~\cite{Mellor2016characterization, mellor2017heterogeneous} which studied a nonlinear voter model with two types of susceptible agents that consult $q_1$ and $q_2$ neighbours in the presence of inflexible zealots respectively. It was shown that the dynamics of such a system are intrinsically out of equilibrium.

To formulate the rate equations for the nonlinear voter model with zealotry we substitute Eqs.~(\ref{eq main: nonlinear vm rates}) into Eq.~(\ref{eq main: general rate eqn}) and find,
\be
    \frac{\dd x}{\dd t} = (1-x-z_{+}-z_{-})(x+z_{+})^{q}-x(1-x-z_{+})^{q}. \label{eq main: nonlinear vm general rate eqn}
\ee
Without zealots, this reduces to
\be\label{eq:nlvm_no_z}
    \frac{\dd x}{\dd t} = (1-x)x^{q}-x(1-x)^{q}.
\ee
For any $q>0$, this equation has three fixed points: $x^{*}=0,\frac{1}{2},1$. When $q<1$ the consensus fixed points ($x^*=0$ and $x^*=1$) are linearly unstable while the central fixed point (describing coexistence) is stable \cite{ramirez2023ordering}. When $q>1$ the stabilities of all fixed points are reversed. The case $q=1$ is the standard voter model (see Sec.~\ref{sec: standard voter model}).

For future reference, we add that sublinear impact $q<1$ favours minorities. We explain this using the nonlinear voter model without zealots, but the argument is similar for the other models we discuss in this paper. For $q<1$, the right-hand side of Eq.~(\ref{eq:nlvm_no_z}) is positive for $x<1/2$ and negative for $x>1/2$. Thus, the flow is towards increasing $x$ when $x<1/2$, and towards decreasing $x$ when $x>1/2$. In this sense, the flow favours the minority, i.e. when fewer than half of the agents are in state $+1$ the flow increases the proportion of state $+1$ agents, and when more than half of all agents are in state $+1$, then the flow increases the proportion of agents in state $-1$. Analogously, the dynamics favours majority opinions for $q>1$.

\subsection{Evolutionary game dynamics} \label{sec: games}
Evolutionary games can be thought of as another extension to the standard voter model. The underlying mechanism is similar to that described in Sec.~\ref{sec: standard voter model} except now agents are also playing a game. Many variations of evolutionary game dynamics have been studied. In absence of zealotry our model is closely related to those in Refs.~\cite{taylor, traulsen2005coevolutionary, kitching2024qdeformed, traulsen2009stochastic, traulsen2007pairwise, nowak2006evolutionary,raducha2022}. The addition, there is also work on zealots in evolutionary games, for example studying their effects on fixation times \cite{masuda2012evolution}, or the evolution of co-operation \cite{nakajima2015evolutionary}. We note there is also previous work on the effect of `co-operation facilitators' in EGT \cite{mobilia2012stochastic, mobilia2013evolutionary, szolnoki2014facilitators}. These agents cooperate with cooperators, but do not interact with defectors.  This increases cooperator payoff, thus cooperation facilitators are similar to zealots in the sense that they promote a particular state.

In this paper we look at two-player games and agents with binary states $\pm 1$, i.e. there are two pure strategies. Thus we have a $2\times 2$ evolutionary game. We focus on a simple case where the game is defined by a payoff matrix with only one parameter $\alpha\in (0,1)$,
\begin{equation} \label{eq main: payoff matrix}
    \begin{blockarray}{cccc}
     & & + & -  \\
    \begin{block}{cc(cc)}
        +~ & & \alpha & 1-\alpha \\
        -~ & & 1-\alpha & \alpha \\
    \end{block}
    \end{blockarray} \hspace{3mm}.
\end{equation}
The different elements of the payoff matrix indicate the payoff to the column player when facing the row player in a single pairwise interaction. For example, if an agent plays strategy $+$ and faces an agent also using strategy $+$, then the payoff to each player is $\alpha$ (upper left entry). If an agent of type $+$ faces a player of type $-$ then the payoff to each player is $1-\alpha$ (upper right and lower left entries). 

In a well-mixed population, every player interacts with every other player, so the expected payoff of a player using strategy $+$ is obtained by weighting the two possible payoffs ($\alpha$ and $1-\alpha$) to the $+$ player by their respective frequencies. The opponent is of type $+$ with probability $x+z_+$, and of type $-$ with probability $s-x+z_-$. Thus the expected payoff of a player playing $+$ is
\be
    \pi_{+} = \alpha(x+z_{+})+(1-\alpha)(s-x+z_{-}). \label{eq main: games expected payoff +}
\ee
An analogous argument for a player playing strategy $-$ yields
\be
    \pi_{-} = (1-\alpha)(x+z_{+})+\alpha(s-x+z_{-}). \label{eq main: games expected payoff -}
\ee
For later purposes, we also introduce the average payoff of the population, which is simply a weighted sum of Eqs.~(\ref{eq main: games expected payoff +}) and (\ref{eq main: games expected payoff -}),
\be
    \phi=(x+z_+)\pi_+ + (1-x+z_-) \pi_-. \label{eq main: average population payoff}
\ee
In our setup all entries in the payoff matrix are non-negative. As a consequence, $\pi_+, \pi_-$ and $\phi$ are also non-negative.

We adopt a commonly used dynamics from evolutionary game theory \cite{traulsen2005coevolutionary, traulsen2009stochastic, ohtsuki2006replicator, traulsen2007pairwise}. Every agent plays the game against all players, and from this every agent establishes an average payoff per game. This is assumed to happen on a timescale that is much faster than the timescale for population evolution. After this, a focal agent is selected for potential update. A random second agent is then selected. If the agents are in different states and the focal agent is not a zealot, the focal agent changes its state with a rate proportional to the payoff for the strategy to be adopted. 

The rates for this model are
\begin{align}
\begin{split} \label{eq main: games rates}
    T^{+}(x) &= (s-x)(x+z_+)\pi_+, \\
    T^{-}(x) &= x(s-x+z_-)\pi_-,
\end{split}
\end{align}
resulting in the rate equation
\be
    \dot x = (s-x)(x+z_+)\pi_+ - x(s-x+z_-)\pi_-. \label{eq main: games general rate eqn}
\ee
For $\alpha=1/2$, we have $\pi^+=\pi^-=1/2$, and this setup reduces to the standard voter model described in Sec.~\ref{sec: standard voter model} up to a change in the timescale. 

Without zealots, Eq.~(\ref{eq main: games general rate eqn}) reduces to
\be
    \dot x = x(1-x)(\pi_+-\pi_-),
\ee
which can also be written in the form
\be
    \dot x = x(\pi_+-\phi). \label{eq main: standard replicator}
\ee
This is the traditional replicator equation, discussed in detail for example in Ref.~\cite{hofbauer2003evolutionary}. For the payoff matrix in Eq.~(\ref{eq main: payoff matrix}) this becomes
\be
    \dot x = (2\alpha-1)x(1-x)(2x-1).
\ee
This equation also has the three fixed points $x^{*}=0,\frac{1}{2},1$. For $\alpha<1/2$ the consensus fixed points are linearly stable while the central coexistence fixed point is unstable. When $\alpha>1/2$ the stabilities of all fixed points reverse.

In evolutionary game theory these types of flows are often referred to as `coordination' ($\alpha<1/2$) and `coexistence' ($\alpha>1/2$) respectively \cite{traulsen2009stochastic}. The case $\alpha=1/2$ is the standard voter model, which would be described as `neutral selection' in evolutionary dynamics. One observes the similarity to the nonlinear voter model, with $q<1$ corresponding to $\alpha>1/2$, and $q>1$ to $\alpha<1/2$.

We note that alternative forms of the replicator equation exist, for example the so-called `adjusted replicator equation', which is obtained by normalising the rates in Eqs.~(\ref{eq main: games rates}) by the average population payoff $\phi$ given in Eq.~(\ref{eq main: average population payoff}) \cite{smith1982evolution, hofbaur2000sophisticated, traulsen2006stochastic, muk2021replicator}. Both formulations leads to the same fixed points and stability for positive payoffs as we have here. We choose to work with the traditional form for simplicity.

\subsection{The partisan voter model} \label{sec: partisan}
The final model we will consider is the partisan voter model originally introduced in Refs.~\cite{masuda2010heterogeneous, masuda2011can}. The dynamics we use are described in Ref.~\cite{llabres2023partisan}, and are outlined below. To the best of our knowledge, the partisan voter model with the addition of zealots has not previously been studied.

In this model all susceptible agents have an internal preference for one of the two states. The preference of an agent does not change in time. There are thus four distinct types of susceptible agents, those with preference for state $-$ but who are currently in state $+$, those with preference for $-$ and currently in state $-$, and so on.  We write $n_a^b$ for the number of susceptible agents in the population who are in state $a$ with preference for state $b$ (with $a,b\in\{\pm 1\}$). We also write $x_{a}^{b}=n_a^b/N$ for the proportion of such agents. 

The strength of the preference for a state is quantified by a parameter $\epsilon\in(0,1)$. The detailed role of this parameter will become clear below when we define the dynamics for the model. We assume that the preference parameter $\epsilon$ is constant in time and the same for all susceptible agents (more general setups have been considered in Ref.~\cite{masuda2011can}). 

We will always assume that half of all susceptible agents prefer state $+$, and the other half state $-$, i.e. we have $n^+_+ + n^+_-=n^-_+ + n^-_-=S/2$. This assumption translates into $
    x_{+}^{+} + x_{-}^{+} = x_{+}^{-}+x_{-}^{-}=\frac{1}{2}(1-z)$. The system therefore has two independent degrees of freedom.

The dynamics are as follows. A random focal agent is selected, then a random second agent. If the focal agent is susceptible, and if the two agents are in different states, the focal agent will change state with a probability that depends on $\epsilon$. More specifically the update is implemented with probability $\frac{1}{2}(1+\epsilon)$ if the change is into the focal agent's preferred state, but only with probability $\frac{1}{2}(1-\epsilon)$ if the change is towards the dispreferred state.  Up to a rescaling of time this reduces to the standard voter model when $\epsilon=0$, i.e. in the absence of any preferences.

The rates for the events in which agents with preference for state $+$ change are
\begin{align}
\begin{split} \label{eq main: partisan x++ rates}
    T_{-\to +}^{+} &= x_{-}^{+}\big[x_{+}^{+}+x_{+}^{-}+z_{+}\big]\frac{1+\epsilon}{2}, \\
    T_{+ \to -}^{+} &= x_{+}^{+}\big[x_{-}^{-}+x_{-}^{+}+z_{-}\big]\frac{1-\epsilon}{2}.
\end{split}
\end{align}
There are analogous rates $T_{+ \to -}^{-}$ and $T_{- \to +}^{-}$ for changes among the susceptible agents with preference for state $-$. 

Recalling that there are only two independent degrees of freedom, the rate equations for $x_{+}^{+}$ and $x_-^-$ become (after rescaling time to remove an overall factor of $1/2$)
\begin{align} \label{eq main: partisan x rate eqns}
\begin{split}
    \frac{\dd x_{+}^{+}}{\dd t} &= x_{-}^{+}\big[x_{+}^{+}+x_{+}^{-}+z_{+}\big](1+\epsilon) \\
    &\quad - x_{+}^{+}\big[x_{-}^{-}+x_{-}^{+}+z_{-}\big](1-\epsilon), \\
    \frac{\dd x_{-}^{-}}{\dd t} &= x_{+}^{-}\big[x_{-}^{+}+x_{-}^{-}+z_{-}\big](1+\epsilon)  \\
    &\quad - x_{-}^{-}\big[x_{+}^{-}+x_{+}^{+}+z_{+}\big](1-\epsilon),
\end{split}
\end{align}
where $x^+_-=\frac{1}{2}(1-z)-x_{+}^{+}$ and $x^{-}_{+}=\frac{1}{2}(1-z)-x_-^-$ on the right hand side. 

Following \cite{masuda2011can, llabres2023partisan} it is convenient to trade $x^+_+$ and $x^-_-$ as independent degrees for the following quantities,
\begin{align}
\begin{split} \label{eq:delta_sigma}
    \Delta &\equiv  x_{+}^{+}-x_{-}^{-}, \\
    \Sigma & \equiv x_{+}^{+}+x_{-}^{-}.
\end{split}
\end{align}
Here, $\Sigma$ is the proportion of satisfied susceptible agents in the total population, and $\Delta$ is related to a magnetisation 
\begin{align}
    m &\equiv (x_{+}^{+}+x_{+}^{-}+z_{+}) - (x_{-}^{+}+x_{-}^{-}+z_{-}) \nonumber \\
    &= 2\Delta + z_{+}-z_{-}. \label{eq main: partisan magnetisation}
\end{align}
Since $x_{a}^{b}\in[0,1-z]$ for all $a,b \in \{\pm 1\}$, the variables $(\Delta, \Sigma)$ are restricted to a diamond shape defined by the following four inequalities,
\begin{align} 
\begin{split} \label{eq main: partisan conditions}
    \Sigma-(1-z) \leq &\Delta \leq -\Sigma+(1-z), \\
    -\Sigma \leq &\Delta \leq \Sigma.
\end{split}
\end{align}
An example is illustrated in Fig.~\ref{fig main: partisan flow} further below. Finally, the dynamics in Eqs.~(\ref{eq main: partisan x rate eqns}) can be written as
\begin{align} 
\begin{split}\label{eq main: partisan general rate eqns}
    \dot{\Delta} &= 2\epsilon\Delta(1-2\Sigma)+z_{+}\Big[(1+\epsilon)(1-2\Delta)-2\epsilon\Sigma\Big]  \\
    &\quad -z_{-}\Big[(1+\epsilon)(1+2\Delta)-2\epsilon\Sigma\Big]  \\ 
    &\quad -(z_{+}^{2}-z_{-}^{2})(1+\epsilon), \\
    \dot{\Sigma} &= (1+\epsilon)(1-z_{+}-z_{-}) \\
    &\quad -2\epsilon\Delta[z_{+}-z_{-}+2\Delta]-2\Sigma. 
\end{split}
\end{align}
When there are no zealots Eqs.~(\ref{eq main: partisan general rate eqns}) reduce to
\begin{align}
\begin{split}
    \dot{\Delta} &= \epsilon\Delta(1-2\Sigma), \\
    \dot{\Sigma} &= \frac{1}{2}(1+\epsilon)-\Sigma-2\epsilon\Delta^{2}.
\end{split}
\end{align}
Similar to the nonlinear voter model and evolutionary games these equations also have two consensus fixed points, which we will call $C^\pm$, and an interior coexistence fixed point $I$. These are given by
\begin{align}\label{eq main: partisan fixed points}
\begin{split}
    C^{+} &= (\Delta^{*}, \Sigma^{*})_{+} = \left(\frac{1}{2}, \frac{1}{2}\right), \\
    C^{-} &= (\Delta^{*}, \Sigma^{*})_{-} = \left(-\frac{1}{2}, \frac{1}{2}\right), \\
    I &= (\Delta^{*}, \Sigma^{*})_{I} = \left(0, \frac{1+\epsilon}{2}\right).
\end{split}
\end{align}
The consensus fixed points are always saddle points while the interior fixed point is always a sink for all $\epsilon$. This is similar to the behaviour of the nonlinear voter model with $q<1$, and to that of evolutionary games when $\alpha<1/2$.

\section{Fixed point analysis for the models with zealotry} \label{sec: fixed point analysis}
We have seen that the nonlinear voter model, the evolutionary game and the partisan voter model can produce similar behaviour in the absence of zealots. Namely, for $q<1$, $\alpha<1/2$ and for all $\epsilon$ we have a stable coexistence fixed point. We now study how adding different configurations of zealots affects coexistence, and/or brings about consensus.

\subsection{Zealots in both states} \label{sec: unbalanced}
\begin{figure*}[t]
    \centering
    \includegraphics[scale=0.18]{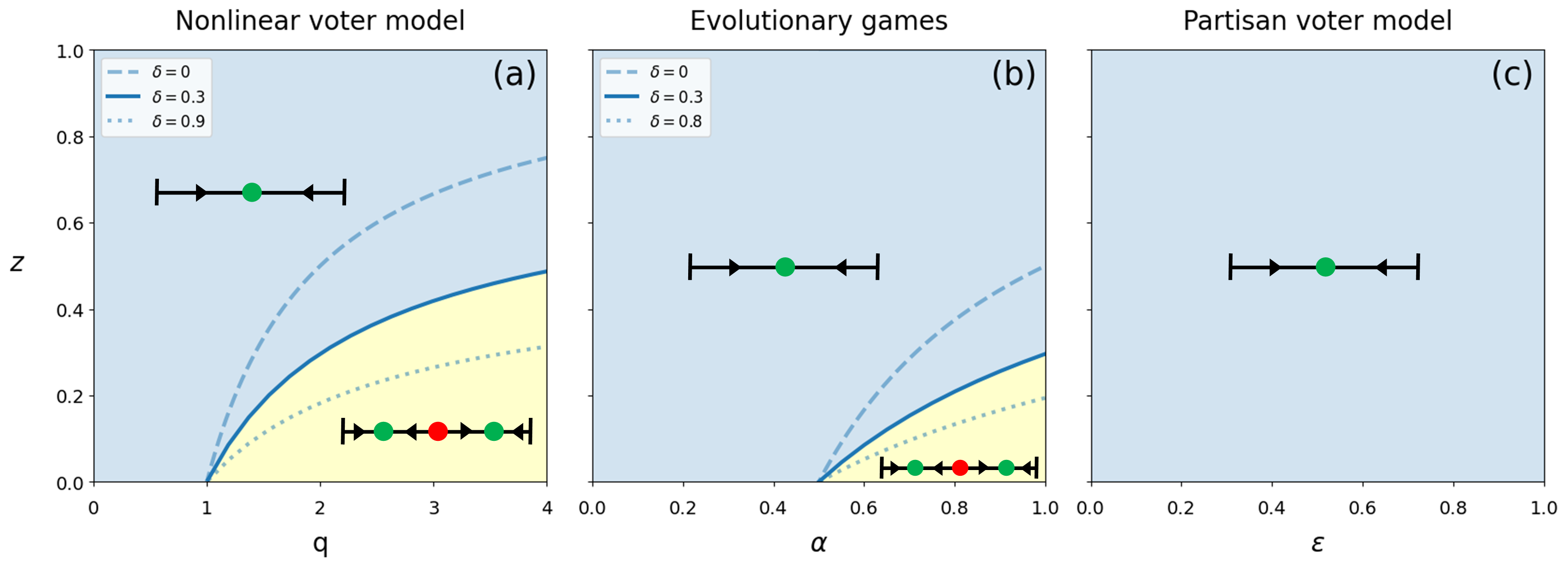}
    \caption{Phase diagrams for unbalanced zealots, i.e. $z_{\pm}=\frac{1}{2}(1\pm\delta)z$ with $z\in(0, 1)$ (not including $z=0$ and $z=1$), for three different models: (a) the nonlinear voter model $q>0$, (b) evolutionary games $\alpha\in(0,1)$, and (c) the partisan voter model $\epsilon\in(0,1)$. The blue lines are the critical zealotries $z_{c}(q)$ [Eq.~(\ref{eq supp: nonlinear balanced zC})] and $z_{c}(\alpha)$ [Eq.~(\ref{eq supp: games balanced zC})]. For the partisan voter model with unbalanced zealots there is no critical zealotry. We show the critical zealotry lines for three different $\delta$ values in panels (a) and (b), but the phase diagram is with respect to the solid line in each case, hence the other lines are faded. In each segment we give an illustrative representation of the flow of the density of $+1$ susceptible agents. These flows are always bound between $[0, 1-z]$. Green/red circles indicate stable/unstable fixed points. The arrows indicate the direction of flow. The background and line colours correspond to the flow classification for easy comparison between panels and other figures. The flow for the partisan voter model is really two-dimensional (see Fig.~\ref{fig main: partisan flow}), but here we show a reduced one-dimensional representation.}
    \label{fig main: unbalanced phase plot}
\end{figure*}

Consensus is possible only if there are zealots in only one of the two states (or if there are no zealots at all). However, for completeness, we briefly consider the case of zealots in both states, $z_{\pm}=\frac{1}{2}(1\pm\delta)z$ with $\delta\in(0, 1)$. Further details of the analysis can be found in Appendices~\ref{appendix: nonlinear unbalanced}, \ref{appendix: games unbalanced} and \ref{appendix: partisan unbalanced}. We also consider the limiting case $\delta=0$ which is known as `balanced zealots' \cite{khalil2018zealots} (see Appendices~\ref{appendix: nonlinear balanced}, \ref{appendix: games balanced} and \ref{appendix: partisan balanced}). 

We note the symmetry with respect to $\delta \to -\delta$ in all three models (this results in a simple renaming of the two states $+$ and $-$). 

In the nonlinear voter model, and for evolutionary games, we find two different types of deterministic flow, as shown in Figs.~\ref{fig main: unbalanced phase plot}(a) and (b).

In the nonlinear voter model the central coexistence fixed point is stable for values of $q$ below some threshold and/or zealotry above some critical value for $z$ [blue region in Fig.~\ref{fig main: unbalanced phase plot}(a)]. Similarly, in the evolutionary game, the coexistence fixed point is stable for sufficiently small values of $\alpha$ and/or sufficiently many zealots. Otherwise, we find it is unstable, and that there are two further stable internal fixed points, see the illustration of the flow in the yellow areas in Fig.~\ref{fig main: unbalanced phase plot}(a) and (b). These fixed points can be characterised as ``polarised coexistence states''. Both opinion states are present in the population but typically in unequal proportions. These fixed points arise as a balance of two competing effects. First, zealotry for both opinions means that the deterministic flow is away from the boundary states $x=0$ and $x=1-z$. Second, superlinear social impact ($q>1$) provides a drive towards the majority opinion and hence to consensus. We calculated the critical values that separate the blue and yellow regions numerically from the rate equations (see Appendices~\ref{appendix: nonlinear unbalanced} and \ref{appendix: games unbalanced}). 

For the partisan voter model with zealots in both states there is only one type of flow [see Fig.~\ref{fig main: unbalanced phase plot}(c), and Appendix~\ref{appendix: partisan balanced}]. In this case there is no critical amount of zealotry. The central fixed point is always the only fixed point of the rate equations, and it is stable.

\subsection{Zealots in one state only} \label{sec: fixed point analysis absorbing}
We now focus on the case of zealotry in one state only. Without loss of generality we make this the $+$ state, and set $z_{-}=0$ and $z_{+}=z$. 

\begin{figure*}[hbtp]
    \centering
    \includegraphics[scale=0.18]{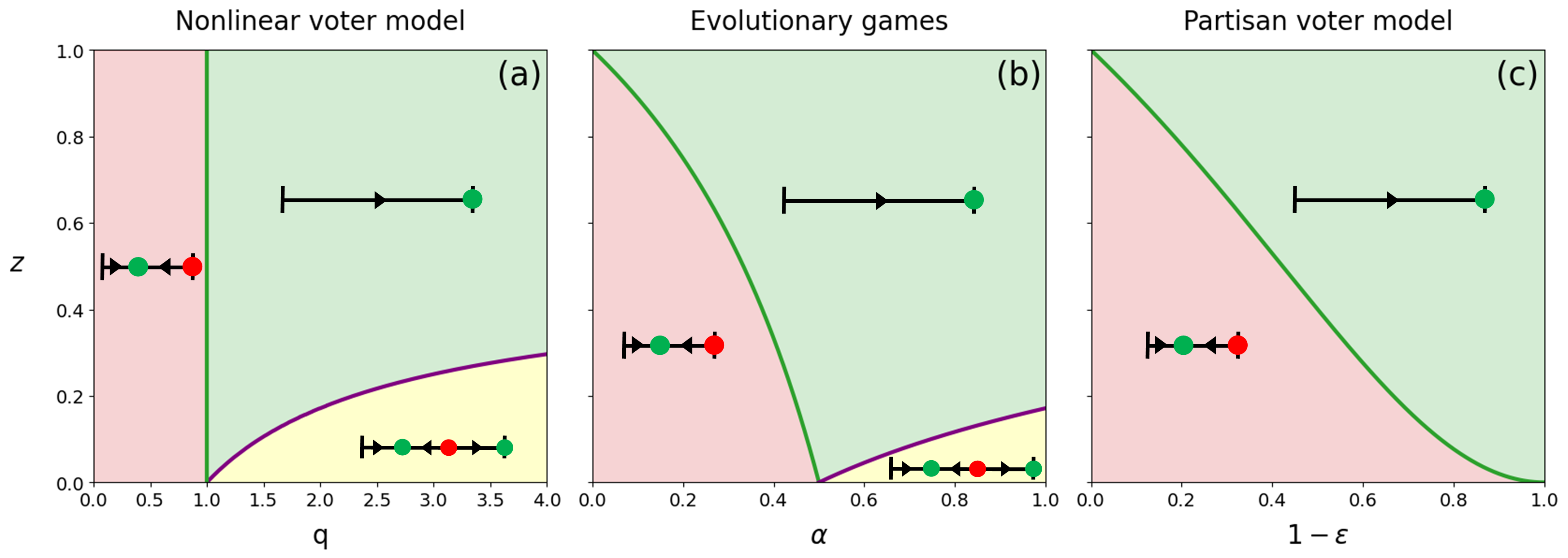}
    \caption{Phase diagrams for one-sided zealots, i.e. $z_{-}=0$ and $z_{+}=z$ with $z\in(0,1)$ (not including $z=0$ and $z=1$), for three different models: (a) the nonlinear voter model $q>0$, (b) evolutionary games $\alpha \in (0, 1)$, and (c) the partisan voter model $\epsilon \in (0,1)$. We note that the $x$-axis in panel (c) is $1-\epsilon$ so the regions can be better compared with panels (a) and (b). The solid lines dividing the regions are the critical zealotries defined in Eqs.~(\ref{eq main: nonlinear zC}), (\ref{eq main: games zC}) and (\ref{eq main: partisan zC}) for panels (a), (b) and (c) respectively. In each segment we give an illustrative representation of the flow of the density of $+1$ susceptible agents. These flows are always bound between $[0, 1-z]$. Green/red circles indicate stable/unstable fixed points. The arrows indicate the direction of flow. The background and line colours correspond to the flow classification for easy comparison between panels and other figures. On the purple lines one finds a marginal flow with a bistable interior fixed point [see e.g. Fig.~\ref{fig supp: nonlinear absorbing flow}(d) in the Appendix]. The flow for the partisan voter model is really two-dimensional (see Fig.~\ref{fig main: partisan flow}), but here we show a one-dimensional representation.}
    \label{fig main: absorbing phase plot}
\end{figure*}

For all three models, and in the deterministic limit governed by the rate equations in Eqs.~(\ref{eq main: nonlinear vm general rate eqn}), (\ref{eq main: games general rate eqn}) and (\ref{eq main: partisan general rate eqns}) with $z_-=0$ and $z_+ = z$, consensus is the only outcome for certain combinations of parameters (green regions in Fig.~\ref{fig main: absorbing phase plot}). For evolutionary games and the partisan voter model [panels (a) and (b)], consensus is the guaranteed outcome in infinite populations for large enough $z$ for any $\alpha$ or $\epsilon$ respectively. In the nonlinear voter model there is an additional requirement, $q\geq1$.
 
The critical values of $z$ can be determined analytically. For evolutionary games we find (see Appendix~\ref{appendix: games absorbing})
\be \label{eq main: games zC}
    z_{c}(\alpha) =
    \begin{cases}
        \frac{1-2\alpha}{1-\alpha} \hspace{15mm} \text{if} \quad \alpha<\frac{1}{2}\\
        \frac{2\alpha+1-2\sqrt{2\alpha}}{2\alpha-1} \hspace{6mm} \text{if} \quad \alpha>\frac{1}{2} 
    \end{cases},
\ee
These conditions describe the lower edge of the green region in Fig.~\ref{fig main: absorbing phase plot}(b). 

For the partisan voter model we have (Appendix~\ref{appendix: partisan absorbing})
\be
    z_{c}(\epsilon) = \frac{2\epsilon^{2}}{1+\epsilon^{2}}.\label{eq main: partisan zC}
\ee
This is the line separating the red and green regions in Fig.~\ref{fig main: absorbing phase plot}(c). We note that $z_{c}(\epsilon=0)=0$ and $z_c(\alpha=1/2)=0$. In these cases the two models reduce to the standard voter model, and any finite fraction of one-sided zealotry is sufficient to enforce consensus in infinite populations.

For the nonlinear voter model we can also analytically derive the critical level of zealotry separating the green and yellow regions in Fig.~\ref{fig main: absorbing phase plot}(b); this critical zealotry only exists if $q>1$. We find a converging infinite series (see Appendix~\ref{appendix: nonlinear absorbing})
 \begin{eqnarray} \label{eq main: nonlinear zC}
    z_{c}(q)&=&\frac{q}{q-1}\sum_{k=0}^{\infty}\left\{\frac{(-1)^{k}}{k!}\left[\frac{q-1}{q^\frac{q}{q-1}}\right]^{k+1} \right. \nonumber \\
    && \left. \times\frac{\Gamma\left(\frac{q(k+1)}{q-1}\right)}{\Gamma\left(\frac{q(k+1)}{q-1}-k+1\right)}\right\}.
\end{eqnarray}
This extends work done in Ref.~\cite{mobilia2015nonlinear} where $z_{c}(q)$ was given numerically for integer $q$. For $q<1$ the rate equations have an unstable fixed point $x^*=1-z$ and a stable internal fixed point, for all levels of zealotry. Again we note that $q=1$ is the standard voter model so the system always reaches consensus for any finite fraction $z>0$ of one-sided zealotry.

We note that Eq.~(\ref{eq main: nonlinear zC}) is not easy to work with. As an alternative, an accurate closed-form approximation is given by (see Appendix~\ref{appendix: derivation of approx} for a derivation)
\be
    z_c(q) = (q-1)(2q-1)^{\frac{q-\frac{1}{3}}{1-q}}.
\ee

Figure~\ref{fig main: absorbing phase plot} shows the phase behaviour of the three models in the case of one-sided zealots. In the nonlinear voter model [panel (a)] a sufficiently large group of one-sided zealots enforces consensus for all initial conditions, but only provided $q>1$. In the evolutionary games we look at, and in the partisan voter model, consensus is always the outcome with sufficiently many zealots (for any $\alpha$ or $\epsilon$ respectively). We note that in the red regions of Fig.~\ref{fig main: absorbing phase plot}, the stable interior fixed point smoothly approaches the unstable consensus fixed point as $z$ increases [see the bifurcation diagram in Fig.~\ref{fig supp: nonlinear absorbing bifurcation}(a) in the Appendix].

\begin{figure}[hbtp]
    \centering
    \includegraphics[scale=0.3]{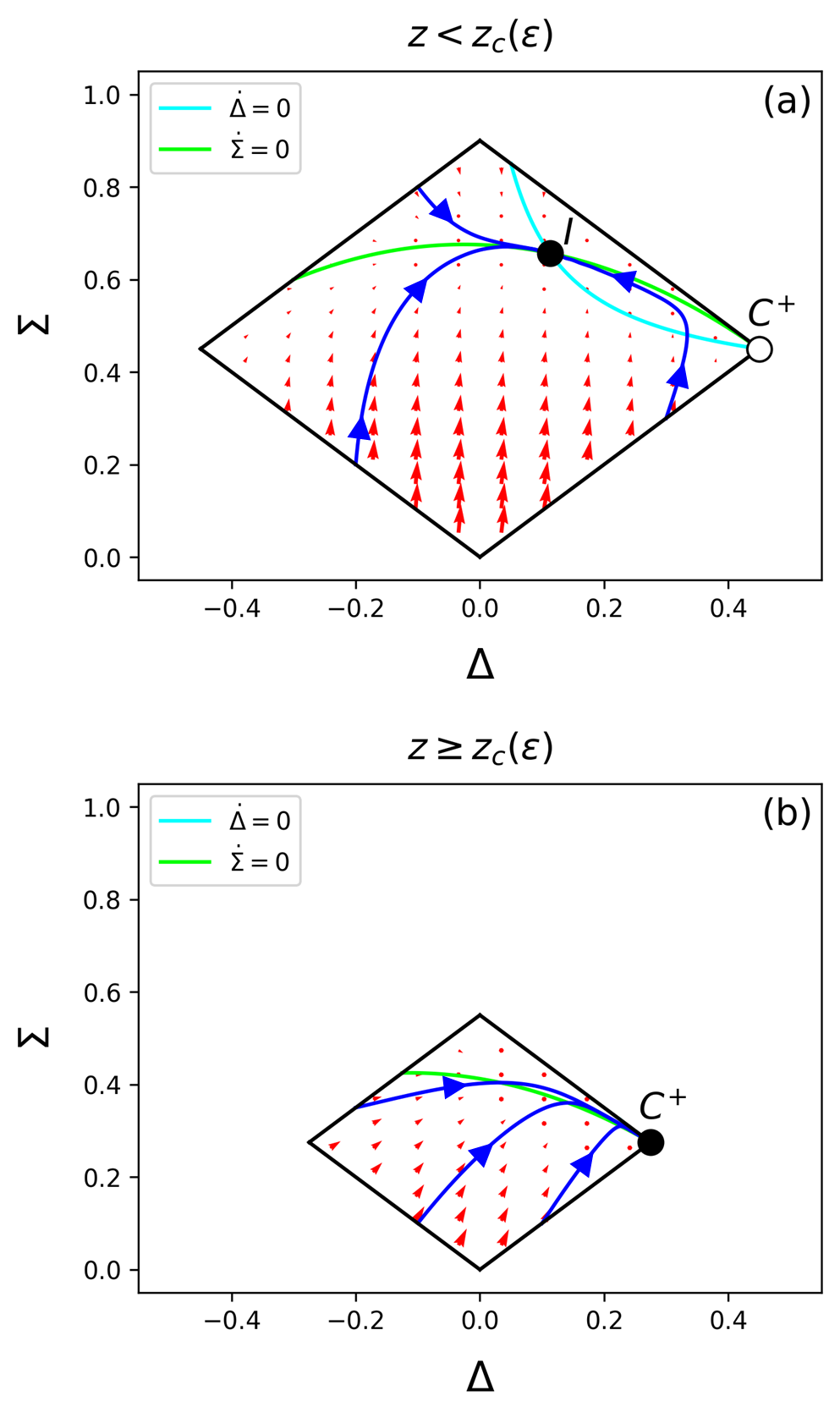}
    \caption{Flow diagrams in the ($\Delta$, $\Sigma$)-space for the partisan voter model with one-sided zealots. $z=(0.1, 0.45)$ in panels (a) and (b), and $\epsilon=0.5$ in both. The rectangular region is the allowed region defined by Eqs.~(\ref{eq main: partisan conditions}). Red arrows indicate the flow field defined by Eqs.~(\ref{eq main: partisan general rate eqns}), the larger the arrow the faster the flow. Empty (filled) circle markers are saddle (sink) fixed points (see \cite{github_zealots} for details). The dark blue lines are example deterministic trajectories starting from various points in the plane, determined by numerically integrating Eqs.~(\ref{eq main: partisan general rate eqns}). The cyan line is the $\dot{\Delta}=0$ nullcline [Eq.~(\ref{eq supp: partisan absorbing delta=0 nullcline})]. The lime green line is the $\dot{\Sigma}=0$ nullcline [Eq.~(\ref{eq supp: partisan absorbing sigma=0 nullcline})].}
    \label{fig main: partisan flow}
\end{figure}

The flow diagrams (black arrows) in Fig.~\ref{fig main: absorbing phase plot}(c) are one-dimensional illustrations of the of the full flows for the two degrees of freedom $\Delta$ and $\Sigma$ of the partisan voter model, broadly indicating if the rate equations converge to consensus or to a state in which both opinions coexist. In Fig.~\ref{fig main: partisan flow} we show examples of the full flow in $(\Delta, \Sigma)$-space. Figures~\ref{fig main: partisan flow}(a) and (b) correspond to the red and green regions respectively in Fig.~\ref{fig main: absorbing phase plot}(c). 

\section{Nonlinearity vs zealotry: the effects of social impact} \label{sec: social impact}
As apparent from Fig.~\ref{fig main: absorbing phase plot}, one-sided zealotry in evolutionary games and the partisan voter model can enforce consensus for any value of their respective  parameters ($\alpha$ and $\epsilon$), provided that $z$ is larger than some critical value. For the nonlinear voter model however, this is only the case for $q>1$. When $q<1$ no amount of zealotry can guarantee consensus for all initial conditions, in fact the consensus fixed point always remains linearly unstable. 

We now ask what it is in the nonlinear voter model that produces this outcome. A partial answer to this question is related to what we will call the `social impact function'. As we will see, the shape of this function determines the local stability of consensus fixed points. To explain this, we focus  on the case of one-sided zealots ($z_{-}=0$ and $z_{+}=z$).

\subsection{Definition of a social impact function} \label{sec: social impact function}

Social impact theory was first developed by Latan\'e \cite{latane}. He describes social impact as `\textit{the effect of other persons on an individual}' and proposed that impact on a target individual `\textit{should be a multiplicative function of the strength, immediacy, and number of other people}'. Latan\'e asserted that the strength of social impact by a group scales as some power of the size of the group. Experiments by Latan\'e suggest that social impact grows with power of less than one in certain circumstances. Specifically, he reports approximate values of $0.24 $ and $0.46 $ for example \cite{latane}. Studies of the dynamics of languages, suggested an exponent of $1.31$, see Ref.~\cite{abrams2003modelling}.

In his original work, Latan\'e did not consider any dynamical models of social impact, and so the empirical values for the exponents he reports cannot directly be interpreted as the exponent $q$ in the nonlinear voter model. We point to early collaborations of Latan\'e with physicists though to model social impact \cite{lewenstein}. We use Latan\'e's empirical findings as a motivation to study the effects of nonlinear impact in voter models and evolutionary games.

The idea of social impact can be implemented using the conventional (linear) and the nonlinear voter models. In the conventional voter model (and again focusing on all-to-all interaction) a focal agent is selected and then adopts the opinion of a randomly selected individual from the population. This means that social impact is linear, i.e. the probability to adopt a given opinion state is proportional to the first power of its prevalence in the population. The rate with which the prevalence $x$ of an opinion increases is thus given by $(1-x) x$, where the first factor describes the probability that the agent selected for update does not hold the opinion, and the second factor describes linear social impact. If we define a social impact function $f(x)$, then one would have $f(x)=x$ in the case of the conventional voter model. These ideas can naturally be extended to dynamics on networks. What is relevant then is the local prevalence of the opposing opinion among the neighbours of the focal agent.

In the nonlinear voter model social impact is of the form $f(x)=x^q$, that is, the probability to adopt a certain opinion state increases with the $q^{\rm th}$ power of its prevalence. For integer $q$ this can be interpreted as consulting with a group of $q$ individuals and adopting their opinion state only if they all agree with one another \cite{castellano2009nonlinear}.

In the following we will ask what features the social impact function $f(\cdot)$ must have so that zealotry can enforce the local stability of consensus in the rate equations. We always require $f(0)=0$, a vanishing group of agents cannot have any impact. We also demand $f(1)=1$, to normalise social impact.

\subsection{The nonlinear voter model} \label{sec: social impact nonlinear vm}
\subsubsection{Stability condition for consensus}
We can generalize the rate equation for the nonlinear voter model [Eq.~(\ref{eq main: nonlinear vm general rate eqn})] with one-sided zealots as follows,
\be 
    \frac{\dd x}{\dd t} = (1-x-z)f(x+z)-xf(1-x-z), \label{eq main: nonlinear social impact rate eqn}
\ee
where $f(x)$ is the social impact function. Given that $f(0)=0$, we find that consensus at opinion state $+$, i.e. $x^{*}=1-z$, is a fixed point for all $q$, as indicated in Fig.~\ref{fig main: absorbing phase plot}(a).

For the consensus fixed point to be locally stable, we require $F'(1-z)<0$, where $F(x)$ is the right-hand side of Eq.~(\ref{eq main: nonlinear social impact rate eqn}) (and where $F'$ is the derivative of $F$). From this we find the condition
\be
    (1-z) f'(0) < 1. \label{eq main: nonlinear vm stable consensus condition}
\ee
Thus, the local stability of the consensus fixed point is governed by two factors: (i) the zealotry driving the system towards consensus must be sufficiently strong ($z$ must be sufficiently large), and (ii) the social impact function must not be too steep at small group sizes.

The first of these factors is intuitive, the stronger the zealotry supporting an opinion the more stable consensus at that opinion will be. To understand the second factor, we imagine the system being close to consensus on one opinion, with a small proportion of agents $y$ supporting the other opinion state. The impact of that group is then of magnitude $f(y)$. The minority group may be small, but if their impact is sufficiently large, they may prevent convergence to consensus. Hence $f(y)$ must be sufficiently small. For small arguments this translates into a condition on $f'(y=0)$.

\subsubsection{Consequences in the stability diagram}
In the nonlinear voter model we have $f(x)=x^{q}$, and therefore $f'(x)=qx^{q-1}$. Thus, for any $q<1$ we find $f'(x)\to+\infty$ for $x\downarrow 0$. Therefore the condition in Eq.~(\ref{eq main: nonlinear vm stable consensus condition}) cannot hold for any $q<1$. This means that the consensus fixed point $x^{*}=1-z$ can never be locally stable for $q<1$, in-line with Fig.~\ref{fig main: absorbing phase plot}(a). For $q>1$ on the other hand, $f'(0)=0$, i.e. the social impact of small contrarian groups is small. Any level of one-sided zealotry then enforces local stability of the consensus fixed point, as also indicated in Fig.~\ref{fig main: absorbing phase plot}(a).

We note that the separations of the green and yellow regions in Fig.~\ref{fig main: absorbing phase plot}(a) (corresponding to the absence or presence of a coexistence fixed point) cannot be obtained from a local analysis of the consensus fixed point. Instead an analysis along the lines of Appendix~\ref{appendix: nonlinear absorbing} is required. Nonetheless, existence and stability of such an internal fixed point is also determined by the shape of social impact function $f(\cdot)$.

\subsection{Evolutionary games}
\subsubsection{Stability condition for consensus}
The rate equation for evolutionary games [Eq.~(\ref{eq main: games general rate eqn})] can be generalised as follows, to account for a social impact function $f$,
\begin{eqnarray}
    \frac{\dd x}{\dd t} &=& (1-x-z)f(x+z)\pi_+(x,z) \nonumber \\
    &&- xf(1-x-z)\pi_-(x,z). \label{eq main: games social impact rate eqn}
\end{eqnarray}
The $\pi_{\pm}$ are given in Eqs.~(\ref{eq main: games expected payoff +}) and (\ref{eq main: games expected payoff -}), with $z_+=z$ and $z_-=0$. 

We have anticipated a general social impact function $f(\cdot)$, that in the evolutionary dynamics can be thought of as follows. A focal agent is chosen for potential update. The update is then considered further with a probability that depends on the prevalence of the strategy opposite to that of the focal agent. The function $f$ quantifies this dependence [in conventional evolutionary games we have $f(x)=x$]. Then in a third step the update is implemented with a rate proportional to the expected payoff of the strategy that the agent considers switching to.

For the game defined by the payoff matrix in Eq.~(\ref{eq main: payoff matrix}) a linear stability analysis shows that the consensus fixed point $x^{*}=1-z$ is locally stable if
\be
    (1-z) \, f'(0) \, \frac{1-\alpha}{\alpha} < 1. \label{eq main: games f'(0) condition}
\ee
Thus, local stability of consensus is now governed by three factors. As in the voter model, there needs to be sufficiently strong zealotry and the social impact of small minority groups must not be too large. Thirdly, we note the factor $(1-\alpha)/\alpha$. This is the ratio of the payoff received from playing against a player with the opposite strategy to one's own, divided by the payoff received from playing against a player with the same strategy as one's own [see the payoff matrix in (\ref{eq main: payoff matrix})]. Thus, the larger the ratio $(1-\alpha)/\alpha$ the stronger the evolutionary pressure away from the consensus (or coordination) point. 

\subsubsection{Evolutionary dynamics with linear social impact} \label{sec: games linear social impact}
In conventional evolutionary games we have linear social impact, $f(x)=x$, and the condition for locally stable consensus reduces to
\be
    (1-z)\frac{1-\alpha}{\alpha}<1. \label{eq main: games stable consensus condition}
\ee
For $\alpha>1/2$ this condition is always fulfilled (for all $z$) as $(1-\alpha)/\alpha$ is then smaller than one. Thus the consensus fixed point $x^*=1-z$ is locally stable. This describes a situation in which the underlying game is of the coordination type. Both the evolutionary dynamics and the zealotry then drive the system towards consensus, provided $x$ is sufficiently close to $1-z$. We note that the game itself also promotes local stability of the fixed point at $x=0$ (corresponding to consensus at the strategy not supported by any zealots). For small zealotry [yellow region in Fig.~\ref{fig main: absorbing phase plot}(b)] the flow is towards $x=0$ for a sufficiently small initial fraction $x$ of players with strategy $+$. This is due to the unstable internal fixed point.

For $\alpha<1/2$, the game is of the coexistence type, i.e. the evolutionary dynamics itself (without zealotry) is towards a state in which both strategies are present. One-sided zealotry on the other hand pulls the population towards the consensus state at the $+$ strategy. The overall outcome is thus governed by the balance of these two `forces'. The condition for locally stable consensus in Eq.~(\ref{eq main: games stable consensus condition}) can be written as $z > (1-2\alpha)/(1-\alpha)$, which is the first equation in Eq.~(\ref{eq main: games zC}). This is the condition separating the green and red regions in Fig.~\ref{fig main: absorbing phase plot}(b), which have locally stable or unstable consensus fixed points respectively.

\subsubsection{Evolutionary games with nonlinear social impact}
We can also consider the case where the social impact function is nonlinear. We focus again on $f(x)=x^q$, noting that $q$-deformed evolutionary dynamics without zealots were previously studied in Ref.~\cite{kitching2024qdeformed}. The $q$-deformation is akin to nonlinear social impact, and as shown in Ref.~\cite{kitching2024qdeformed} this can produce types of deterministic dynamics not seen in conventional evolutionary $2\times 2$ games. For example one finds  bistable coexistence, or mixed coordination and coexistence. Here, we focus on scenarios with (one-sided) zealotry, i.e. we always assume $z>0$.

\begin{figure}[hbtp]
    \centering
    \includegraphics[scale=0.2]{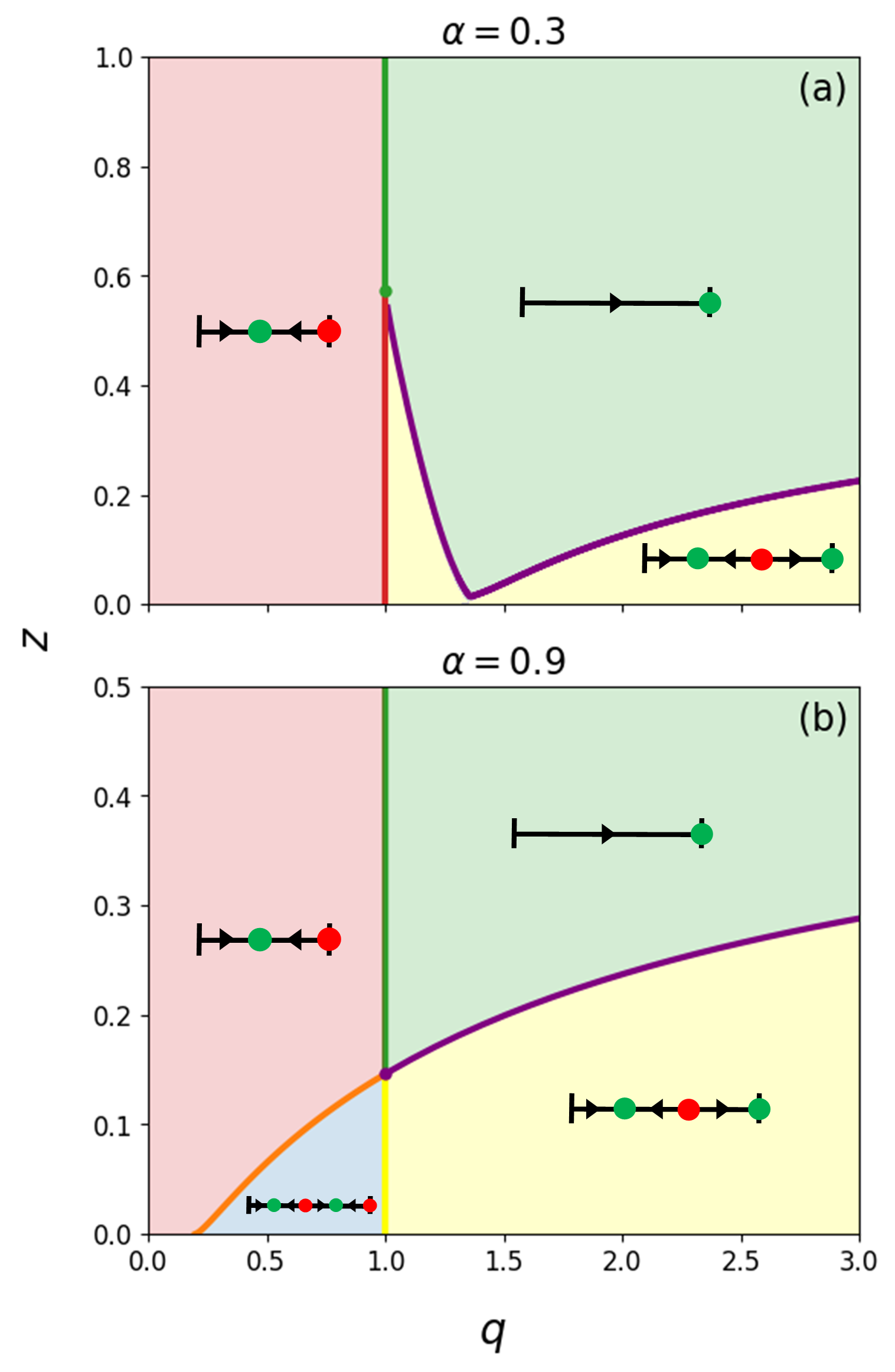}
    \caption{Phase diagrams in ($q$, $z$)-space [with $z\in(0,1)$] for nonlinear evolutionary games with one-sided zealots at (a) $\alpha=0.3$ and (b) $\alpha=0.9$. The solid lines dividing the regions are determined numerically (see Appendix~\ref{appendix: nonlinear games}). In each segment we give an illustrative representation of the flow of the density of $+1$ susceptible agents. These flows are always bound between $[0, 1-z]$. Green/red circles indicate stable/unstable fixed points. The arrows indicate the direction of flow. The background and line colours correspond to the flow classification for easy comparison between panels and other figures. When sitting exactly on the purple and orange lines one gets a marginal flow with a bistable interior fixed point [see e.g. Fig.~\ref{fig supp: nonlinear absorbing flow}(d) in the Appendix]. We also note that there is a small region near the minimum of the purple line in panel (a) where there exist four interior fixed points. Further marginal flows exist at the boundary of this region.}
    \label{fig main: nonlinear games}
\end{figure}

The condition in Eq.~(\ref{eq main: games f'(0) condition}) is violated when $q<1$ for all $\alpha$, so the consensus fixed point is always locally unstable when $q<1$. Except for the additional consensus fixed point, going from the blue to red region in Fig.~\ref{fig main: nonlinear games} is topologically similar to the transition in Figs.~\ref{fig main: unbalanced phase plot}(a) and (b) where the coexistence fixed point changes stability and two additional stable fixed points appear. The interpretation here is that sublinear impact amplifies the minority state, and prevents the system reaching consensus at small $z$. As $z$ increases the one-sided zealotry overpowers this minority amplification resulting in a single coexistence fixed point that smoothly approaches the upper boundary.

When $q>1$, Eq.~(\ref{eq main: games f'(0) condition}) is satisfied for all $\alpha$ and the consensus fixed point is always locally stable. This is indeed what we see in Fig.~\ref{fig main: nonlinear games}. 

The shape of the deterministic flow beyond the stability of consensus cannot be determined from linear stability analysis alone, for further details see Appendix~\ref{appendix: nonlinear games}.

\subsection{The partisan voter model} \label{sec: social impact partisan vm}
\subsubsection{Stability criterion for general social impact}
The analysis of the effects of nonlinear social impact for the partisan voter model is more involved as there are now two degrees of freedom. Allowing for a general social impact function, Eqs.~(\ref{eq main: partisan x rate eqns}) turn into
\begin{align}
    \frac{\dd x_{+}^{+}}{\dd t} &= x_{-}^{+}f\big(x_{+}^{+}+x_{+}^{-}+z\big)(1+\epsilon) \nonumber \\
    &\quad - x_{+}^{+}f\big(x_{-}^{-}+x_{-}^{+}\big)(1-\epsilon),
\end{align}
and a similar relation for $x_-^-$.  The rate equations for $\Delta$ and $\Sigma$ [Eqs.~(\ref{eq main: partisan general rate eqns})] with one-sided zealots become
\begin{align}
\begin{split} \label{eq main: partisan social impact rate eqn}
    \dot{\Delta} &= \Big[(1+\epsilon)(1-z)-2(\Delta+\epsilon\Sigma)\Big]\times f\left(\frac{1+z}{2}+\Delta\right) \\
    &\quad - \Big[(1+\epsilon)(1-z)+2(\Delta-\epsilon\Sigma)\Big] \times f\left(\frac{1-z}{2}-\Delta\right), \\
    \dot{\Sigma} &= \Big[(1+\epsilon)(1-z)-2(\epsilon\Delta+\Sigma)\Big]  \times f\left(\frac{1+z}{2}+\Delta\right) \\
    &\quad + \Big[(1+\epsilon)(1-z)+2(\epsilon\Delta-\Sigma)\Big] \times f\left(\frac{1-z}{2}-\Delta\right),
\end{split}
\end{align}
where we have rescaled time to remove an overall factor of $1/2$. Using the assumption that $f(0)=0$, one finds the consensus fixed point
\be\label{eq main: partisan C+}
    C^{+} = \left(\frac{1-z}{2}, \frac{1-z}{2}\right),
\ee
as in the model with linear social impact [for details see Eq.~(\ref{eq supp: partisan absorbing C+})]. To determine its stability we evaluate the eigenvalues of the Jacobian at $C^+$ using Eqs.~(\ref{eq main: partisan social impact rate eqn}). We find the eigenvalues have strictly negative real parts, and hence $C^+$ is stable, when (see Appendix~\ref{appendix: partisan social impact})
\be
    (1-z) f'(0) \frac{1+\epsilon^2}{1-\epsilon^2} < 1. \label{eq main: partisan f'(0) condition}
\ee
Again we note that strong zealotry (large $z$) and a slowly increasing social impact function [$f'(0)$ small] promote stability of the consensus point. We also note that $(1+\epsilon^2)/(1-\epsilon^2)$ is an increasing function of $\epsilon$. Thus strong affinity towards a preferred state limits local stability of consensus. We attribute this to agents with a preference opposite to the state of the zealots becoming increasingly resistant with increasing $\epsilon$, thus preventing consensus.  

For linear social impact, $f(x)=x$, the condition (\ref{eq main: partisan f'(0) condition}) reduces to
\be\label{eq:condition_nlpvm}
    z > \frac{2\epsilon^2}{1+\epsilon^2},
\ee
which reproduces the critical zealotry from Eq.~(\ref{eq main: partisan zC}).

\subsubsection{Nonlinear partisan voter model with \texorpdfstring{$f(x)=x^q$}{}} 
Focusing on a nonlinear partisan voter model with impact function $f(x)=x^q$ (see also Ref.~\cite{jaume_future}), 
Eq.~(\ref{eq main: partisan f'(0) condition}) shows that $C^+$ is unstable for $q<1$, and stable for $q>1$, irrespective of the preference parameter $\epsilon$, and the amount of zealotry. An analysis of the coexistence fixed point (see Appendix~\ref{appendix: nonlinear partisan}) reveals that coexistence is always a local attractor when $q<1$. For $q>1$ the coexistence fixed point is linearly stable when $\epsilon>\sqrt{(q-1)/q}$, as also pointed out in Ref.~\cite{jaume_future}.

\section{Finite populations}\label{sec:finite}
So far we have focused on infinite populations. However, the stochastic nature of dynamics in finite populations gives rise to potentially quite different behaviour. For example, in the red regions of Fig.~\ref{fig main: absorbing phase plot} the deterministic system has an interior stable fixed point, and the consensus fixed point at $1-z$ is unstable. However, for one-sided zealots, the consensus point is the only absorbing state in finite populations. Thus, we expect individual trajectories for finite systems, starting with no susceptible agents in the $+$ state, to move to states near the stable interior fixed point, fluctuate around this fixed point for some time, before a large fluctuation causes absorption at the positive consensus state. This is indeed what is seen in simulations. 

We now proceed to characterise the mean time-to-fixation in these circumstances, and the quasi-stationary distribution. This is the long-term distribution of opinions in the population, conditioned on not having reached an absorbing consensus state.

\subsection{Fixation times} \label{sec: fixation times}
\subsubsection{Analytical calculation of mean fixation times} 
We now present analytical calculations of mean fixation time. This is restricted to the nonlinear voter model and evolutionary games. This is because for these models there is only one integer degree of freedom for the stochastic population dynamics. We do however present numerical results for the partisan voter model.

We define $t_i$ as the average time a system takes to reach fixation starting with $i$ agents in the state $+1$. An analytical expression for $t_i$ can be found by adapting the calculation reported for example in Ref.~\cite{traulsen2009stochastic} to a scenario with only one absorbing state at $n=N$. One finds (see Appendix~\ref{appendix: fixation times deriv})
\be
    t_i = \frac{1}{T_0^+}\sum_{k=i}^{N-Z-1}\prod_{m=1}^{k}\gamma_m + \sum_{k=i}^{N-Z-1}\sum_{l=1}^{k}\frac{1}{T_l^+}\prod_{m=l+1}^{k}\gamma_m, \label{eq main: fixation time}
\ee
where $\gamma_n\equiv T_{n}^{-}/T_{n}^{+}$, and where the $T_{n}^{\pm}$ are the rates for transitions $n\to n\pm 1$. 

\subsubsection{Nonlinear voter model with one-sided zealots} 
For the nonlinear voter model with one-sided zealots the transition rates are defined analogously to Eqs.~(\ref{eq main: nonlinear vm rates})
\begin{align}
\begin{split} \label{eq main: nonlinear vm finite rates}
    T_{n}^{+} &= N\frac{S-n}{N}\left(\frac{n+Z}{N-1}\right)^{q}, \\
    T_{n}^{-} &= N\frac{n}{N}\left(\frac{N-Z-n}{N-1}\right)^{q}.
\end{split}
\end{align}

\begin{figure}[htbp]
    \centering
    \includegraphics[scale=0.57]{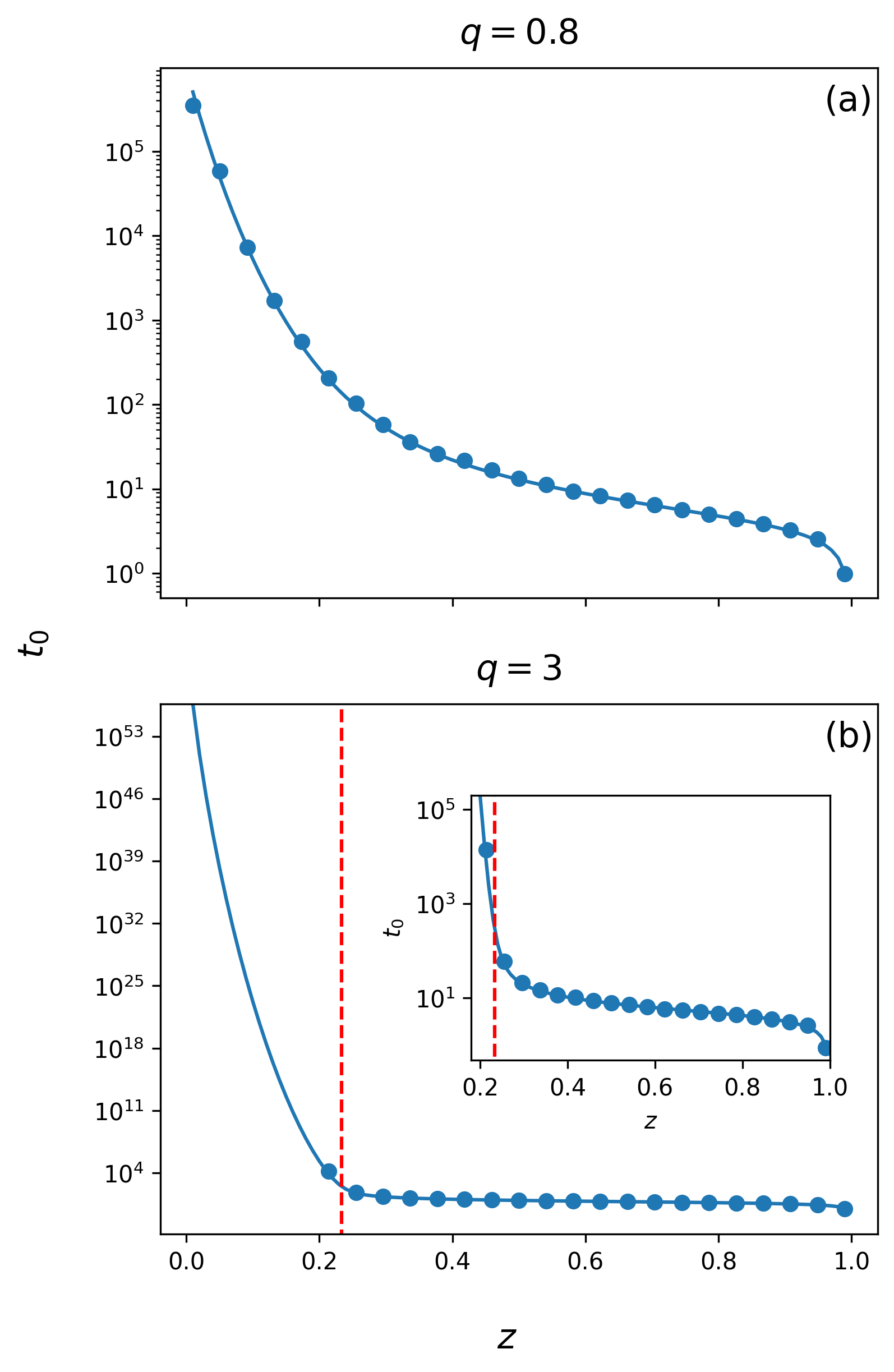}
    \caption{Plots of fixation time $t_0$ against $z$ for the nonlinear voter model with zealots in a finite population of $N=100$. $t_0$ is defined as the time it takes the system to reach fixation, i.e. all susceptibles are type $+1$, starting from an initial configuration where all susceptibles are type $-1$. We note that the $t_0$ axis is logarithmic. The blue dots are from averaging 1000 independent simulations using the Gillespie algorithm \cite{gillespie1976general} and the rates in Eqs.~(\ref{eq main: nonlinear vm finite rates}). We note that in panel (b) we are only able to perform simulations above a particular $z$, as the fixation times become too extreme below this. The solid blue lines are the analytical solutions from Eq.~(\ref{eq main: fixation time}). Panels (a) and (b) are for $q=0.8$ and $q=3$ respectively. In panel (b) we indicate the critical zealotry $z_c(q)$ [Eq.~(\ref{eq main: nonlinear zC})] for a corresponding infinite population by a vertical dashed red line. Panel (b) also contains an inset which is a zoomed-in version of the main panel (b).}
    \label{fig main: nonlinear absorption}
\end{figure}

While Eq.~(\ref{eq main: fixation time}) provides the mean time to fixation for an arbitrary initial condition, we focus on $t_0$ in Fig.~\ref{fig main: nonlinear absorption}. The starting point $i=0$ is the state furthest away from consensus at $n=N$, hence $t_0$ gives an upper bound to the mean fixation time from any initial state. In Fig.~\ref{fig main: nonlinear absorption} we plot $t_0$ against $z$ for different values of $q$. In panel (a) we have $q<1$, so the flow in an infinite population is of the type as shown in the red region of Fig.~\ref{fig main: absorbing phase plot}(a). There is one stable internal fixed point, the consensus fixed point is locally unstable. Increasing the zealotry (i.e. $z$) moves the interior fixed point closer to the consensus one. Thus, we expect the fixation time $t_0$ to smoothly decrease with $z$. This is indeed  what we see in Fig.~\ref{fig main: nonlinear absorption}(a). 

In Fig.~\ref{fig main: nonlinear absorption}(b) we choose $q>1$. In an infinite population there is then a critical level of zealotry $z_c(q)$ [Eq.~(\ref{eq main: nonlinear zC})]. Above this level of zealotry the deterministic flow is towards the consensus state [green region in Fig.~\ref{fig main: absorbing phase plot}(a)]. For $z<z_c$, the consensus state is also locally stable, but there is an unstable internal fixed point [yellow region in Fig.~\ref{fig main: absorbing phase plot}(a)]. We thus expect the fixation time $t_0$ for $z>z_c(q)$ to be much lower than that that for $z<z_c(q)$. In the former case the system is attracted straight to consensus, whereas in the latter the dynamics in finite populations remains in a meta-stable state around the interior fixed point for a long time. Indeed, in Fig.~\ref{fig main: nonlinear absorption}(b) we observe a steep increase of the mean time to fixation below $z=z_c(q)$.

\subsubsection{Evolutionary games}
A similar analysis for evolutionary games is detailed in Appendix~\ref{appendix: fixation times games}, see in particular Fig.~\ref{fig supp: games absorption}. The behaviour is similar to that in Fig.~\ref{fig main: nonlinear absorption}. Broadly, the fixation time in the evolutionary games for $\alpha<1/2$ behaves like that in the nonlinear voter model for $q<1$, and similarly, the qualitative shape of the fixation time for $\alpha>1/2$ is akin to that for $q>1$ in the nonlinear voter model.

We stress that in the evolutionary game there is always (for all $\alpha$) a critical level of zealotry $z_{c}(\alpha)$ [Eq.~(\ref{eq main: games zC})], above which the consensus fixed point becomes attracting in the rate equations. This is in contrast with the nonlinear voter model, where a critical zealotry only exists for $q>1$. Nonetheless, the mean fixation time shows relatively smooth behaviour as a function of $z$ in the evolutionary games for $\alpha<1/2$ [see Appendix, Fig.~\ref{fig supp: games absorption}(a)]. This is because for $\alpha<1/2$ [moving from the red to green region in Fig.~\ref{fig main: absorbing phase plot}(b)] the stable interior fixed point smoothly approaches the consensus fixed point before disappearing [a bifurcation diagram is shown in Fig.~\ref{fig supp: games absorbing bifurcation}(a) in the Appendix]. For $\alpha>1/2$ on the other hand [i.e. moving from the yellow to green region in Fig.~\ref{fig main: absorbing phase plot}(b)] the interior fixed points come together and then vanish [see  Fig.~\ref{fig supp: games absorbing bifurcation}(b)], thus there is a distinct change in the fixation time.

\subsubsection{Partisan voter model}
Since the partisan voter model has two degrees of freedom, we cannot determine the fixation time in closed form analytically. We report simulation results in Fig.~\ref{fig supp: partisan absorption} in Appendix~\ref{appendix: partisan}, noting that there is only one type of transition as a function of the zealotry $z$, for any fixed value of the preference parameter $\epsilon$ [this transition occurs when moving from the red to green region in Fig.~\ref{fig main: absorbing phase plot}(c)]. The qualitative behaviour of the deterministic system is similar to that for the evolutionary game for $\alpha<1/2$. There is a critical level of zealotry $z_{c}(\epsilon)$ [Eq.~(\ref{eq main: partisan zC})] but this has no discernable effect on the fixation time since there is no  discontinuity in the location of the stable fixed point.

\subsection{Quasi-stationary distribution}
\subsubsection{Definition}
The fixation time can become rather large if there are attracting internal fixed points of the deterministic rate equations. For example in Fig.~\ref{fig main: nonlinear absorption}(b), for $q>1$ and small $z$, fixation times are on the order of $10^{50}$ generations (or equivalent, $10^{50}$ Monte Carlo sweeps). Thus, in practice, the system will never reach absorption, but will reside in a quasi-stationary metastable state. We now seek to determine the quasi-stationary distribution characterising this state. 

Formally, and following Refs.~\cite{naasell1996quasi,collet2013quasi}, we define the quasi-stationary distribution as $Q_n^*\equiv\lim_{t\to\infty}Q_{n}(t)$, where 
\be
Q_{n}(t)=P(n|n\neq N-Z; t)
\ee
is the probability for the system to be in state $n$ at time $t$ conditioned on not having been absorbed at $N-Z$ (the latter is the only absorbing state for one-sided zealots).

One can show that this conditional distribution fulfils the following evolution equation \cite{naasell1996quasi, llabres2023partisan}
\begin{align}
\begin{split} \label{eq main: nasell}
    \frac{\dd Q_0(t)}{\dd t} &= T_1^-Q_1(t) \\
    &\quad -\left[T_0^+ - T_{N-Z-1}^{+} Q_{N-Z-1}(t)\right]Q_0(t),\\
    \frac{\dd Q_n(t)}{\dd t} &= T_{n-1}^{+}Q_{n-1}(t) -(T_n^+ + T_n^-)Q_n(t) \\
    &\quad + T_{n+1}^{-}Q_{n+1}(t)  \\
    &\quad + T_{N-Z-1}^{+}Q_{N-Z-1}(t)Q_{n}(t). 
\end{split}
\end{align}
We note that this is not a regular master equation, as the last term in the second equation contains the quadratic expression $Q_{N-Z-1}(t)Q_{n}(t)$.

To determine the quasi-stationary distribution we integrate these equations numerically and then identify the asymptotic values $Q_n^*$ at long times (see Appendix~\ref{appendix: quasi-stationary distributions} for further details). 

\subsubsection{Nonlinear voter model}
\begin{figure}[htbp]
    \centering
    \includegraphics[scale=0.5]{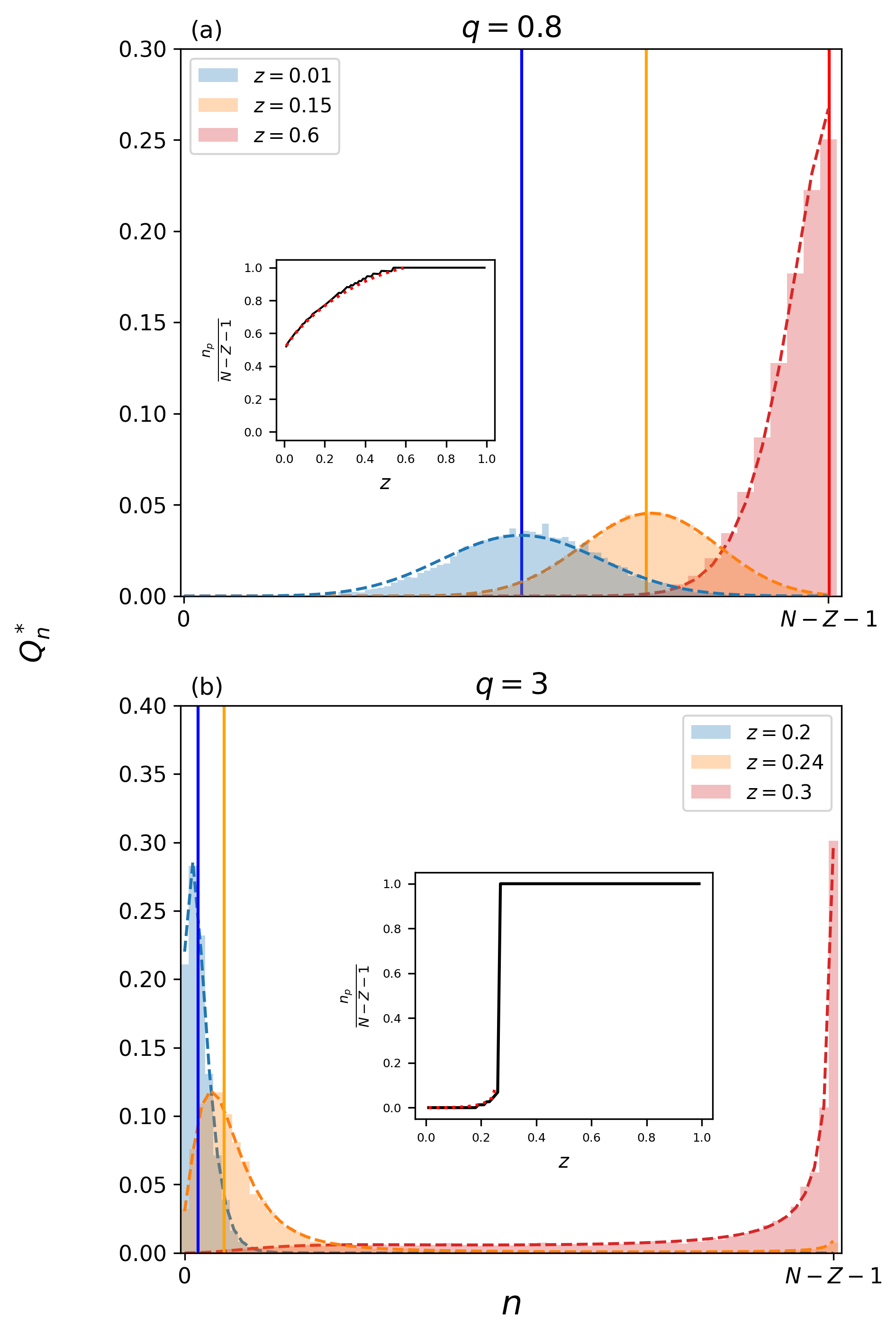}
    \caption{Plots of the quasi-stationary state distribution $Q_n^*$ for the nonlinear voter model ($N=100$) with varying densities of zealots $z$. $q=0.8$ in panel (a), and $q=3$ in panel (b). We note that $n\in[0,N-Z-1]$. The bars are from stochastic simulation using the Gillespie algorithm \cite{gillespie1976general} and the rates in Eqs.~(\ref{eq main: nonlinear vm finite rates}), averaging over $10^4$ trajectories. We note that a resampling method, where any absorbed trajectories continue from a randomly selected non-absorbed trajectory, has been used. The vertical lines are stable internal fixed points from the deterministic analysis, i.e. numerically solving Eq.~(\ref{eq main: nonlinear vm general rate eqn}). The dashed envelopes are from numerical integration of Eqs.~(\ref{eq main: nasell}). The solid black lines of the inset panels indicate the position of the peak $n_p$ of the quasi-stationary distribution for $z\in[\frac{1}{N}, 1-\frac{1}{N}]$; for better comparison across different values of $Z$, we plot $n_p/(N-Z-1)$. The dotted red line is the stable internal fixed point position from the deterministic analysis.}
    \label{fig main: nonlinear vm quasi}
\end{figure}

In Fig.~\ref{fig main: nonlinear vm quasi} we plot the quasi-stationary distribution for the nonlinear voter model with varying numbers of zealots. In panel (a) $q=0.8$ and in panel (b) $q=3$. We know from the deterministic analysis (Sec.~\ref{sec: nonlinear vm}) that for $q<1$ there is always a stable internal fixed point that smoothly approaches the upper boundary as $z$ increases. We expect the quasi-stationary distribution to be concentrated near this deterministic fixed point, and for the distribution's mode to slowly shift to the right as more zealots are added. This is indeed what we see in Fig.~\ref{fig main: nonlinear vm quasi}(a), see in particular the inset.

For $q>1$ the deterministic analysis shows that there are two interior fixed points for low $z$, one stable and one unstable, as well as a stable fixed point at the upper boundary. Then at a critical $z_c$ the interior fixed points come together and vanish (see Fig.~\ref{fig supp: nonlinear absorbing bifurcation} in the Appendix). Thus we expect the quasi-distribution to be peaked internally when the number of zealots is low, and then move sharply to the right once the critical zealotry is crossed. This is exactly what we see in Fig.~\ref{fig main: nonlinear vm quasi}(b).

Given the bifurcation analysis in the Appendix [see Fig.~\ref{fig supp: games absorbing bifurcation}], we expect that the quasi-stationary distribution for the evolutionary games in Eq.~(\ref{eq main: payoff matrix}) behaves similarly to that of the nonlinear voter model, with $\alpha<1/2$ corresponding to $q<1$, and $\alpha>1/2$ to $q>1$.

\subsubsection{Critical zealotry in finite populations}
\begin{figure}[htbp]
    \centering
    \includegraphics[scale=0.52]{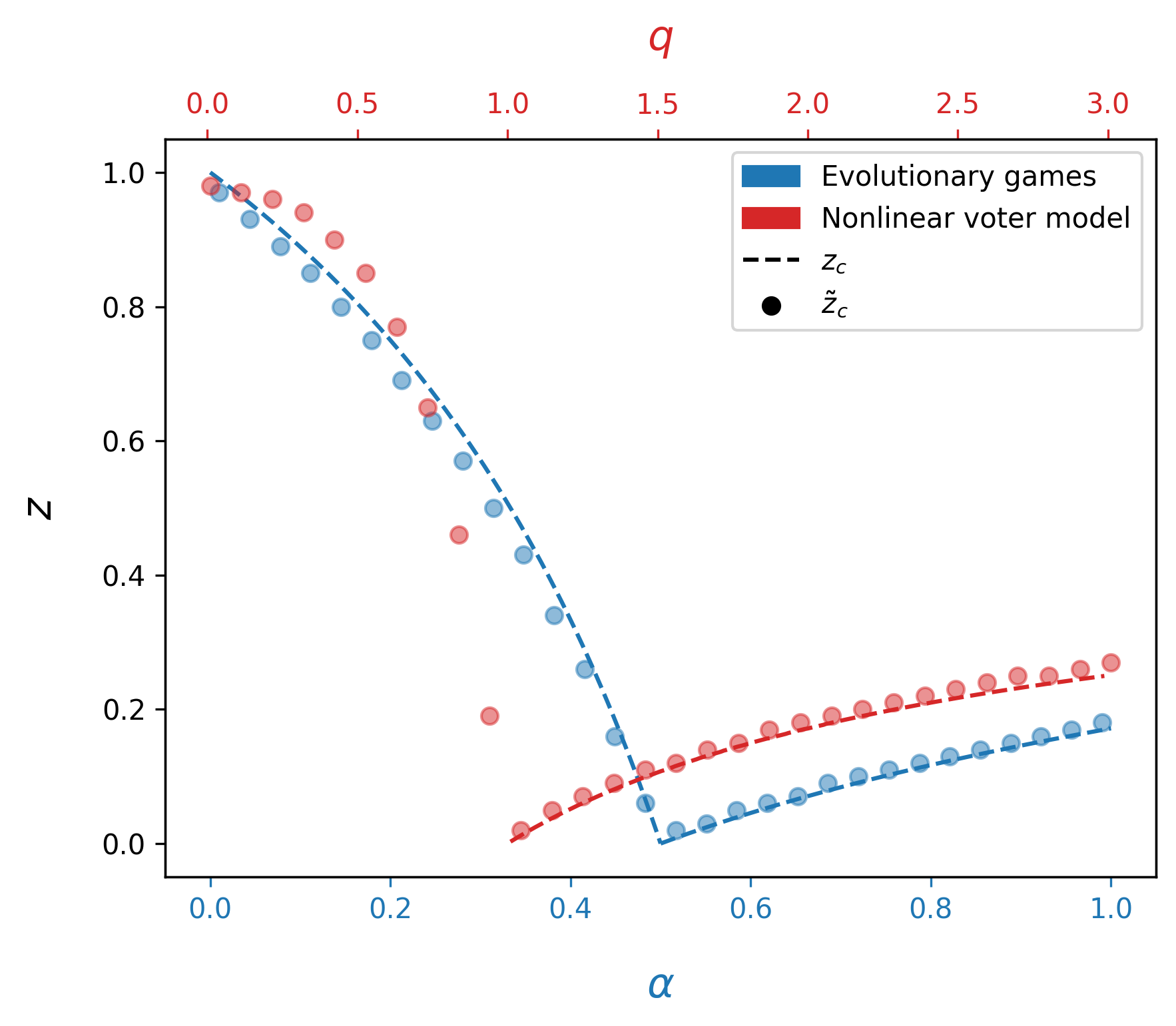}
    \caption{Plots of the critical zealotry $\tilde z_c$ (see text) for a finite system with $N=100$ (circle markers), alongside the deterministic result $z_c$ for infinite systems shown as dashed lines (Sec.~\ref{sec: fixed point analysis absorbing}). Plotted against $q$ for the nonlinear voter model (top axis, red) and $\alpha$ for evolutionary games (bottom axis, blue).}
    \label{fig main: finite system Zc}
\end{figure}

For finite systems we define $\tilde z_c$ as the proportion of zealots at which the quasi-stationary distribution becomes peaked at $n=N-Z-1$. For fixed model parameters (e.g. $q$ or $\alpha$) we can determine this quantity from a numerical integration of Eq.~(\ref{eq main: nasell}). Results are shown in Fig.~\ref{fig main: finite system Zc}. 

For evolutionary games (blue), the critical zealotry for finite systems ($\tilde z_c$) and for infinite systems [$z_c$ from Eq.~(\ref{eq main: games zC})] agree well with one another across all $\alpha$. In the nonlinear voter model, the critical zealotries for finite and infinite systems also coincide for $q>1$.  For $q<1$ the consensus fixed point is always locally unstable in the rate equations for infinite populations (and thus, there is no critical zealotry). However, in finite systems, and for sufficiently high zealotry, the quasi-stationary distribution has its peak at the consensus state. This is because the stable interior fixed point of the rate equations moves towards the upper boundary as $z$ increase [see Fig.~\ref{fig supp: nonlinear absorbing bifurcation}(a) in the Appendix].

\section{Summary and discussion}\label{sec:concl}
In summary we have carried out a detailed analysis of the combined effects of nonlinearity and zealotry in nonlinear voter models, simple evolutionary games and the partisan voter model. These are three models in which each agent can be in one of two symmetric states, and with similar fixed points of the rate equations in the absence of zealots (one coexistence fixed point, and two symmetric consensus states). When there are no zealots, the consensus states are also absorbing states for stochastic dynamics in finite populations.

We have studied if there are critical values of zealotry required to drive the system to consensus. The stability of the fixed points depend on the type and level of zealotry, but also on innate parameter of the models. While the parameter $q$ in the nonlinear voter model is a measure of nonlinear social impact in the sense of Latan\'e \cite{latane}, the parameters $\alpha$ in the evolutionary games, and $\epsilon$ in the partisan voter model describe different details of nonlinear interactions. 

The different types of nonlinearity can drive the population to coexistence, whereas one-sided zealotry generally favours consensus. We thus ask, which one or ones of these nonlinearities compete with the effects of zealotry. Extending models of evolutionary games and the partisan voter model to include nonlinear social impact, we conclude that the nonlinearity associated with a social impact parameter $q<1$ can counter the effects of zealotry.

 We conjecture that sublinear social impact can override the effects of zealotry more generally. When $q<1$ the probability of adopting a state of low prevalence is higher than in a model in which agents imitate a randomly chosen interaction partner. Thus, for $q<1$ the impact of small minorities is strong enough to prevent the formation of consensus on the state promoted by the zealots, even if there is a large majority for that state among susceptible agents. This is particularly relevant, as Latan\'e reported evidence for precisely such sublinear impact in experiments \cite{latane}.

Our work also reveals analogies and differences between the three models we have studied. We believe this a step towards a more systematic classification of the different types of transition and behaviour exhibited by individual-based models (see also Ref.~\cite{jaume_future}). Future work could extend the present study to networked systems and/or to heterogeneous agents each subject to different social impact.

\begin{acknowledgments}
This work was supported by the the Agencia Estatal de Investigaci\'on and Fondo Europeo de Desarrollo Regional (FEDER, UE) under project APASOS (PID2021-122256NB-C21, PID2021-122256NB-C22), the Mar\'ia de Maeztu programme for Units of Excellence, CEX2021-001164-M funded by  MCIN/AEI/10.13039/501100011033. We  also acknowledge a studentship by the Engineering and Physical Sciences Research Council (EPSRC, UK), reference EP/T517823/1.
\end{acknowledgments}

\section*{Data Availability Statement}
All data is generated from direct simulation of the models, or from the theoretical analysis. Ref.~\cite{github_zealots} is a GitHub repository that contains the relevant codes and a Mathematica notebook, as well as additional interactive plots. Further details can be made upon reasonable request.

\appendix

\section{The nonlinear voter model} \label{appendix: nonlinear vm}

\subsection{Balanced zealots} \label{appendix: nonlinear balanced}
For balanced zealots Eq.~(\ref{eq main: nonlinear vm general rate eqn}) becomes
\be
    \frac{\dd x}{\dd t} = (1-x-z)\left(x+\frac{z}{2}\right)^{q}-x\left(1-x-\frac{z}{2}\right)^{q}. \label{eq supp: qvm balanced zealots rate eqn}
\ee
This equation always has the central fixed point $x^{*}=\frac{1}{2}(1-z)$.

There exists potentially two more fixed points. These fixed points are roots of the function
\be
    B(x,z,q) \equiv \ln\left(\frac{2x+z}{2(1-x)-z}\right)-\frac{1}{q}\ln\left(\frac{x}{1-x-z}\right), \label{eq supp: unbalanced f}
\ee
found by setting the RHS of Eq.~(\ref{eq supp: qvm balanced zealots rate eqn}) to zero and re-arranging. This function is continuous over the allowed domain $x\in(0,1-z)$. We have
\begin{gather} \label{eq supp: qvm balanced limits}
\begin{split}
    \lim_{x\to 0^{+}} B(x,z,q) = +\infty, \\
    \lim_{x\to (1-z)^{-}} B(x,z,q) = -\infty.
\end{split}
\end{gather}
So the function $B(x,z,q)$ goes from $+\infty$ to $-\infty$ as $x$ goes from $0$ to $1-z$, crossing the $x$-axis at least once. The number of crosses can be determined by solving for the stationary points (extrema). $B(x,z,q)$ can only have 2 stationary points:
\be
    x_{1,2} = x^{*} \pm \frac{\sqrt{\left[1+q(z-1)\right](z-1)(q+z-1)}}{2(q+z-1)}, \label{eq supp: qvm balanced x12}
\ee
which are only physical when $q>1$ and $z<z_c(q)$ where
\be
    z_{c}(q) \equiv 1-\frac{1}{q}. \label{eq supp: nonlinear balanced zC}
\ee
This equation reproduces the values reported for specific integer $q$ in Ref.~\cite{mobilia2015nonlinear} [we note that the author of Ref.~\cite{mobilia2015nonlinear} uses the convention $z_+=z_-=z$ with $z\in (0,1/2)$, i.e., our definition of $z$ differs from that in Ref.~\cite{mobilia2015nonlinear} by a factor of 2]. We also highlight Ref.~\cite{mellor2017heterogeneous} in which the authors studied the nonlinear voter model with balanced zealots and an arbitrary number of sub-populations using different integer-valued exponents $q_1,q_2,\dots$, and where $z_c$ was also computed analytically, again with the alternative convention as mentioned above.

The stationary points in Eqs.~(\ref{eq supp: qvm balanced x12}), if they exist, always occur on opposite sides of the $x$-axis. We conclude that if $q>1$ and $z<z_c(q)$ Eq.~(\ref{eq supp: qvm balanced zealots rate eqn}) has three interior fixed points (the outer fixed points must be solved for numerically). Otherwise, if $q>1$ and $z\geq z_{c}(q)$, there is only the single fixed point $x^{*}=\frac{1}{2}(1-z)$. 

\begin{figure*}[hbtp]
    \centering
    \includegraphics[scale=0.35]{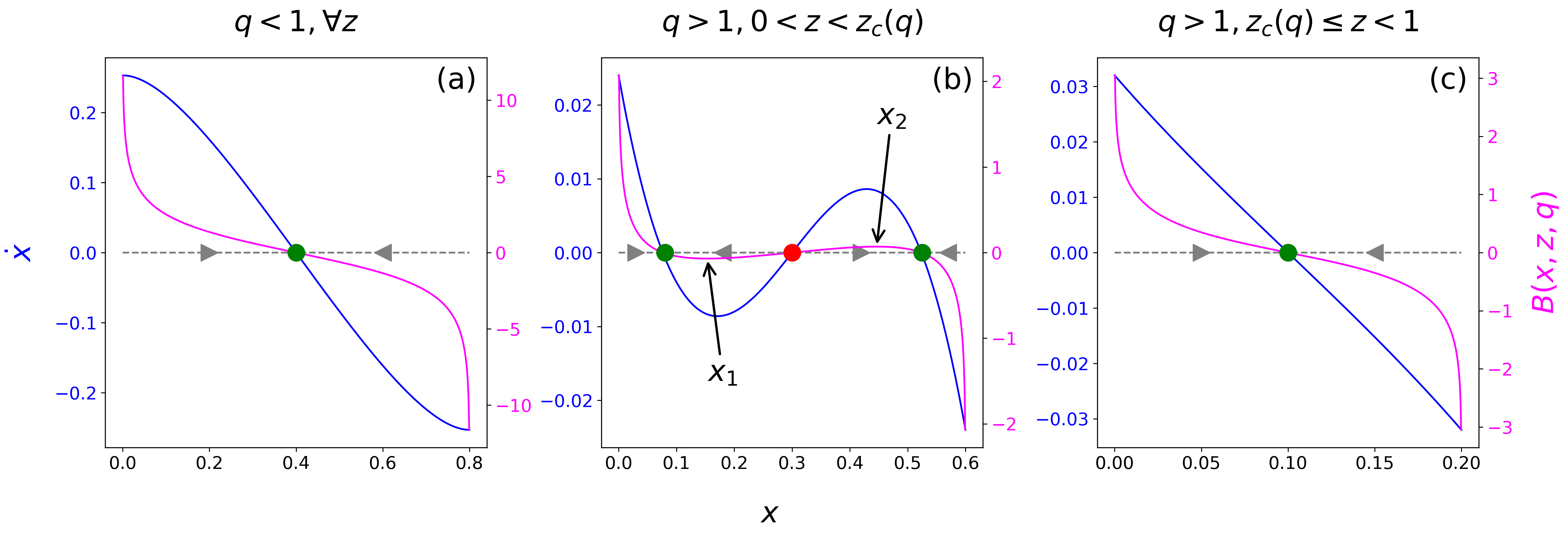}
    \caption{Plots of $\dot{x}$ [Eq.~(\ref{eq supp: qvm balanced zealots rate eqn})] in blue using the left hand axis. Plots of $B(x,z,q)$ [Eq.~(\ref{eq supp: unbalanced f})] in pink using the right hand axis. We note that $x\in[0, 1-z]$. Green markers are stable fixed points. Red markers are unstable fixed points. Grey arrows indicate the direction of the flow. In the case $q>1$ and $z<z_{c}(q)$ we have annotated the stationary points of $B(x,z,q)$, i.e. $x_{1,2}$ from Eqs.~(\ref{eq supp: qvm balanced x12}). $z_{c}(q)$ is defined in Eq.~(\ref{eq supp: nonlinear balanced zC}). The parameters used for the different scenarios are $q=(0.5, 2, 2)$ and $z=(0.2, 0.4, 0.8)$. In all panels there is a fixed point at $\frac{1}{2}(1-z)$. In panel (b) there are two additional outer fixed points which have been determined numerically.}
    \label{fig supp: nonlinear balanced flow}
\end{figure*}

The stability of the fixed points can be determined by first noting that $\dot{x}$ and $B(x,z,q)$ always have the same sign. Together with the limits from Eqs.~(\ref{eq supp: qvm balanced limits}), this means that when there is a single fixed point, it must always be stable. When there are three fixed points the central one must be unstable and the outer ones stable. This is illustrated in Fig.~\ref{fig supp: nonlinear balanced flow} where we plot $\dot{x}$ and $B(x,z,q)$ for three different scenarios. 

\begin{figure}[hbtp]
    \centering
    \includegraphics[scale=0.5]{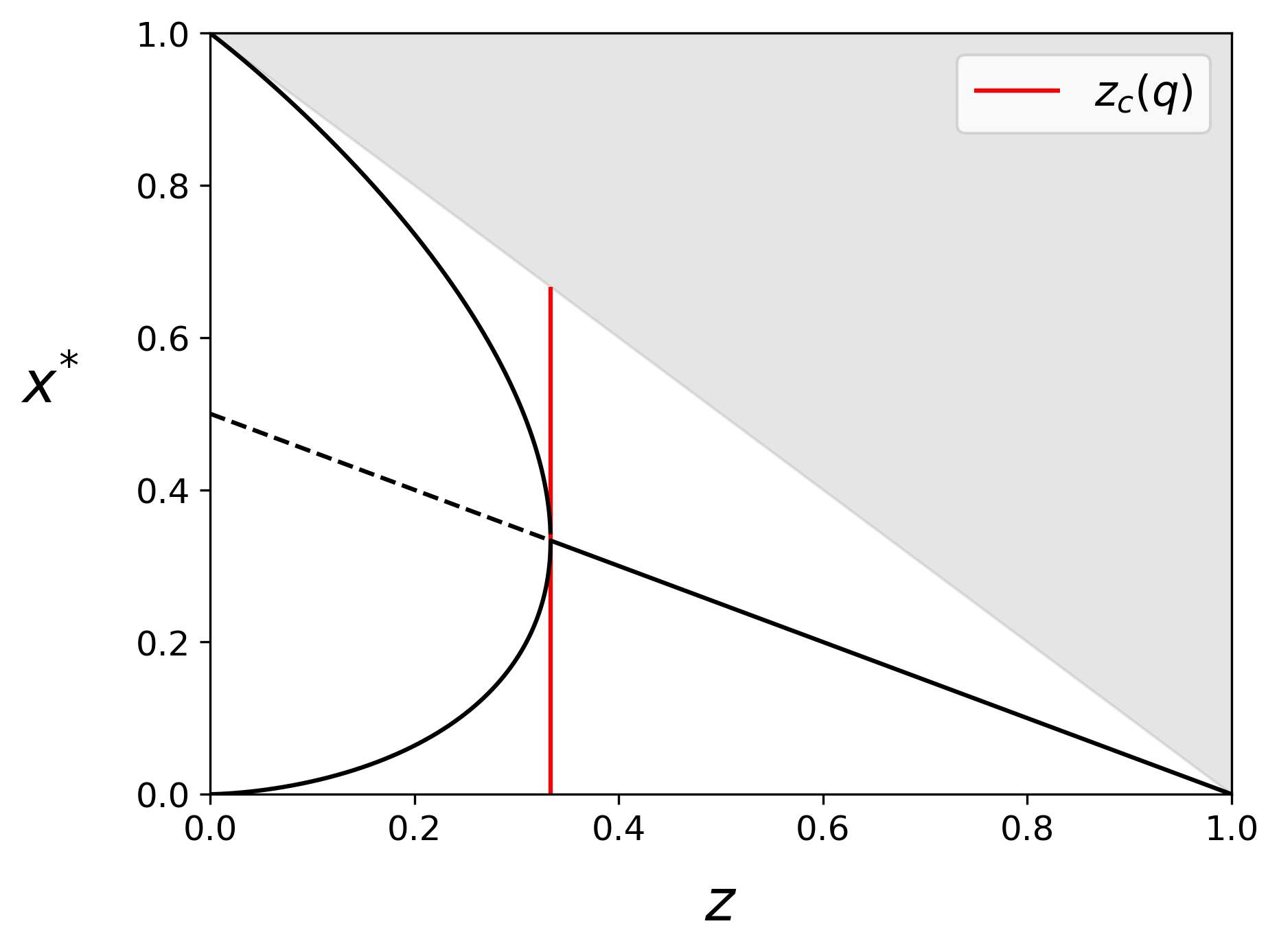}
    \caption{Bifurcation diagram for the nonlinear voter model with balanced zealots showing how the fixed points of $\dot{x}$ [Eq.~(\ref{eq supp: qvm balanced zealots rate eqn})] change as we vary $z$ at $q=1.5$. Solid/dashed black lines correspond to stable/unstable fixed points. The vertical red line corresponds to $z_{c}(q)=1/3$ [Eq.~(\ref{eq supp: nonlinear balanced zC})]. There is always a fixed point at $(1-z)/2$. The additional two outer fixed points that appear when $z<z_{c}(q)$ can be determined numerically. The shaded grey region is not a valid region since $x\in[0,1-z]$.}
    \label{fig supp: nonlinear balanced bifurcation}
\end{figure}

In Fig.~\ref{fig supp: nonlinear balanced bifurcation} we plot the bifurcation diagram for the case $q=1.5$ as we vary $z$. We see for $z<z_{c}(q)$ we have situations like in Fig.~\ref{fig supp: nonlinear balanced flow}(b), and for $z\geq z_{c}(q)$ we have cases like in Fig.~\ref{fig supp: nonlinear balanced flow}(c).

\subsection{One-sided zealots} \label{appendix: nonlinear absorbing}
For one-sided zealots Eq.~(\ref{eq main: nonlinear vm general rate eqn}) becomes
\be
    \frac{\dd x}{\dd t} = (1-x-z)(x+z)^{q}-x(1-x-z)^{q}. \label{eq supp: qvm absorbing rate eqn}
\ee
This equation always has the consensus fixed point $x_{c}^{*}=1-z$.

There are potentially further fixed points. To determine these we carry out a similar procedure as in Appendix~\ref{appendix: nonlinear balanced} by first defining
\begin{align}
    A(x,z,q) = \ln\left(\frac{x+z}{1-x-z}\right) -\frac{1}{q}\ln\left(\frac{x}{1-x-z}\right). \label{eq supp: absorbing zealots g}
\end{align}
We note that $\dot{x}$ and $A(x,z,q)$ always have the same sign. The limits of $A(x,z,q)$ are
\begin{subequations}
\begin{gather}
    \lim_{x\to 0^{+}} A(x,z,q) = +\infty, \\
    \begin{align}
    \lim_{x\to (1-z)^{-}} A(x,z,q) 
    &= 
    \begin{cases}
        +\infty, & q > 1 \\
        c > 0,   & q = 1 \\
        -\infty, & q < 1
    \end{cases}.
    \end{align}
\end{gather}
\end{subequations}

In the case $q=1$, we can show that $A(x,z,q)$ has no roots in the interval $x\in (0, 1-z)$, so the only fixed point will be $x_{c}^{*}$.

\begin{figure*}[hbtp]
    \centering
    \includegraphics[scale=0.3]{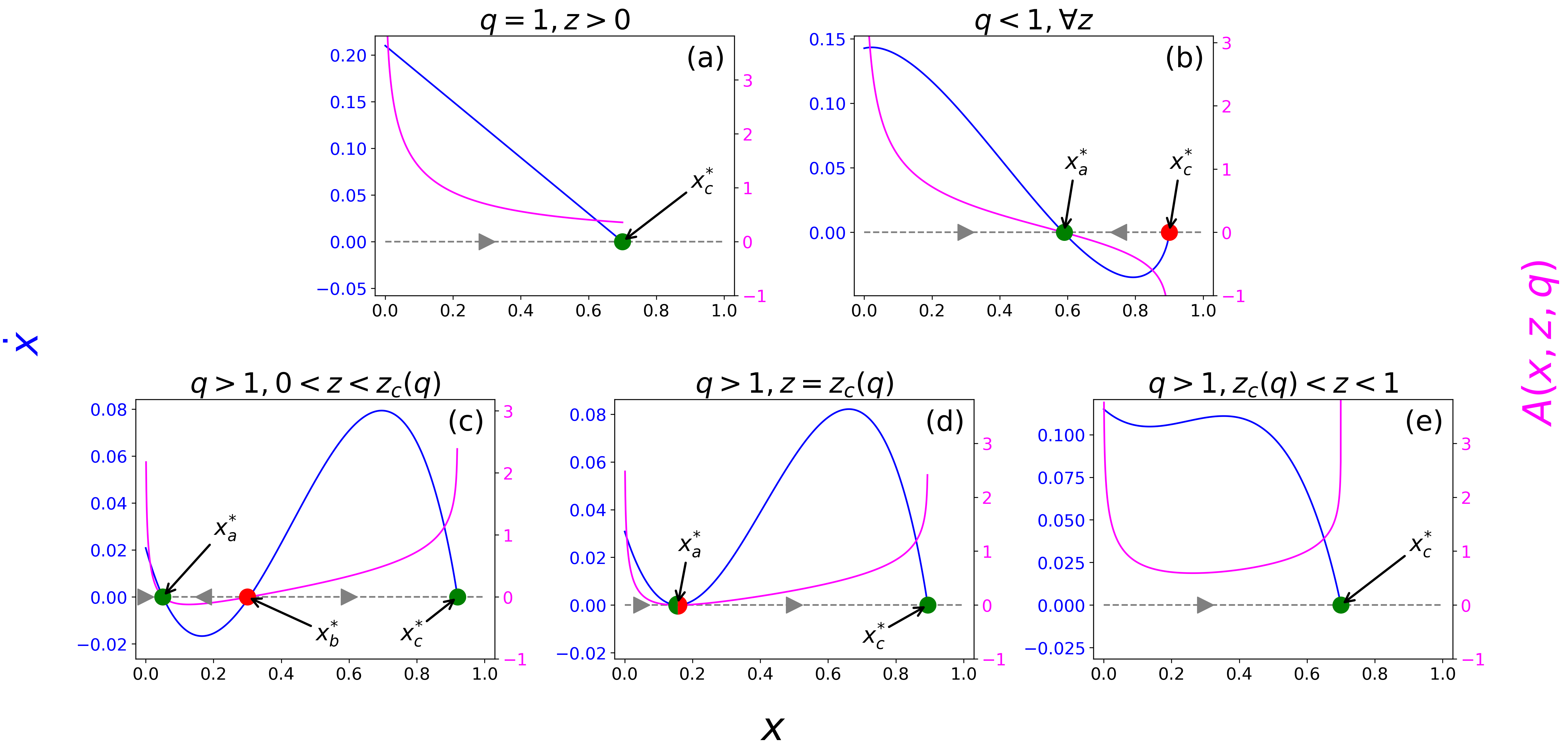}
    \caption{Plots of $\dot{x}$ [Eq.~(\ref{eq supp: qvm absorbing rate eqn})] in blue using the left hand axis, and of $A(x,z,q)$ [Eq.~(\ref{eq supp: absorbing zealots g})] in pink using the right hand axis. We note that $x\in[0,1-z]$. Green/red markers are stable/unstable fixed points. Grey arrows indicate the direction of the flow. The parameters used for the different scenarios (a)-(e) are $q=(1, 0.5, 1.5, 1.5, 1.5)$ and $z=(0.3, 0.1, 0.08, 0.106, 0.3)$ respectively. The fixed points $x_{a}^{*}$, $x_{b}^{*}$ and $x_{c}^{*}$ have been annotated (see text). We know analytically that $x_{c}^{*}=1-z$, while the other interior fixed points must be solved for numerically.}
    \label{fig supp: nonlinear absorbing flow}
\end{figure*}

For $q<1$, the limits of $A(x,z,q)$ are $\pm\infty$, thus $A(x,z,q)$ must cross the $x$-axis once, hence one root is always guaranteed. Because the signs of $\dot{x}$ and $A(x,z,q)$ are always the same, and $A(x,z,q)$ goes from $-\infty$ to $+\infty$ as $x$ increases, the interior fixed point must be stable [flow is positive to the left and negative to the right, see Fig.~\ref{fig supp: nonlinear absorbing flow}(b)].

It is possible for there to be additional roots, which we can find by determining the stationary points of $A(x,z,q)$. There turns out to be only one stationary point,
\be
    x_{1} = \frac{z(z-1)}{1-z-q}. \label{eq supp: absorbing g stat point}
\ee
A necessary condition for this stationary point to be physical is $q>1$. Therefore, for $q<1$ there are no valid stationary points. Thus $A(x,z,q)$ only has a single root. In summary, for $q<1$, there is the unstable consensus fixed point $x_{c}^{*}$ and an additional stable interior fixed point $x_{a}^*$. This scenario is illustrated in Fig.~\ref{fig supp: nonlinear absorbing flow}(b). The interior fixed point $x_a^*$ can be solved for numerically.

The case $q>1$ is the most complex. Now the stationary point of $A(x,z,q)$ [Eq.~(\ref{eq supp: absorbing g stat point})] is physically valid. Depending on its value $A(x,z,q)$ can have zero, one or two roots, meaning zero, one or two additional fixed points for Eq.~(\ref{eq supp: qvm absorbing rate eqn}). We note that in all cases the consensus fixed point $x_{c}^{*}$ from before exists and is stable. We now summarise the cases of any additional fixed points:
\begin{enumerate}
    \item $A(x_{1}, z, q) < 0$: There are two roots, hence Eq.~(\ref{eq supp: qvm absorbing rate eqn}) has two additional fixed points $x_{a}^{*}$ and $x_b^{*}$ where $x_{a}^{*}<x_b^{*}<x_{c}^{*}$. $x_{a}^{*}$ is always stable and $x_b^{*}$ is always unstable. This is illustrated in Fig.~\ref{fig supp: nonlinear absorbing flow}(c).
    \item $A(x_{1}, z, q) = 0$: There is one root that occurs exactly on the $x_a$-axis, hence Eq.~(\ref{eq supp: qvm absorbing rate eqn}) has one additional fixed point $x_a^{*}<x_{c}^{*}$ that is stable from the left and unstable from the right. This is illustrated in Fig.~\ref{fig supp: nonlinear absorbing flow}(d).
    \item $A(x_{1}, z, q) > 0$: There are no roots, hence Eq.~(\ref{eq supp: qvm absorbing rate eqn}) has no additional fixed points. This is illustrated in Fig.~\ref{fig supp: nonlinear absorbing flow}(e).
\end{enumerate}
We note that the second scenario above separates the first and the third. We wish to determine the critical value of the zealotry, $z_c(q)$, at which this second scenario occurs. This translates to solving $A(x_{1}, z_{c}(q), q) = 0$ for $z_{c}(q)$. Using Eqs.~(\ref{eq supp: absorbing zealots g}) and (\ref{eq supp: absorbing g stat point}) we have
\be
    \ln\left[\frac{qz_{c}}{(1-q)(z_{c}-1)}\right]-\frac{1}{q}\ln\left(\frac{z_{c}}{q-1}\right) = 0.
\ee
Re-arranging we find the above is equivalent to  
\begin{equation}
    z_c - \left[-\frac{(1-n)^{n-1}}{n^n}\right]z_c^n - 1 = 0, \label{eq supp: absorbing zealots z_c eq to solve}
\end{equation}
where $n=1-\frac{1}{q}$. The above is a trinomial equation that has been analysed in-depth, see for example Eq.~(8.1) in Ref.~\cite{belkic2019trinominal}. The solution can be represented in many ways, probably the most useful for evaluating is
\begin{align}
    z_{c}(q)&=\frac{q}{q-1}\sum_{k=0}^{\infty}\frac{(-1)^{k}}{k!}\left[\frac{q-1}{q^\frac{q}{q-1}}\right]^{k+1} \nonumber \\
    &\quad \times\frac{\Gamma\left(\frac{q(k+1)}{q-1}\right)}{\Gamma\left(\frac{q(k+1)}{q-1}-k+1\right)}. \label{eq supp: nonlinear vm absorbing zcq sum}
\end{align}

Alternatively, solving trinomial equations like Eq.~(\ref{eq supp: absorbing zealots z_c eq to solve}) numerically is fast with standard root-finding algorithms as the left-hand side is continuously increasing in $z$, and only has one root.

We also note that a highly accurate closed form approximation to Eq.~(\ref{eq supp: absorbing zealots z_c eq to solve}) is given by (see Appendix~\ref{appendix: derivation of approx} for a derivation)
\be
    z_{c}(q) = (q-1)(2q-1)^{\frac{q-\frac{1}{3}}{1-q}}. \label{eq supp: absorbing zealots approx solution}
\ee

We note that Eq.~(\ref{eq supp: nonlinear vm absorbing zcq sum}) generalises results of Ref.~\cite{mobilia2007role} (which only gave numerical solutions for a select integer $q$) by providing a complete analysis for all $q\in \mathbb{R}_{>0}$. 

\begin{figure}[hbtp]
    \centering
    \includegraphics[scale=0.53]{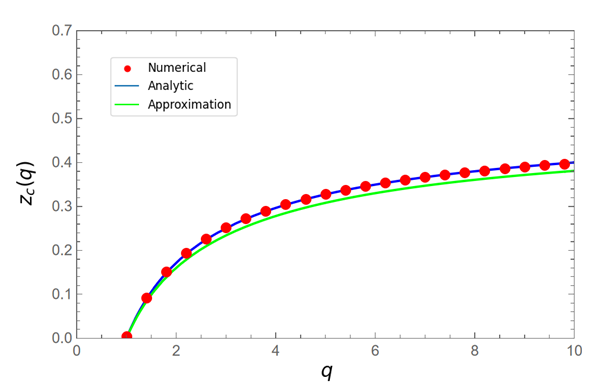}
    \caption{Plot of the critical zealotry $z_{c}(q)$ against $q$ in the case of one-sided zealots in the nonlinear voter model. The blue line is the converging series solution [Eq.~(\ref{eq supp: nonlinear vm absorbing zcq sum})]. The green line is the closed form approximation [Eq.~(\ref{eq supp: absorbing zealots approx solution})]. The red markers are from numerically solving Eq.~(\ref{eq supp: absorbing zealots z_c eq to solve}).}
    \label{fig supp: qvm absorbing z_c}
\end{figure}

In Fig.~\ref{fig supp: qvm absorbing z_c} we demonstrate the validity of Eqs.~(\ref{eq supp: nonlinear vm absorbing zcq sum}) and (\ref{eq supp: absorbing zealots approx solution}) by comparing them to the result from numerically solving Eq.~(\ref{eq supp: absorbing zealots z_c eq to solve}). 

\begin{figure}[hbtp]
    \centering
    \includegraphics[scale=0.5]{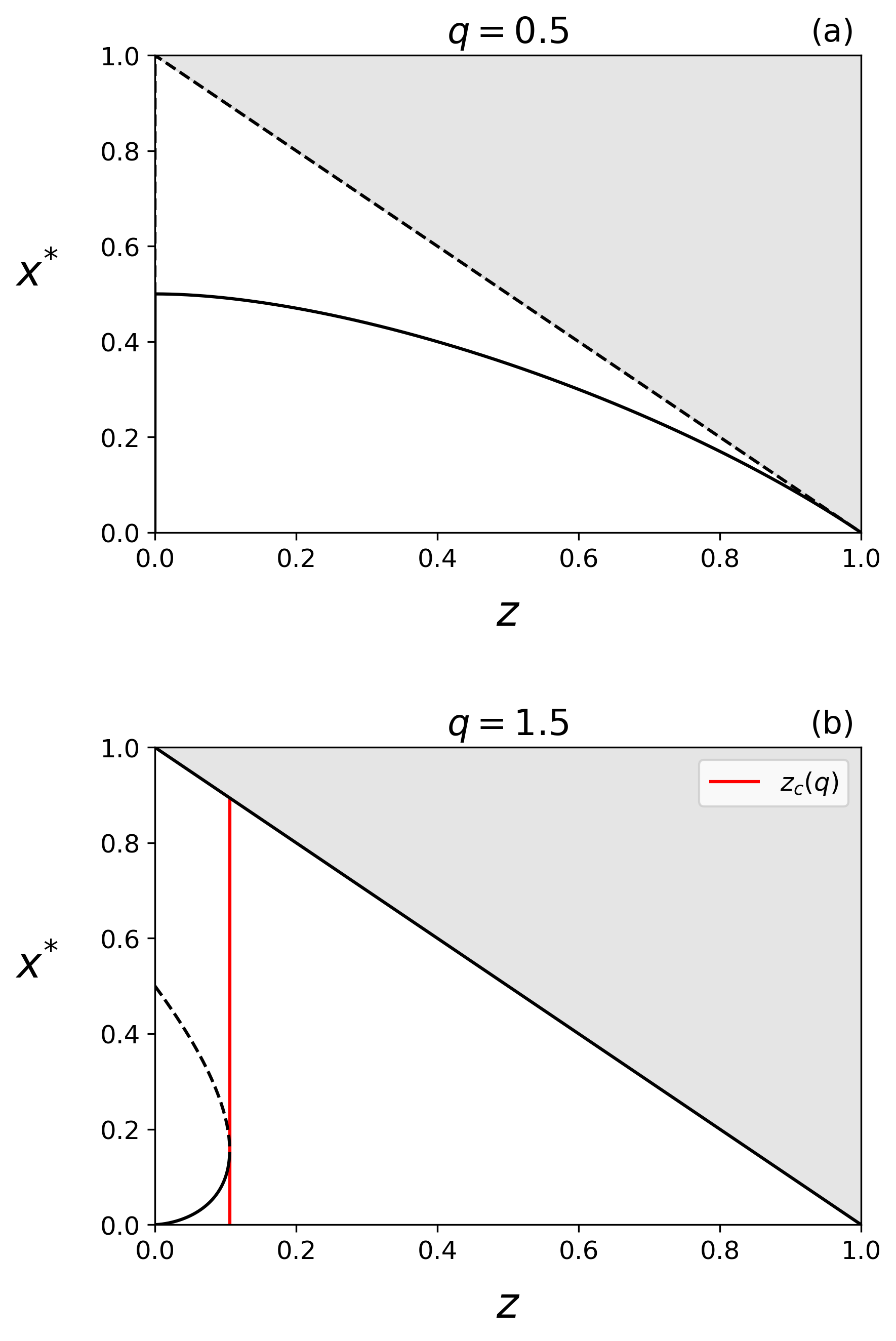}
    \caption{Bifurcation diagram for the nonlinear voter model with one-sided zealots, showing how the fixed points of $\dot{x}$ [Eq.~(\ref{eq supp: qvm absorbing rate eqn})] change as we vary $z$ at (a) $q=0.5$ and (b) $q=1.5$. Solid/dashed black lines represent stable/unstable fixed points. The vertical red line corresponds to $z_{c}(q)= 0.106$ [Eq.~(\ref{eq supp: nonlinear vm absorbing zcq sum})]. Fig.~\ref{fig supp: nonlinear absorbing flow}(c)-(e) shows snapshots at particular $z$ values. There is always a fixed point at $1-z$. The additional interior fixed points that appear when $z\leq z_{c}(q)$ can be determined numerically. The grey shaded region is not a valid region since $x\in[0,1-z]$.}
    \label{fig supp: nonlinear absorbing bifurcation}
\end{figure}

In Fig.~\ref{fig supp: nonlinear absorbing bifurcation} we plot the bifurcation diagram for the cases $q=0.5$ and $q=1.5$ as we vary $z$. Fig.~\ref{fig supp: nonlinear absorbing flow} then shows snapshots of these bifurcation diagrams at specific values of $z$.

\subsection{Unbalanced zealots} \label{appendix: nonlinear unbalanced}
For unbalanced zealots the rate equation [Eq.~(\ref{eq main: nonlinear vm general rate eqn})] becomes
\begin{align}
    \dot{x}&=(1-x-z)\left[x+\frac{1}{2}(1+\delta)z\right]^{q} \nonumber \\
    &\quad -x\left[1-x-\frac{1}{2}(1+\delta)z\right]^{q}. \label{eq supp: unbalanced flow}
\end{align}
Exactly like in Appendices.~\ref{appendix: nonlinear balanced} and \ref{appendix: nonlinear absorbing} we define a new function $U(x,z,q,\delta)$ whose zeros correspond to the fixed points of Eq.~(\ref{eq supp: unbalanced flow}),
\begin{align}
    U(x,z,q,\delta) &= \ln\left[\frac{2x+(1+\delta)z}{2(1-x)-(1+\delta)z}\right] \nonumber \\
    &\quad -\frac{1}{q}\ln\left(\frac{x}{1-x-z}\right). \label{eq supp: qvm unbalanced U}
\end{align}
$\dot{x}$ and $U(x,z,q,\delta)$ have the same sign. The limits of $U(x,z,q,\delta)$ are
\begin{subequations}
\begin{align}
    \lim_{x\to 0^{+}}U(x,z,q,\delta) &= +\infty, \\
    \lim_{x\to (1-z)^{-}}U(x,z,q,\delta) &= -\infty.
\end{align}
\end{subequations}
Thus we have a situation analogous to that of balanced zealots (Appendix~\ref{appendix: nonlinear balanced}). We will therefore only consider the interesting scenario where there can be multiple internal fixed points.

$U(x,z,q,\delta)$ only has two physical stationary points ($x_{1,2}$ where we assume $x_{1}<x_{2}$) when
\begin{equation} \label{eq supp: unbalanced conditions}
\begin{gathered}
    0 < z < \frac{(q-1)^{2} - \abs{q-1}\sqrt{(q+1)^{2} - 4q\delta^{2}}}{2q(\delta^{2} - 1)},\\
    q > 1
\end{gathered}
\end{equation}
where we recall $-1 < \delta < 1$. In order to have three interior fixed points [i.e. a scenario as in Fig.~\ref{fig supp: nonlinear balanced flow}(c)] we require $U(x_{1}, z, q, \delta)<0$ and $U(x_{2}, z, q, \delta) > 0$. 

Focusing on $0<\delta<1$ (the situation is symmetric), we find $U(x_{2}, z, q, \delta) > 0$ is always true. Therefore, we define the critical zealotry $z_{c}(q)$ as the $z$ value which solves $U(x_{1}, z_{c}(q), q, \delta)=0$. It is difficult to solve this equation analytically for $z_{c}(q)$. However, under the conditions of Eq.~(\ref{eq supp: unbalanced conditions}), $U(x_{1}, z, q, \delta)$ is an increasing function with exactly one root, $z_{c}(q)$, for a given $(q, \delta)$ within the range specified by the first condition in Eq.~(\ref{eq supp: unbalanced conditions}). Under such a setup, common root finding techniques like Newton's Method converge almost instantly to the solution. Thus for any valid $(q, \delta)$ pair we can determine $z_{c}(q)$ easily. Fig.~\ref{fig main: unbalanced phase plot}(a) in the main paper contains plots of $z_c(q)$ against $q$ for different $\delta$.

The flow diagrams for unbalanced zealots are analogous to those of balanced zealots as presented in Fig.~\ref{fig supp: nonlinear balanced flow}.

\subsection{Derivation of Eq.~(\ref{eq supp: absorbing zealots approx solution})} \label{appendix: derivation of approx}
Here we derive a closed form approximation to Eq.~(\ref{eq supp: absorbing zealots z_c eq to solve}). We first define
\be
    f(z) = z + \left[\frac{(1-n)^{n-1}}{n^n}\right]z^n - 1, \label{eq supp: trinomial original}
\ee
where $0<n<1$ and the domain is $z\in(0, 1)$. The goal is to find the root of Eq.~(\ref{eq supp: trinomial original}). 

With the substitution
\be
    z = \frac{n}{1-n}t, \label{eq supp: z->t transform}
\ee
Eq.~(\ref{eq supp: trinomial original}) becomes
\be
    f(t) = t^n + nt + (n-1), \label{eq supp: trinomial in t}
\ee
where $t\in(0,+\infty)$. Since $f'(t)=nt^{n-1}+n>0$, $f(t)$ is strictly increasing. Moreover, since $f(0) = n-1<0$ and $f(1)=2n>0$, thus there is a unique root in the domain $t\in(0, 1)$.

Now introduce a function similar to Eq.~(\ref{eq supp: trinomial in t}),
\be
    g(t) = t^n + nt^n + (n-1). \label{eq supp: simpler t trinomial}
\ee
This function is also strictly increasing with a unique root in the domain $t\in(0, 1)$, which can be seen by doing an analysis similar to what was done for $f(t)$. The root of $g(t)$ is
\be
    t_g(n) = \left(\frac{1-n}{1+n}\right)^{\frac{1}{n}}.
\ee
Importantly, in the domain $t\in(0, 1)$ we have $t<t^n$, thus $g(t)>f(t)$ in this domain. Therefore the root of $g(t)$ is a lower bound for the root of $f(t)$. Based on this we make the ansatz that the root of $f(t)$ can be written as
\be
    t_f(n) \equiv \left(\frac{1-n}{1+n}\right)^{\frac{\alpha(n)}{n}}, \label{eq supp: t_f defn}
\ee
where $\alpha(n) \in (0,1)$. The question now becomes if can we determine $\alpha(n)$, or at least find an approximation for it?

We can re-write Eq.~(\ref{eq supp: t_f defn}) as
\be
    \alpha(n) = \frac{n\ln[t_f(n)]}{\ln\left(\frac{1-n}{1+n}\right)}. \label{eq supp: alpha n}
\ee
Now we look to determine the limits of $\alpha(n)$ as $n\downarrow 0$ and $n\uparrow 1$. This requires determining the limits of $t_f(n)$, i.e. the limits of the true solution to Eq.~(\ref{eq supp: trinomial in t}).

\subsubsection{The limit \texorpdfstring{$n\downarrow 0$}{}} \label{appendix: n->0}
For the limit $n\downarrow 0$ we write $n = \epsilon \ll 1$, Eq.~(\ref{eq supp: trinomial in t}) is then,
\begin{align}
    f(t; n=\epsilon) &= t^{\epsilon} + \epsilon t + (\epsilon - 1) \nonumber \\
    &= e^{\epsilon\ln(t)} + \epsilon t + (\epsilon - 1) \nonumber \\
    &= 1+\epsilon\ln(t) + \epsilon t + (\epsilon - 1) +\order{\epsilon^2} \nonumber \\
    &= \epsilon\ln(t) + \epsilon t + \epsilon + \order{\epsilon^2}.
\end{align}
Setting this equation to zero, then dividing through by $\epsilon$ and letting $\epsilon\to 0$, we have
\be
    \ln(t) + t + 1 = 0,
\ee
and then rearranging we find
\be
    te^t = e^{-1}.
\ee
The solution to the above is given by the Lambert-$W$ function. Thus we have.
\be
    \lim_{n\downarrow 0}\; t_f(n) = W(e^{-1}) = 0.278\dots. \label{eq supp: t lower}
\ee
Now we expand the denominator of Eq.~(\ref{eq supp: alpha n}) at small $n=\epsilon$,
\begin{align}
    \ln\left(\frac{1-\epsilon}{1+\epsilon}\right) &= \ln(1-\epsilon) - \ln(1+\epsilon) \nonumber \\
    &= \left(-\epsilon-\frac{\epsilon^2}{2}-\dots\right) - \left(\epsilon-\frac{\epsilon^2}{2}+\dots\right) \nonumber \\
    &= -2\epsilon + \order{\epsilon^3}. \label{eq supp: alpha denom lower}
\end{align}
Using Eqs.~(\ref{eq supp: t lower}) and (\ref{eq supp: alpha denom lower}), we can evaluate the limit of Eq.~(\ref{eq supp: alpha n}) as $n\downarrow 0$,
\begin{align}
    \lim_{n\downarrow 0}\;\alpha(n) &= \lim_{\epsilon \downarrow 0}\; \alpha(\epsilon) \nonumber \\
    &= \frac{\epsilon \ln[W(e^{-1})]}{-2\epsilon + \order{\epsilon^3}} \nonumber \\
    &=-\frac{\ln(0.278\dots)}{2} = 0.639\dots. \label{eq supp: alpha lower}
\end{align}

\subsubsection{The limit \texorpdfstring{$n\uparrow 1$}{}} \label{appendix: n->1}
For the limit $n\uparrow 1$ we write $n=1-\delta$ with $\delta \to 0$. Substituting this into Eq.~(\ref{eq supp: trinomial in t}) we have
\be
    f(t; n =1-\delta) = t^{1-\delta}+(1-\delta)t - \delta. \label{eq supp: f(1-delta)}
\ee
We now show that the root of Eq.~(\ref{eq supp: f(1-delta)}) is in the interval $(\delta/4,\delta/2)$.

We evaluate the first two terms in Eq.~(\ref{eq supp: f(1-delta)}) divided by $\delta$ at $t=\delta/4$:
\be
    \frac{t^{1-\delta}+(1-\delta)t}{\delta}\Big|_{t=\frac{\delta}{4}} = \frac{\delta^{-\delta}}{4^{1-\delta}} + \frac{1-\delta}{4}. \label{eq supp: two terms at delta/4}
\ee
Since $\delta \ll 1$, we have $\delta^{-\delta}<\sqrt{2}$, and $4^{1-\delta} > 4^{\frac{1}{2}} = 2$. Using these two facts in Eq.~(\ref{eq supp: two terms at delta/4}) we have 
\be
    \frac{t^{1-\delta}+(1-\delta)t}{\delta}\Big|_{t=\frac{\delta}{4}} < \frac{\sqrt{2}}{2} + \frac{1}{4} < 1.
\ee
So the first two terms of Eq.~(\ref{eq supp: f(1-delta)}) are $<\delta$, meaning that $f(\delta/4) < 0$, so the root of Eq.~(\ref{eq supp: f(1-delta)}) is bounded from below by $\delta/4$.

Now do the same thing but with $t=\delta/2$:
\begin{align}
    \frac{t^{1-\delta}+(1-\delta)t}{\delta}\Big|_{t=\frac{\delta}{2}} &= \frac{\delta^{-\delta}}{2^{1-\delta}} + \frac{1-\delta}{2} \nonumber \\
    &= \frac{1}{2}\left[\left(\frac{\delta}{2}\right)^{-\delta}+1-\delta\right] \nonumber \\
    &= \frac{1}{2}\left[1-\delta\ln\frac{\delta}{2}+1-\frac{\delta}{2}\right] + \order{\delta^2} \nonumber \\
    &= 1-\delta\left(\ln\frac{\delta}{2}+\frac{1}{2}\right) + \order{\delta^2}
\end{align}
For $\delta\downarrow 0$ the bracket in the last expression is negative, and hence $ \frac{\delta^{-\delta}}{2^{1-\delta}} + \frac{1-\delta}{2}>1$. Thus the first two terms of Eq.~(\ref{eq supp: f(1-delta)}) are $>\delta$, meaning that $f(\delta/2) > 0$, so the root of Eq.~(\ref{eq supp: f(1-delta)}) is bounded from above by $\delta/2$.

We have shown then that the root of Eq.~(\ref{eq supp: f(1-delta)}) is in the domain $(\delta/4, \delta/2)$. Therefore we make the ansatz $t=a\delta$ in Eq.~(\ref{eq supp: f(1-delta)}). We have
\begin{align}
    f(a\delta; n=1-\delta) &= (a\delta)^{1-\delta}+(1-\delta)a\delta - \delta \nonumber \\
    &= a\delta(a\delta)^{-\delta} + a\delta - \delta + \order{\delta^2} \nonumber \\
    &= a\delta e^{-\delta\ln(a\delta)} + a\delta - \delta + \order{\delta^2} \nonumber \\
    &= \delta(2a-1) + \order{\delta^2}, \label{eq supp: f(a delta)}
\end{align}
where in going from the third to fourth line we used $e^{-\delta\ln(a\delta)}=1-\delta\ln(a\delta)+\order{\delta^2}$. Setting Eq.~(\ref{eq supp: f(a delta)}) to zero and taking $\delta\to 0$ we find $a\to 1/2$. Thus we have found that 
\be
    \lim_{n\uparrow 1}\; t_f(n) = \frac{\delta}{2} + \order{\delta^2}. \label{eq supp: t upper}
\ee

Now we expand the denominator of Eq.~(\ref{eq supp: alpha n}),
\begin{align}
    \ln\left(\frac{1-n}{1+n}\right) &= \ln\left(\frac{\delta}{2-\delta}\right) \nonumber \\
    &= \ln \delta - \ln(2-\delta) \nonumber \\
    &= \ln\delta - \ln 2 + \frac{\delta}{2} + \order{\delta^2}. \label{eq supp: alpha denom upper}
\end{align}

Finally, using Eqs.~(\ref{eq supp: t upper}) and (\ref{eq supp: alpha denom upper}), we can evaluate the limit of Eq.~(\ref{eq supp: alpha n}) as $n\uparrow 1$,
\begin{align}
    \lim_{n\uparrow 1}\; \alpha(n) &= \lim_{\delta \downarrow 0}\; \alpha(1-\delta) \nonumber \\
    &= \frac{(1-\delta)\left[\ln\delta - \ln2 + \order{\delta}\right]}{\ln\delta - \ln 2 + \frac{\delta}{2} + \order{\delta^2}} \nonumber \\
    &\to \frac{\ln\delta}{\ln\delta} = 1. \label{eq supp: alpha upper}
\end{align}

\subsubsection{A simple linear interpolation for \texorpdfstring{$\alpha(n)$}{}}
Combining the results from Eqs.~(\ref{eq supp: alpha lower}) and (\ref{eq supp: alpha upper}) we see that as $n$ varies from $0\to 1$, $\alpha(n)$ varies smoothly from $0.639\dots$ to $1$. We do not know the exact functional form of $\alpha(n)$ but based on these limits we can approximate it with a linear interpolation between the points $2/3$ and $1$, i.e.
\be
    \alpha(n) = \frac{n}{3} + \frac{2}{3}.
\ee
Thus our approximation for the root of Eq.~(\ref{eq supp: trinomial in t}) is [using Eq.~(\ref{eq supp: t_f defn})]
\be
    t_f(n) = \left(\frac{1-n}{1+n}\right)^{\frac{n+2}{3n}}.
\ee
This approximation is simple but surprisingly accurate (see the interactive plot at Ref.~\cite{github_zealots}). 

Transforming back to the original variable $z$ using Eq.~(\ref{eq supp: z->t transform}) we obtain an approximation for the root of Eq.~(\ref{eq supp: trinomial original}),
\be
    z_f(n) \equiv \frac{n}{1-n}\left(\frac{1-n}{1+n}\right)^{\frac{n+2}{3n}}.
\ee
Finally, we change variables using $n=1-\frac{1}{q}$,
\be
    z_f(q) = (q-1)(2q-1)^{\frac{q-\frac{1}{3}}{1-q}},
\ee
which is Eq.~(\ref{eq supp: absorbing zealots approx solution}).

\subsubsection{A more accurate interpolation}
We know the endpoints,
\begin{align}
\begin{split} \label{eq supp: alpha bounds}
    \alpha(0) &= \frac{1+W(e^{-1})}{2}, \\
    \alpha(1) &= 1.
\end{split}
\end{align}
Consider a generic nonlinear interpolation between these endpoints,
\be
    \alpha(n) = \frac{1}{2}\left[(1+n) + W(e^{-1})(1-n)^p\right], \label{eq supp: alpha n nonlinear interp}
\ee
where $p$ controls the curvature. The above equation satisfies the boundary conditions in Eq.~(\ref{eq supp: alpha bounds}). A simple search suggests that a value of $p\approx 1.43$ works well (see the interactive plot at Ref.~\cite{github_zealots}).

Using Eq.~(\ref{eq supp: alpha n nonlinear interp}) in Eq.~(\ref{eq supp: t_f defn}), transforming back to $z_f(n)$ with Eq.~(\ref{eq supp: z->t transform}), then replacing $n=1-\frac{1}{q}$, we find
\be
    z_f(q) = (q-1)(2q-1)^{-\frac{1}{2(q-1)}\left[2q-1+q^{1-p}W(e^{-1})\right]}.
\ee

\section{Evolutionary games} \label{appendix: games}

\subsection{Balanced zealots} \label{appendix: games balanced}
For balanced zealots Eq.~(\ref{eq main: games general rate eqn}) becomes
\begin{align}
    \frac{\dd x}{\dd t} &= (1-x-z)\left(x+\frac{z}{2}\right) \nonumber \\ 
    &\quad \times\Bigg[\alpha\left(x+\frac{z}{2}\right)+(1-\alpha)\left(1-x-\frac{z}{2}\right)\Bigg] \nonumber \\
    &\quad -x\left(1-x-\frac{z}{2}\right) \nonumber \\
    &\quad \times \Bigg[(1-\alpha)\left(x+\frac{z}{2}\right)+\alpha\left(1-x-\frac{z}{2}\right)\Bigg]. \label{eq supp: games balanced rate eq}
\end{align}
This rate equation has up to three fixed points. The central fixed point $x_{2}^{*}=\frac{1}{2}(1-z)$ always exists. The other two fixed points, $x_{1}^{*}$ and $x_{3}^{*}$, sit on either side of the central fixed point, i.e. $x_{1}^{*}<x_{2}^{*}<x_{3}^{*}$. These are
\be
    x_{1, 3}^{*} = x_{2}^{*} \pm \frac{1}{2}\sqrt{1+\frac{2\alpha z}{1-2\alpha}},
\ee
and are only physical when $\alpha>1/2$ and $z<z_{c}(\alpha)$, where
\be
    z_{c}(\alpha) = 1-\frac{1}{2\alpha}. \label{eq supp: games balanced zC}
\ee

The stabilities of the various fixed points can be deduced by looking an interactive plot of Eq.~(\ref{eq supp: games balanced rate eq}) \cite{github_zealots}. There are three scenarios:
\begin{itemize}
    \item If $\alpha < 1/2$, only $x_{2}^{*}$ is physical and it is stable.
    \item If $\alpha > 1/2$: 
    \begin{itemize}
        \item If $z<z_{c}(\alpha)$, all three fixed points are physical, $x_{2}^{*}$ is stable while $x_{1,3}^{*}$ are unstable.
        \item If $z \geq z_c(\alpha)$, only $x_{2}^{*}$ is physical and it is stable.
    \end{itemize}
\end{itemize}

Flow and bifurcation diagrams for evolutionary games with balanced zealots are analogous to those for the nonlinear voter model, i.e. Figs.~\ref{fig supp: nonlinear balanced flow} and \ref{fig supp: nonlinear balanced bifurcation}.

\subsection{One-sided zealots} \label{appendix: games absorbing}
For one-sided zealots Eq.~(\ref{eq main: games general rate eqn}) becomes
\begin{align}
    \frac{\dd x}{\dd t} &= (1-x-z)(x+z) \nonumber \\
    &\quad \times \big[\alpha(x+z)+(1-\alpha)(1-x-z)\big] \nonumber \\ 
    &\quad -x(1-x-z) \nonumber \\
    &\quad \times \big[(1-\alpha)(x+z)+\alpha(1-x-z)\big]. \label{eq supp: games absorbing rate eq}
\end{align}
This rate equation always has the consensus fixed point $x_{c}^{*}=1-z$. There are potentially two other fixed points, we find these to be
\be
    x_{\pm}^{*} = \frac{1}{4}\Bigg[1-3z\pm\sqrt{(1-z)^{2}+\frac{4z}{1-2\alpha}}\Bigg].
\ee
The situation splits based on the value of $\alpha$ relative to $1/2$:
\begin{itemize}
    \item If $\alpha<1/2$, only $x_+^*$ has the potential to be physical, i.e. to be located in $(0,1-z)$, and this will only be the case provided that
    \be
        z < z_c^{(1)} \equiv \frac{2\alpha -1}{\alpha -1}. \label{eq supp: games absorbing zc1}
    \ee
    We note that when $z=z_c^{(1)}$ exactly, $x_+^*=x_c^*$.
    \item If $\alpha>1/2$, both $x_{\pm}^*$ are physical provided that
    \be
        z < z_c^{(2)} \equiv \frac{2\alpha+1-2\sqrt{2\alpha}}{2\alpha-1}. \label{eq supp: games absorbing zc2} 
    \ee
    We note that when $z=z_c^{(2)}$ exactly, $x_+^* = x_-^*$.
\end{itemize}
Fig.~\ref{fig main: absorbing phase plot}(b) in the main paper contains plots of $z_{c}^{(1,2)}(\alpha)$ against $\alpha$.

The stabilities of the various fixed point $x_{c, \pm}^{*}$ can be deduced using the an interactive plot of Eq.~(\ref{eq supp: games absorbing rate eq}) \cite{github_zealots}. Flow diagrams for evolutionary games with one-sided zealots are analogous to those for the nonlinear voter model, i.e. Fig.~\ref{fig supp: nonlinear absorbing flow}.

\begin{figure}[hbtp]
    \centering
    \includegraphics[scale=0.62]{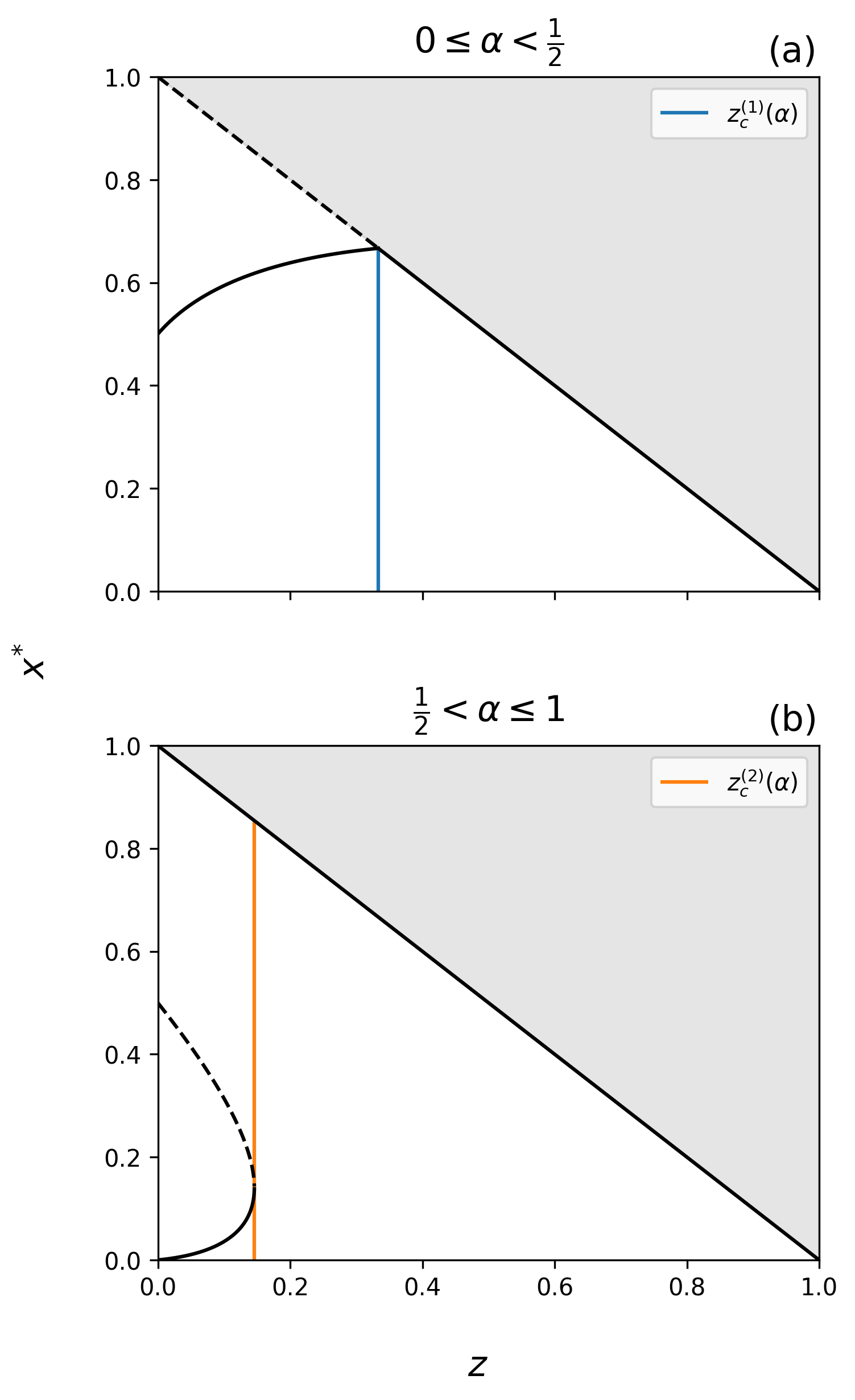}
    \caption{Bifurcation diagrams for evolutionary games with one-sided zealots, showing how the fixed points of $\dot{x}$ [Eq.~(\ref{eq supp: games absorbing rate eq})] change as we vary $z$ at (a) $\alpha=0.4$ and (b) $\alpha=0.9$. Solid/dashed black lines mean the fixed point is stable/unstable. The vertical orange and blue lines are $z_{c}^{(1)}(\alpha)$ and $z_{c}^{(2)}(\alpha)$ respectively from Eqs.~(\ref{eq supp: games absorbing zc1}) and (\ref{eq supp: games absorbing zc2}). The grey shaded region is not a valid region as only $x\in[0,1-z]$ is allowed.}
    \label{fig supp: games absorbing bifurcation}
\end{figure}

In Figs.~\ref{fig supp: games absorbing bifurcation}(a) and \ref{fig supp: games absorbing bifurcation}(b) we show bifurcation diagrams for the cases $\alpha=0.4$ and $\alpha=0.9$ respectively, as we vary $z$.

\subsection{Unbalanced zealots} \label{appendix: games unbalanced}
For unbalanced zealots Eq.~(\ref{eq main: games general rate eqn}) becomes
\begin{align}
    \frac{\dd x}{\dd t} &= (1-x-z)\left(x+\frac{1+\delta}{2}z\right) \nonumber \\
    &\quad \times \Bigg[\alpha\left(x+\frac{1+\delta}{2}z\right) \nonumber \\
    &\quad +(1-\alpha)\left(1-x-\frac{1+\delta}{2}z\right)\Bigg] \nonumber \\ 
    &\quad -x\left(1-x-\frac{1+\delta}{2}z\right) \nonumber \\
    &\quad \times\Bigg[(1-\alpha)\left(x+\frac{1+\delta}{2}z\right) \nonumber \\
    &\quad +\alpha\left(1-x-\frac{1+\delta}{2}z\right)\Bigg], \label{eq supp: games unbalanced rate eq}
\end{align}

When $\alpha < 1/2$, there is always only one fixed point in the interior and it is always stable, much like Fig.~\ref{fig supp: nonlinear balanced flow}(a).

When $\alpha > 1/2$ we can numerically determine $z_{c}(\alpha)$. The flow diagrams are analogous to the nonlinear voter model with unbalanced zealots, i.e. Fig.~\ref{fig supp: nonlinear balanced flow}. Fig.~\ref{fig main: unbalanced phase plot}(b) in the main paper contains plots of $z_{c}(\alpha)$ for different values of $\delta$.

\subsection{Nonlinear evolutionary dynamics} \label{appendix: nonlinear games}
The rate equation for nonlinear evolutionary dynamics with one-sided zealots is as in Eq.~(\ref{eq supp: games absorbing rate eq}) except with a nonlinear social impact function $x^{q}$,
\begin{align}
    \frac{\dd x}{\dd t} &= (1-x-z)(x+z)^{q} \nonumber \\
    &\quad \times \big[\alpha(x+z)+(1-\alpha)(1-x-z)\big] \nonumber \\ 
    &\quad -x(1-x-z)^{q} \nonumber \\
    &\quad \times \big[(1-\alpha)(x+z)+\alpha(1-x-z)\big]. \label{eq supp: nonlinear games rate eq}
\end{align}
We wish to determine the fixed points and the corresponding stabilities for this equation. We see that the consensus fixed point $x_c^{*}=1-z$ is always a fixed point of Eq.~(\ref{eq supp: nonlinear games rate eq}) as expected, as this is a one-sided zealots scenario. To determine the stability, and to find the remaining interior fixed points and their stabilities, we use similar techniques to those in Appendix~\ref{appendix: nonlinear vm}. We first define a new function, by setting the RHS of Eq.~(\ref{eq supp: nonlinear games rate eq}) to zero and then rearranging,
\begin{align}
    E &\equiv \log\left(\frac{x+z}{1-x-z}\right) \nonumber \\
    &\quad -\frac{1}{q}\log\left[{\frac{\alpha(x+z)+(1-\alpha)(1-x-z)}{(1-\alpha)(x+z)+\alpha(1-x-z)}}\right].
\end{align}
The roots of this function correspond to the interior fixed points. $E$ is always of opposite sign to $\dot{x}$, which can be shown by starting with $\dot{x}>0$ and rearranging to find $E<0$. 

As was done in Appendix~\ref{appendix: nonlinear vm} it is possible to determine the flow for a given $\alpha$, $q$ and $z$ by looking at the limits of $E$ as $x\downarrow 0$ and $x\uparrow (1-z)$, as well as its stationary points. See also the interactive plot at Ref.~\cite{github_zealots}. Fig.~\ref{fig main: nonlinear games} shows phase diagrams that result from this analysis. The purple boundary lines are the critical zealotry lines $z_c(q, \alpha)$, which are determined numerically.

\section{The partisan voter model} \label{appendix: partisan}

\subsection{Balanced zealots} \label{appendix: partisan balanced}
For balanced zealots the restrictions on $(\Delta, \Sigma)$ from Eqs.~(\ref{eq main: partisan conditions}) become
\begin{align} \label{eq supp: partisan conditions balanced}
    \begin{split}
        \Sigma-(1-z) \leq &\Delta \leq -\Sigma+(1-z), \\
        -\Sigma \leq &\Delta \leq \Sigma.
    \end{split}
\end{align}

The rate equations from Eqs.~(\ref{eq main: partisan general rate eqns}) become
\begin{subequations}
\begin{align}
    \dot{\Delta} &= -\Delta\Big[(1+\epsilon)z-\epsilon(1-2\Sigma)\Big], \label{eq supp: partisan delta dot balanced} \\
    \dot{\Sigma} &= \frac{1}{2}(1+\epsilon)(1-z)-2\epsilon\Delta^{2}-\Sigma. \label{eq supp: partisan sigma dot balanced}
\end{align}
\end{subequations}

For $0<\epsilon<1$ we find that these rate equations only have one physical fixed point which is in the interior,
\be 
    I \equiv \left(0, \frac{1}{2}(1+\epsilon)(1-z)\right). \label{eq supp: partisan balanced zealots FP}
\ee
Linear stability analysis reveals this fixed point is always a sink.

The nullclines of the system, i.e. the lines in $(\Delta, \Sigma)$-space where $\dot{\Delta}=0$ and $\dot{\Sigma}=0$ can be found using Eqs.~(\ref{eq supp: partisan delta dot balanced}) and (\ref{eq supp: partisan sigma dot balanced}). One has $\dot\Delta=0$ when
\be
    \Delta = 0,~\mbox{or}~~ \\
    \Sigma = \frac{\epsilon-(1+\epsilon)z}{2\epsilon},  \label{eq supp: partisan balanced delta=0 nullcline2}
\ee
and $\dot\Sigma=0$ when
\be
    \Sigma = \frac{1}{2}(1+\epsilon)(1-z)-2\epsilon\Delta^{2}. \label{eq supp: partisan balanced sigma=0 nullcline}
    \ee

\begin{figure*}[hbtp]
    \centering
    \includegraphics[scale=0.19]{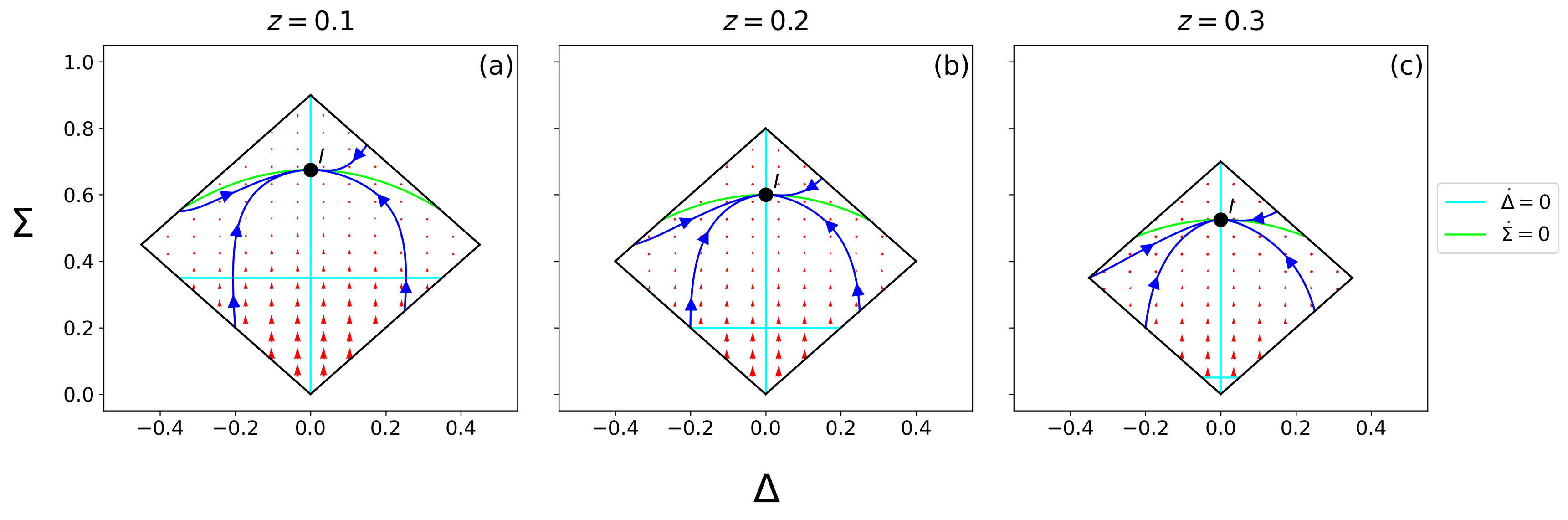}
    \caption{Flow diagrams in $(\Delta, \Sigma)$-space for the partisan voter model with balanced zealots. $z=(0.1, 0.2, 0.3)$ in panels (a)-(c), and $\epsilon=0.5$ in all. The rectangular region is the allowed region defined by Eqs.~(\ref{eq supp: partisan conditions balanced}). Red arrows indicate the flow field defined by Eqs.~(\ref{eq supp: partisan delta dot balanced}) and (\ref{eq supp: partisan sigma dot balanced}), the larger the arrow the faster the flow. The filled circle marker is the single stable interior fixed point [Eq.~(\ref{eq supp: partisan balanced zealots FP})]. The dark blue lines are example deterministic trajectories starting from various points in the plane, determined by numerically integrating Eqs.~(\ref{eq supp: partisan delta dot balanced}) and (\ref{eq supp: partisan sigma dot balanced}). The cyan line is the $\dot{\Delta}=0$ nullcline [Eqs.~(\ref{eq supp: partisan balanced delta=0 nullcline2})]. The lime green line is the $\dot{\Sigma}=0$ nullcline [Eq.~(\ref{eq supp: partisan balanced sigma=0 nullcline})].}
    \label{fig supp: partisan balanced zealots flow}
\end{figure*}

In Fig.~\ref{fig supp: partisan balanced zealots flow} we show the $(\Delta, \Sigma)$-space for different values of $z$ and fixed $\epsilon$, along with some deterministic trajectories from numerically integrating Eqs.~(\ref{eq supp: partisan delta dot balanced}) and (\ref{eq supp: partisan sigma dot balanced}). There is also an interactive plot at Ref.~\cite{github_zealots}.

\subsection{One-sided zealots} \label{appendix: partisan absorbing}
For one-sided zealotry the restrictions on $(\Delta, \Sigma)$ [Eqs.~(\ref{eq main: partisan conditions})] are identical to the conditions for balanced zealots [Eqs.~(\ref{eq supp: partisan conditions balanced})].

The rate equations from Eqs.~(\ref{eq main: partisan general rate eqns}) become
\begin{subequations}
\begin{align}
    \dot{\Delta} &= \epsilon\Delta(1-2\Sigma)+\frac{1}{2}z\Big[(1+\epsilon)(1-2\Delta)-2\epsilon\Sigma\Big] \nonumber \\
    &\quad -\frac{1}{2}z^{2}(1+\epsilon), \label{eq supp: partisan delta dot absorbing} \\
    \dot{\Sigma} &= \frac{1}{2}(1+\epsilon)(1-z)-\epsilon\Delta(z+2\Delta)-\Sigma. \label{eq supp: partisan sigma dot absorbing}
\end{align}
\end{subequations}
These rate equations have at most two fixed points. There is a consensus fixed point which always exists,
\be
    C^{+} = \left(\frac{1-z}{2}, \frac{1-z}{2}\right). \label{eq supp: partisan absorbing C+}
\ee
This corresponds to the scenario where all agents are in the $+1$ state [we note that in the case $z_{+}=0$, $z_{-}=z$, we would have found a consensus fixed point at $C^{-}=\frac{1}{2}(z-1, 1-z)$, corresponding to all agents in the $-1$ state]. There is also an interior fixed point $I$ (the full expression can be found at \cite{github_zealots}) which exists provided that
\be
    z < z_{c}(\epsilon) \equiv \frac{2\epsilon^{2}}{1+\epsilon^{2}}. \label{eq supp: partisan absorbing zC}
\ee
Fig.~\ref{fig main: absorbing phase plot}(c) in the main paper contains a plot of $z_c(\epsilon)$ against $\epsilon$.

We can determine the stability of these fixed points with linear stability analysis. When the interior fixed point $I$ exists, i.e. when $z<z_{c}(\epsilon)$, it is always a sink, while the consensus fixed point $C^{+}$ is always a saddle point. When $z\geq z_{c}(\epsilon)$, i.e. only the consensus fixed point $C^{+}$ exists, it is always a sink.

The nullclines of the system, found using Eqs.~(\ref{eq supp: partisan delta dot absorbing}) and (\ref{eq supp: partisan sigma dot absorbing}), are
\begin{subequations}
\begin{gather}
    \Sigma = \frac{1}{2}\left(1-z-\frac{z(z-1+2\Delta)}{\epsilon(z+2\Delta)}\right), \label{eq supp: partisan absorbing delta=0 nullcline} \\
    \Sigma = \frac{1}{2}(1+\epsilon)(1-z)-\epsilon\Delta(z+2\Delta). \label{eq supp: partisan absorbing sigma=0 nullcline} 
\end{gather}
\end{subequations}

In Fig.~\ref{fig main: partisan flow} in the main paper we show the $(\Delta, \Sigma)$-space for the two possible scenarios. A further interactive plot can be found at Ref.~\cite{github_zealots}.

\subsection{Unbalanced zealots} \label{appendix: partisan unbalanced}
For unbalanced zealots the restrictions on $(\Delta, \Sigma)$ from Eqs.~(\ref{eq main: partisan conditions}) are again the same as in the balanced and one-sided zealots cases, i.e. Eq.~(\ref{eq supp: partisan conditions balanced}).

The rate equations from Eqs.~(\ref{eq main: partisan general rate eqns}) reduce to
\begin{align}
\begin{split}
    \dot{\Delta} &= \epsilon\Delta(1-2\Sigma) \\
    &\quad +\frac{1}{2}\delta z\Bigg\{(1+\epsilon)(1-z) - 2\Big[\Delta+\epsilon(\Delta+\Sigma)\Big]\Bigg\}, \\
    \dot{\Sigma} &= \frac{1}{2}(1+\epsilon)(1-z) -\epsilon\Delta(\delta z+2\Delta)-\Sigma.
\end{split}
\end{align}

These rate equations are complicated but it can be shown that they only have one physical fixed point. This fixed point is always an interior fixed point and, by evaluating the eigenvalues of the Jacobian numerically at this fixed point, it appears to always be a sink. This case is then very similar to balanced zealots (Appendix~\ref{appendix: partisan balanced}), except the interior fixed point has more freedom to move around the allowed region.

\subsection{Social impact function} \label{appendix: partisan social impact}
Using Eqs.~(\ref{eq main: partisan social impact rate eqn}) we form the Jacobian matrix evaluated at $C^+$ [Eq.~(\ref{eq main: partisan C+})],
\be
    \bJ =
    \begin{bmatrix}
        -1+(1-z)f'(0) & -\epsilon \\
        -\epsilon[1+(1-z)f'(0)] & -1
    \end{bmatrix},
\ee
where we have used the boundary conditions $f(0)=0$ and $f(1)=1$. The derivative $f'(0)$ is to be understood as the right-hand derivative. 

The eigenvalues of this Jacobian are
\begin{align}
    \lambda_{\pm} &= \frac{1}{2}\Big\{-2+(1-z)f'(0) \nonumber \\
    &\quad \pm \sqrt{(1-z)^2f'(0)^2+4\epsilon^2[1+(1-z)f'(0)]}\Big\},
\end{align}
which are always real. Further, $\lambda_-<0$ always. We will have $\lambda_+<0$ also provided that
\be
    (1-z)\frac{1+\epsilon^2}{1-\epsilon^2}f'(0) < 1.
\ee
When the above condition holds true, both eigenvalues are real and negative, hence the consensus fixed point is locally stable.

\subsection{Nonlinear partisan voter model} \label{appendix: nonlinear partisan}
For the partisan voter model with a nonlinear social impact function $f(x)=x^q$ we know from Eq.~(\ref{eq main: partisan f'(0) condition}) that the consensus fixed point $C^+$ is stable when $q<1$ and unstable when $q>1$.

\begin{figure}[hbtp]
    \centering
    \includegraphics[scale=0.25]{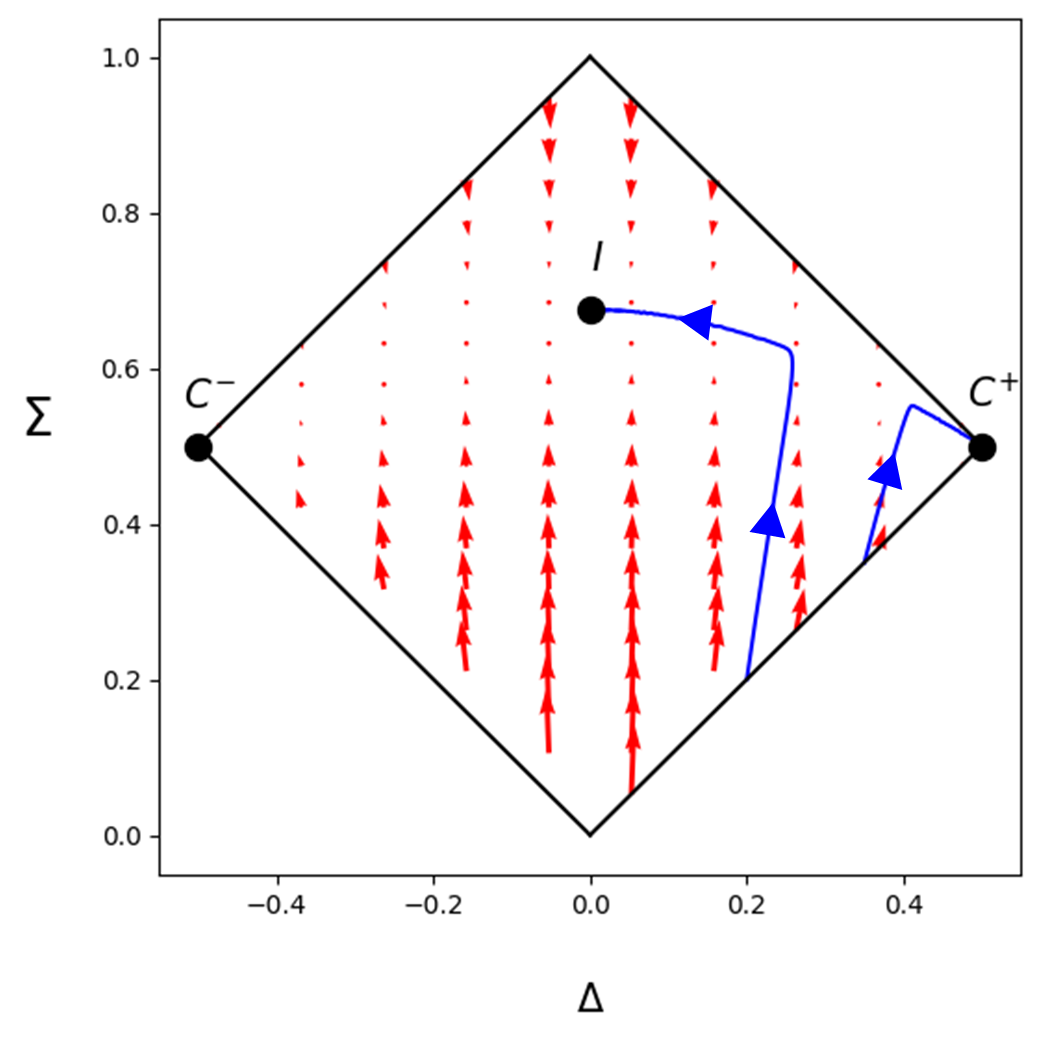}
    \caption{Flow diagram in the $(\Delta, \Sigma)$-space for the nonlinear partisan voter model with no zealots, $q=1.1$, $\epsilon=0.35$. The rectangular region is the allowed region defined by Eqs.~(\ref{eq supp: partisan conditions balanced}). Red arrows indicate the flow field defined by Eqs.~(\ref{eq main: partisan social impact rate eqn}) with $f(x)=x^q$, the larger the arrow the faster the flow. The filled circle markers are stable fixed points (see text). The dark blue lines are example deterministic trajectories starting from various points in the plane, determined by numerically integrating Eqs.~(\ref{eq main: partisan social impact rate eqn}).}
    \label{fig supp: nonlinear partisan no zealots}
\end{figure}

We note however that when $q>1$, consensus is not guaranteed even though $C^+$ is stable, as there are potentially other stable fixed points in the interior. We omit a full analysis with general $z$. However, consider the simple case $z=0$, where there  is a coexistence fixed point $I=[0, (1+\epsilon)/2]$. The eigenvalues of the Jacobian evaluated at this point are
\begin{align}
\begin{split}
    \lambda_1 &= -2^{1-q}, \\
    \lambda_2 &= -2^{1-q}[1-q(1-\epsilon^2)].
\end{split}
\end{align}
Hence $I$ is stable if $\epsilon > \sqrt{1-\frac{1}{q}}$. Note that in this scenario $C^+$ (and $C^-$ which now exists since $z=0$) are also stable. So trajectories started from different points in the $(\Delta, \Sigma)$-plane absorb at different fixed points. This is illustrated in Fig.~ \ref{fig supp: nonlinear partisan no zealots}.

\section{Fixation times} \label{appendix: fixation times}

\subsection{Derivation of Eq.~(\ref{eq main: fixation time})} \label{appendix: fixation times deriv}
We broadly follow the steps in Ref.~\cite{traulsen2009stochastic}. The unconditional fixation time $t_j$ starting from state $j$ is determined by
\begin{gather}
    t_{j} = 1+T^{+}_{j}t_{j+1} + T^-_{j}t_{j-1} + (1-T_j^+ - T_j^-)t_j, \nonumber \\
    \implies t_{j+1} - t_j = \gamma_j(t_j-t_{j-1})-\frac{1}{T_j^+}, \label{eq supp: time 1}
\end{gather}
where $\gamma_j \equiv T_j^-/T_j^+$. We now define $z_j\equiv t_j-t_{j-1}$. Eq.~(\ref{eq supp: time 1}) can then be written as
\be
    z_{j+1} = \gamma_j z_j - \frac{1}{T_j^+}. \label{eq supp: absorb time z recurrence}
\ee
Eqs.~(\ref{eq supp: time 1}) and (\ref{eq supp: absorb time z recurrence}) only hold for $0<j<N-Z$. Noting that $j=0$ is not an absorbing state, we also have
\begin{gather}
    t_0 = 1 + T_0^+ t_1 + (1-T_0^+)t_0, \nonumber \\
    \implies t_0 = t_1 + \frac{1}{T_0^+}. \label{eq supp: absorb time t0}
\end{gather}
Now we use the definition of $z_j$ and Eq.~(\ref{eq supp: absorb time t0}) to compute $z_1$:
\be
    z_1 = t_1 - t_0 = -\frac{1}{T_0^+}.
\ee
We can now use Eq.~(\ref{eq supp: absorb time z recurrence}) to compute all other $z_j$ (for $k<N-Z$),
\be
    z_k = -\frac{1}{T_0^+}\prod_{m=1}^{k-1}\gamma_m -\sum_{l=1}^{k-1}\frac{1}{T_l^+}\prod_{m=l+1}^{k-1}\gamma_m. 
    \label{eq supp: absorb time zk}
\ee
Next we see
\begin{align}
    \sum_{k=i+1}^{N-Z}z_k = -t_{i}, \label{eq supp: absorb time zk sum}
\end{align}
where we use $t_{N-Z}=0$ (which holds by definition). The unconditional fixation times can then be obtained by combining Eqs.~(\ref{eq supp: absorb time zk})  and (\ref{eq supp: absorb time zk sum}),
\be
    t_i = \frac{1}{T_0^+}\sum_{k=i}^{N-Z-1}\prod_{m=1}^{k}\gamma_m + \sum_{k=i}^{N-Z-1}\sum_{l=1}^{k}\frac{1}{T_l^+}\prod_{m=l+1}^{k}\gamma_m,
\ee
which is Eq.~(\ref{eq main: fixation time}).

\subsection{Evolutionary games and the partisan voter model} \label{appendix: fixation times games}
For evolutionary games the rates in finite populations are defined analogously to Eqs.~(\ref{eq main: games rates}),
\begin{align}
\begin{split} \label{eq supp: games finite rates}
    T^{+}_n &= N\frac{S-n}{N}\frac{n+Z}{N-1}\Pi_+, \\
    T^{-}_n &= N\frac{n}{N}\frac{S-n}{N-1}\Pi_-,
\end{split}
\end{align}
where the $\Pi_{\pm}$ are the analogue Eqs.~(\ref{eq main: games expected payoff +}) and (\ref{eq main: games expected payoff -}) for finite populations,
\begin{align}
\begin{split}
    \Pi_+ &= \alpha\frac{n+Z}{N-1} + (1-\alpha)\frac{S-n}{N-1}, \\
    \Pi_- &= (1-\alpha)\frac{n+Z}{N-1} + \alpha\frac{S-n}{N-1}.
\end{split}
\end{align}

\begin{figure}[hbtp]
    \centering
    \includegraphics[scale=0.6]{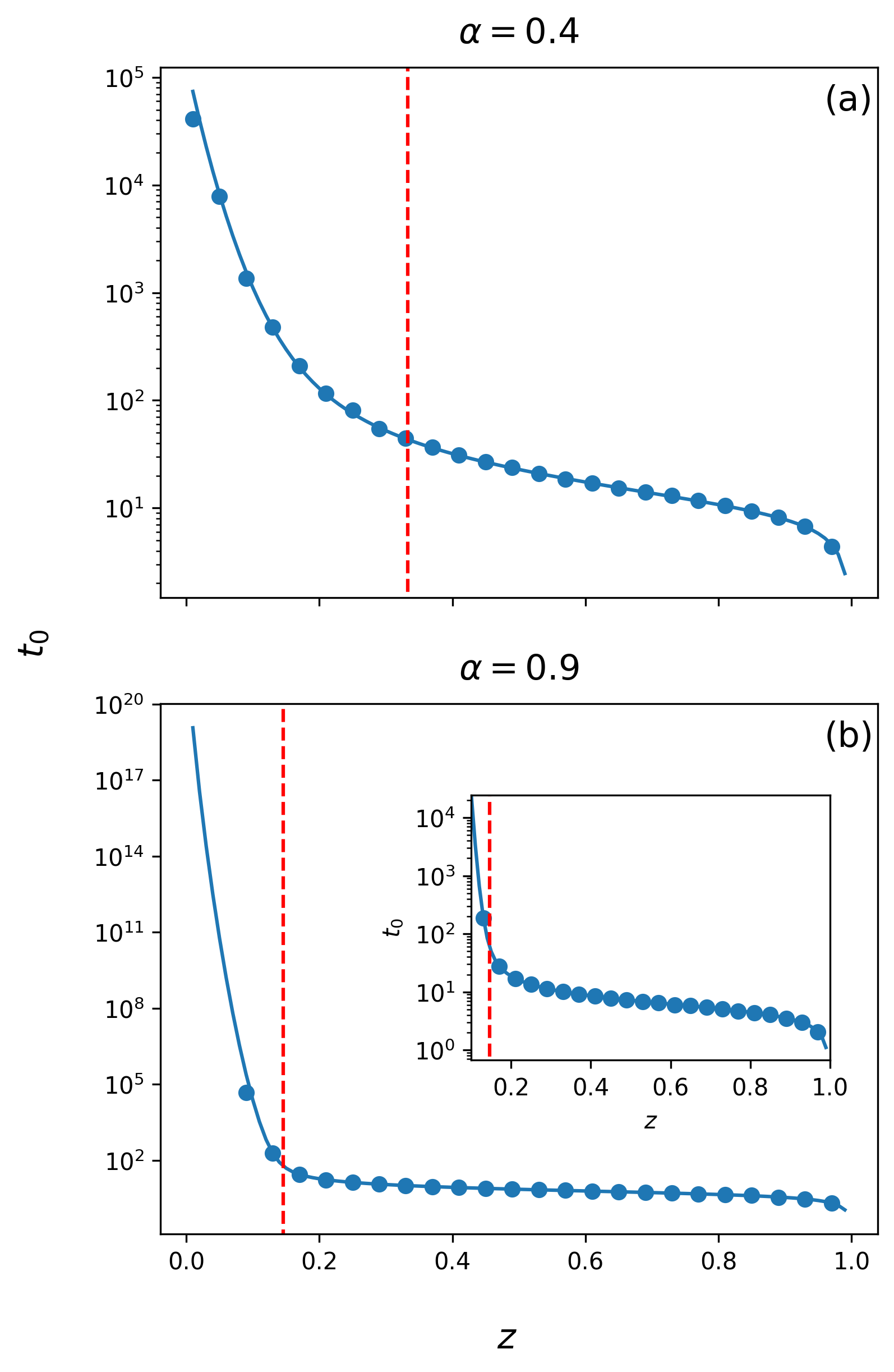}
    \caption{Plots of fixation time $t_0$ against $z$ for evolutionary games with one-sided zealots in a finite population of $N=100$. $t_0$ is defined as the time it takes the system to reach fixation, i.e. all susceptibles are type $+1$, starting from an initial configuration where all susceptibles are type $-1$. We note that the $t_0$ axis is logarithmic. The blue dots are from averaging 1000 independent simulations using the Gillespie algorithm \cite{gillespie1976general}. We note that in panel (b) we are only able to perform simulations above a particular $z$, as the fixation times become too extreme below this. The solid blue lines are the analytical solutions from Eq.~(\ref{eq main: fixation time}). Panels (a) and (b) are for $\alpha=0.4$ and $\alpha=0.9$ respectively. We indicate the critical zealotry $z_c(\alpha)$ [Eq.~(\ref{eq main: games zC})] for a corresponding infinite population by a vertical dashed red line. Panel (b) contains an inset which is a zoomed in version of the main panel.}
    \label{fig supp: games absorption}
\end{figure}

We now use Eq.~(\ref{eq main: fixation time}) with the rates in Eqs.~(\ref{eq supp: games finite rates}) to calculate the fixation time $t_{0}$ for different $\alpha$. This is plotted in Fig.~\ref{fig supp: games absorption}. See the main text at the end of Sec.~\ref{sec: fixation times} for an explanation of these results.

\begin{figure}[hbtp]
    \centering
    \includegraphics[scale=0.48]{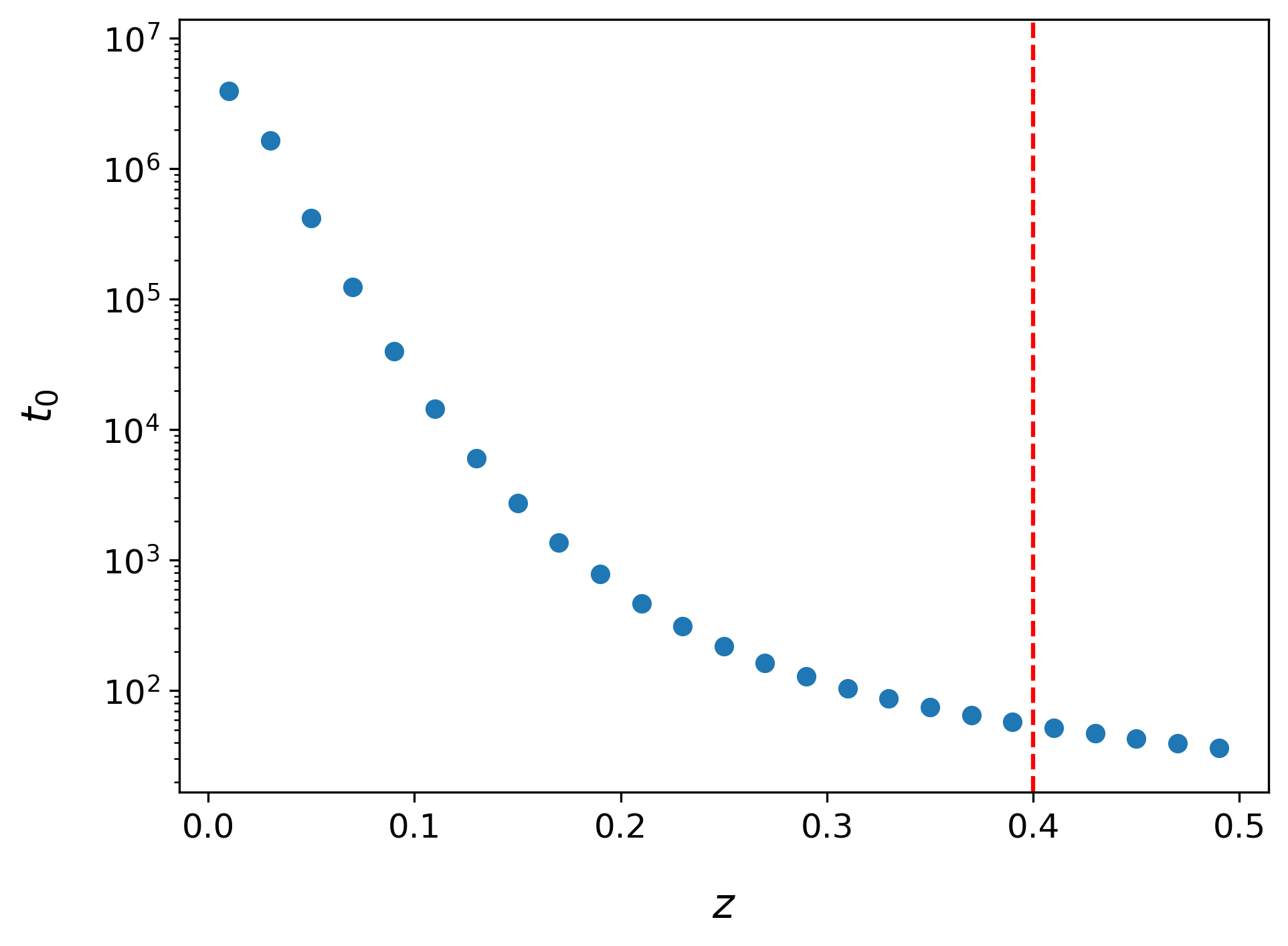}
    \caption{Plots of fixation time $t_0$ against $z$ for the partisan voter model ($\epsilon=0.5$) with one-sided zealots in a finite population of $N=100$. $t_0$ is defined as the time it takes the system to reach fixation, i.e. all susceptibles are type $+1$, starting from an initial configuration where all susceptibles are type $-1$. We note that the $t_0$ axis is logarithmic. The blue dots are from averaging $10^{5}$ independent simulations using the Gillespie algorithm \cite{gillespie1976general}. We indicate the critical zealotry $z_c(\epsilon)$ [Eq.~(\ref{eq main: partisan zC})] for a corresponding infinite population by a vertical dashed red line.}
    \label{fig supp: partisan absorption}
\end{figure}

In Fig.~\ref{fig supp: partisan absorption} we plot the average fixation time against $z$ for the partisan voter model with $\epsilon=0.5$. The population has size $N=100$ and is initialised with no $+1$ state agents and an equal proportion of agents preferring the $\pm 1$ states, i.e. $x_{+}^{\pm}=0$ and $x_{-}^{\pm}=\frac{1}{2}(1-z)$.

\section{Quasi-stationary distributions} \label{appendix: quasi-stationary distributions}
Here we look to calculate the quasi-stationary probability distribution of the number $n$ of susceptible individuals in state $+1$.

$P_{n}(t)$ is the probability for the system to be in state $n$ at time $t$. We are considering a continuous time Markov process for $n$ with transition rates $T^{\pm}_n$ (for events $n\to n\pm 1$) and a single absorbing state at $n=N-Z$. The master equations for this process are
\begin{subequations}
\begin{align}
    \frac{\dd P_0}{\dd t} &= T_{1}^-P_1(t) - T_0^+ P_0(t), \label{eq supp: master eqn 0} \\
    \frac{\dd P_n}{\dd t} &= T_{n-1}^{+}P_{n-1}(t) + T_{n+1}^{-}P_{n+1}(t) \nonumber \\
    &\quad - (T_{n}^{+}+T_{n}^{-})P_{n}(t), \label{eq supp: master eqn n} \\
    \frac{\dd P_{N-Z}}{\dd t} &= T_{N-Z-1}^{+}P_{N-Z-1}(t), \label{eq supp: master eqn N-Z}
\end{align}
\end{subequations}
where Eq.~(\ref{eq supp: master eqn n}) is valid for $n=1,2,\dots, N-Z-1$.

Following e.g. Ref.~\cite{naasell1996quasi}, we now define the probability of the system to be in state $n$ after time $t$ conditioned on not having absorbed at $n=N-Z$, i.e. $Q_n(t)=P(n|n\neq N-Z; t)$. This is given by
\be
    Q_n(t) = \frac{P_n(t)}{1-P_{N-Z}(t)}, \label{eq supp: Qn}
\ee
for $n=0,1,2,\dots,N-Z-1$. Differentiation with respect to time and using Eqs.~(\ref{eq supp: master eqn 0})-(\ref{eq supp: master eqn N-Z}) we find
\begin{align}
    \frac{\dd Q_0(t)}{\dd t} &= T_1^-Q_1(t) \nonumber \\
    &\quad -\left[T_0^+ - T_{N-Z-1}^{+} Q_{N-Z-1}(t)\right]Q_0(t), \label{eq supp: quasi dQ0/dt}
\end{align}
and
\begin{align}
    \frac{\dd Q_n(t)}{\dd t} &= T_{n-1}^{+}Q_{n-1}(t) -(T_n^+ + T_n^-)Q_n(t) \nonumber \\
    &\quad + T_{n+1}^{-}Q_{n+1}(t) \nonumber \\
    &\quad + T_{N-Z-1}^{+}Q_{N-Z-1}(t)Q_{n}(t). \label{eq supp: quasi dQn/dt initial}
\end{align}
We wish to determine the quasi-stationary distribution, i.e $Q_n^*\equiv \lim_{t\to\infty}Q_n(t)$. This can be done by numerically integrating Eqs.~(\ref{eq supp: quasi dQ0/dt}) and (\ref{eq supp: quasi dQn/dt initial}) until convergence. It is helpful to recast Eqs.~(\ref{eq supp: quasi dQ0/dt}) and (\ref{eq supp: quasi dQn/dt initial}) as a matrix equation. To do this, we define $\bq(t)$ as the column vector containing all $Q_n(t)$ for $n=0,\dots,N-Z-1$, and $\bA$ as the $(N-Z)\times (N-Z)$ tridiagonal matrix
\small
\be
\bA = 
\begin{pmatrix}
    -T_0^+ & T_1^- &  &  \\
     T_0^+ & -(T_1^+ + T_1^-) & T_2^- & \\
      & T_1^+ & -(T_2^+ + T_2^-) & \ddots  \\
      & & \ddots & \ddots \\
      &  &  & 
\end{pmatrix}.
\ee
\normalsize
We can then write Eqs.~(\ref{eq supp: quasi dQ0/dt}) and (\ref{eq supp: quasi dQn/dt initial}) in matrix form,
\be
    \mathbf{\dot{q}}(t) = \bA\bq(t) + T_{N-Z-1}^{+}Q_{N-Z-1}(t)\bq(t). \label{eq supp: q master eqn matrix form}
\ee
This equation can be numerically integrated until convergence (for example using a simple Euler-forward scheme).

%


\begin{thebibliography}{67}%
\makeatletter
\providecommand \@ifxundefined [1]{%
 \@ifx{#1\undefined}
}%
\providecommand \@ifnum [1]{%
 \ifnum #1\expandafter \@firstoftwo
 \else \expandafter \@secondoftwo
 \fi
}%
\providecommand \@ifx [1]{%
 \ifx #1\expandafter \@firstoftwo
 \else \expandafter \@secondoftwo
 \fi
}%
\providecommand \natexlab [1]{#1}%
\providecommand \enquote  [1]{``#1''}%
\providecommand \bibnamefont  [1]{#1}%
\providecommand \bibfnamefont [1]{#1}%
\providecommand \citenamefont [1]{#1}%
\providecommand \href@noop [0]{\@secondoftwo}%
\providecommand \href [0]{\begingroup \@sanitize@url \@href}%
\providecommand \@href[1]{\@@startlink{#1}\@@href}%
\providecommand \@@href[1]{\endgroup#1\@@endlink}%
\providecommand \@sanitize@url [0]{\catcode `\\12\catcode `\$12\catcode `\&12\catcode `\#12\catcode `\^12\catcode `\_12\catcode `\%12\relax}%
\providecommand \@@startlink[1]{}%
\providecommand \@@endlink[0]{}%
\providecommand \url  [0]{\begingroup\@sanitize@url \@url }%
\providecommand \@url [1]{\endgroup\@href {#1}{\urlprefix }}%
\providecommand \urlprefix  [0]{URL }%
\providecommand \Eprint [0]{\href }%
\providecommand \doibase [0]{http://dx.doi.org/}%
\providecommand \selectlanguage [0]{\@gobble}%
\providecommand \bibinfo  [0]{\@secondoftwo}%
\providecommand \bibfield  [0]{\@secondoftwo}%
\providecommand \translation [1]{[#1]}%
\providecommand \BibitemOpen [0]{}%
\providecommand \bibitemStop [0]{}%
\providecommand \bibitemNoStop [0]{.\EOS\space}%
\providecommand \EOS [0]{\spacefactor3000\relax}%
\providecommand \BibitemShut  [1]{\csname bibitem#1\endcsname}%
\let\auto@bib@innerbib\@empty
\bibitem [{\citenamefont {Anderson}(1972)}]{anderson1972more}%
  \BibitemOpen
  \bibfield  {author} {\bibinfo {author} {\bibfnamefont {P.~W.}\ \bibnamefont {Anderson}},\ }\href {\doibase 10.1126/science.177.4047.393} {\bibfield  {journal} {\bibinfo  {journal} {Science}\ }\textbf {\bibinfo {volume} {177}},\ \bibinfo {pages} {393} (\bibinfo {year} {1972})}\BibitemShut {NoStop}%
\bibitem [{\citenamefont {May}(1972)}]{may1972will}%
  \BibitemOpen
  \bibfield  {author} {\bibinfo {author} {\bibfnamefont {R.~M.}\ \bibnamefont {May}},\ }\href {\doibase doi.org/10.1038/238413a0} {\bibfield  {journal} {\bibinfo  {journal} {Nature}\ }\textbf {\bibinfo {volume} {238}},\ \bibinfo {pages} {413} (\bibinfo {year} {1972})}\BibitemShut {NoStop}%
\bibitem [{\citenamefont {Simon}(1962)}]{herbert1962architecture}%
  \BibitemOpen
  \bibfield  {author} {\bibinfo {author} {\bibfnamefont {H.~A.}\ \bibnamefont {Simon}},\ }\href@noop {} {\bibfield  {journal} {\bibinfo  {journal} {Proceedings of the American Philosophical Society}\ }\textbf {\bibinfo {volume} {106}} (\bibinfo {year} {1962})}\BibitemShut {NoStop}%
\bibitem [{\citenamefont {Mitchell}(2009)}]{mitchell2009complexity}%
  \BibitemOpen
  \bibfield  {author} {\bibinfo {author} {\bibfnamefont {M.}~\bibnamefont {Mitchell}},\ }\href {\doibase 10.1093/oso/9780195124415.001.0001} {\emph {\bibinfo {title} {Complexity: A guided tour}}}\ (\bibinfo  {publisher} {Oxford University Press},\ \bibinfo {address} {Oxford, UK},\ \bibinfo {year} {2009})\BibitemShut {NoStop}%
\bibitem [{\citenamefont {Strogatz}(2001)}]{strogatz2001exploring}%
  \BibitemOpen
  \bibfield  {author} {\bibinfo {author} {\bibfnamefont {S.~H.}\ \bibnamefont {Strogatz}},\ }\href {\doibase 10.1038/35065725} {\bibfield  {journal} {\bibinfo  {journal} {Nature}\ }\textbf {\bibinfo {volume} {410}},\ \bibinfo {pages} {268} (\bibinfo {year} {2001})}\BibitemShut {NoStop}%
\bibitem [{\citenamefont {Barab{\'a}si}\ and\ \citenamefont {Albert}(1999)}]{barabasi_albert}%
  \BibitemOpen
  \bibfield  {author} {\bibinfo {author} {\bibfnamefont {A.-L.}\ \bibnamefont {Barab{\'a}si}}\ and\ \bibinfo {author} {\bibfnamefont {R.}~\bibnamefont {Albert}},\ }\href {\doibase 10.1126/science.286.5439.509} {\bibfield  {journal} {\bibinfo  {journal} {{S}cience}\ }\textbf {\bibinfo {volume} {286}},\ \bibinfo {pages} {509} (\bibinfo {year} {1999})}\BibitemShut {NoStop}%
\bibitem [{\citenamefont {Bak}\ and\ \citenamefont {Paczuski}(1995)}]{bak1995complexity}%
  \BibitemOpen
  \bibfield  {author} {\bibinfo {author} {\bibfnamefont {P.}~\bibnamefont {Bak}}\ and\ \bibinfo {author} {\bibfnamefont {M.}~\bibnamefont {Paczuski}},\ }\href {\doibase 10.1073/pnas.92.15.6689} {\bibfield  {journal} {\bibinfo  {journal} {Proceedings of the National Academy of Sciences}\ }\textbf {\bibinfo {volume} {92}},\ \bibinfo {pages} {6689} (\bibinfo {year} {1995})}\BibitemShut {NoStop}%
\bibitem [{\citenamefont {Castellano}, \citenamefont {Fortunato},\ and\ \citenamefont {Loreto}(2009)}]{castellano2009statistical}%
  \BibitemOpen
  \bibfield  {author} {\bibinfo {author} {\bibfnamefont {C.}~\bibnamefont {Castellano}}, \bibinfo {author} {\bibfnamefont {S.}~\bibnamefont {Fortunato}}, \ and\ \bibinfo {author} {\bibfnamefont {V.}~\bibnamefont {Loreto}},\ }\href {\doibase 10.1103/RevModPhys.81.591} {\bibfield  {journal} {\bibinfo  {journal} {Reviews of Modern Physics}\ }\textbf {\bibinfo {volume} {81}},\ \bibinfo {pages} {591} (\bibinfo {year} {2009})}\BibitemShut {NoStop}%
\bibitem [{\citenamefont {Mezard}\ and\ \citenamefont {Montanari}(2009)}]{mezard2009information}%
  \BibitemOpen
  \bibfield  {author} {\bibinfo {author} {\bibfnamefont {M.}~\bibnamefont {Mezard}}\ and\ \bibinfo {author} {\bibfnamefont {A.}~\bibnamefont {Montanari}},\ }\href {\doibase 10.1093/acprof:oso/9780198570837.001.0001} {\emph {\bibinfo {title} {Information, physics, and computation}}}\ (\bibinfo  {publisher} {Oxford University Press},\ \bibinfo {address} {Oxford, UK},\ \bibinfo {year} {2009})\BibitemShut {NoStop}%
\bibitem [{\citenamefont {Sornette}(2006)}]{sornette2006critical}%
  \BibitemOpen
  \bibfield  {author} {\bibinfo {author} {\bibfnamefont {D.}~\bibnamefont {Sornette}},\ }\href {\doibase 10.1007/3-540-33182-4} {\emph {\bibinfo {title} {Critical phenomena in natural sciences: chaos, fractals, selforganization and disorder: concepts and tools}}},\ \bibinfo {edition} {2nd}\ ed.\ (\bibinfo  {publisher} {Springer Berlin, Heidelberg},\ \bibinfo {year} {2006})\BibitemShut {NoStop}%
\bibitem [{\citenamefont {Nowak}(2006)}]{nowak2006evolutionary}%
  \BibitemOpen
  \bibfield  {author} {\bibinfo {author} {\bibfnamefont {M.~A.}\ \bibnamefont {Nowak}},\ }\href {\doibase 10.2307/j.ctvjghw98} {\emph {\bibinfo {title} {Evolutionary dynamics: exploring the equations of life}}}\ (\bibinfo  {publisher} {Harvard University Press},\ \bibinfo {address} {Harvard, MA, USA},\ \bibinfo {year} {2006})\BibitemShut {NoStop}%
\bibitem [{\citenamefont {Hofbauer}\ and\ \citenamefont {Sigmund}(2003)}]{hofbauer2003evolutionary}%
  \BibitemOpen
  \bibfield  {author} {\bibinfo {author} {\bibfnamefont {J.}~\bibnamefont {Hofbauer}}\ and\ \bibinfo {author} {\bibfnamefont {K.}~\bibnamefont {Sigmund}},\ }\href {https://www.ams.org/journals/bull/2003-40-04/S0273-0979-03-00988-1/S0273-0979-03-00988-1.pdf} {\bibfield  {journal} {\bibinfo  {journal} {Bulletin of the American Mathematical Society}\ }\textbf {\bibinfo {volume} {40}},\ \bibinfo {pages} {479} (\bibinfo {year} {2003})}\BibitemShut {NoStop}%
\bibitem [{\citenamefont {Clifford}\ and\ \citenamefont {Sudbury}(1973)}]{clifford1973model}%
  \BibitemOpen
  \bibfield  {author} {\bibinfo {author} {\bibfnamefont {P.}~\bibnamefont {Clifford}}\ and\ \bibinfo {author} {\bibfnamefont {A.}~\bibnamefont {Sudbury}},\ }\href@noop {} {\bibfield  {journal} {\bibinfo  {journal} {Biometrika}\ }\textbf {\bibinfo {volume} {60}},\ \bibinfo {pages} {581} (\bibinfo {year} {1973})}\BibitemShut {NoStop}%
\bibitem [{\citenamefont {Liggett}(1994)}]{liggett1994coexistence}%
  \BibitemOpen
  \bibfield  {author} {\bibinfo {author} {\bibfnamefont {T.~M.}\ \bibnamefont {Liggett}},\ }\href {\doibase 10.1214/aop/1176988729} {\bibfield  {journal} {\bibinfo  {journal} {The Annals of Probability}\ }\textbf {\bibinfo {volume} {22}},\ \bibinfo {pages} {764} (\bibinfo {year} {1994})}\BibitemShut {NoStop}%
\bibitem [{\citenamefont {Redner}(2019)}]{redner2019reality}%
  \BibitemOpen
  \bibfield  {author} {\bibinfo {author} {\bibfnamefont {S.}~\bibnamefont {Redner}},\ }\href {\doibase 10.1016/j.crhy.2019.05.004} {\bibfield  {journal} {\bibinfo  {journal} {Comptes Rendus Physique}\ }\textbf {\bibinfo {volume} {20}},\ \bibinfo {pages} {275} (\bibinfo {year} {2019})}\BibitemShut {NoStop}%
\bibitem [{\citenamefont {Fotouhi}\ \emph {et~al.}(2019)\citenamefont {Fotouhi}, \citenamefont {Momeni}, \citenamefont {Allen},\ and\ \citenamefont {Nowak}}]{fotouhi2019evolution}%
  \BibitemOpen
  \bibfield  {author} {\bibinfo {author} {\bibfnamefont {B.}~\bibnamefont {Fotouhi}}, \bibinfo {author} {\bibfnamefont {N.}~\bibnamefont {Momeni}}, \bibinfo {author} {\bibfnamefont {B.}~\bibnamefont {Allen}}, \ and\ \bibinfo {author} {\bibfnamefont {M.~A.}\ \bibnamefont {Nowak}},\ }\href {\doibase 10.1098/rsif.2018.0677} {\bibfield  {journal} {\bibinfo  {journal} {Journal of the Royal Society Interface}\ }\textbf {\bibinfo {volume} {16}},\ \bibinfo {pages} {20180677} (\bibinfo {year} {2019})}\BibitemShut {NoStop}%
\bibitem [{\citenamefont {Allen}\ \emph {et~al.}(2012)\citenamefont {Allen}, \citenamefont {Traulsen}, \citenamefont {Tarnita},\ and\ \citenamefont {Nowak}}]{allen2012mutation}%
  \BibitemOpen
  \bibfield  {author} {\bibinfo {author} {\bibfnamefont {B.}~\bibnamefont {Allen}}, \bibinfo {author} {\bibfnamefont {A.}~\bibnamefont {Traulsen}}, \bibinfo {author} {\bibfnamefont {C.~E.}\ \bibnamefont {Tarnita}}, \ and\ \bibinfo {author} {\bibfnamefont {M.~A.}\ \bibnamefont {Nowak}},\ }\href {\doibase 10.1016/j.jtbi.2011.03.034} {\bibfield  {journal} {\bibinfo  {journal} {Journal of Theoretical Biology}\ }\textbf {\bibinfo {volume} {299}},\ \bibinfo {pages} {97} (\bibinfo {year} {2012})}\BibitemShut {NoStop}%
\bibitem [{\citenamefont {Shakarian}, \citenamefont {Roos},\ and\ \citenamefont {Johnson}(2012)}]{shakarian2012review}%
  \BibitemOpen
  \bibfield  {author} {\bibinfo {author} {\bibfnamefont {P.}~\bibnamefont {Shakarian}}, \bibinfo {author} {\bibfnamefont {P.}~\bibnamefont {Roos}}, \ and\ \bibinfo {author} {\bibfnamefont {A.}~\bibnamefont {Johnson}},\ }\href {\doibase 10.1016/j.biosystems.2011.09.006} {\bibfield  {journal} {\bibinfo  {journal} {Biosystems}\ }\textbf {\bibinfo {volume} {107}},\ \bibinfo {pages} {66} (\bibinfo {year} {2012})}\BibitemShut {NoStop}%
\bibitem [{\citenamefont {Galam}(1997)}]{galam1997}%
  \BibitemOpen
  \bibfield  {author} {\bibinfo {author} {\bibfnamefont {S.}~\bibnamefont {Galam}},\ }\href@noop {} {\bibfield  {journal} {\bibinfo  {journal} {Physica A: Statistical Mechanics and its Applications}\ }\textbf {\bibinfo {volume} {238}},\ \bibinfo {pages} {66} (\bibinfo {year} {1997})}\BibitemShut {NoStop}%
\bibitem [{\citenamefont {Galam}\ and\ \citenamefont {Jacobs}(2007)}]{galam2007}%
  \BibitemOpen
  \bibfield  {author} {\bibinfo {author} {\bibfnamefont {S.}~\bibnamefont {Galam}}\ and\ \bibinfo {author} {\bibfnamefont {F.}~\bibnamefont {Jacobs}},\ }\href@noop {} {\bibfield  {journal} {\bibinfo  {journal} {Physica A: Statistical Mechanics and its Applications}\ }\textbf {\bibinfo {volume} {381}},\ \bibinfo {pages} {366} (\bibinfo {year} {2007})}\BibitemShut {NoStop}%
\bibitem [{\citenamefont {Mobilia}(2015)}]{mobilia2015nonlinear}%
  \BibitemOpen
  \bibfield  {author} {\bibinfo {author} {\bibfnamefont {M.}~\bibnamefont {Mobilia}},\ }\href {\doibase 10.1103/PhysRevE.92.012803} {\bibfield  {journal} {\bibinfo  {journal} {Physical Review E}\ }\textbf {\bibinfo {volume} {92}},\ \bibinfo {pages} {012803} (\bibinfo {year} {2015})}\BibitemShut {NoStop}%
\bibitem [{\citenamefont {Masuda}(2012)}]{masuda2012evolution}%
  \BibitemOpen
  \bibfield  {author} {\bibinfo {author} {\bibfnamefont {N.}~\bibnamefont {Masuda}},\ }\href {\doibase 10.1038/srep00646} {\bibfield  {journal} {\bibinfo  {journal} {Scientific Reports}\ }\textbf {\bibinfo {volume} {2}},\ \bibinfo {pages} {646} (\bibinfo {year} {2012})}\BibitemShut {NoStop}%
\bibitem [{\citenamefont {Nakajima}\ and\ \citenamefont {Masuda}(2015)}]{nakajima2015evolutionary}%
  \BibitemOpen
  \bibfield  {author} {\bibinfo {author} {\bibfnamefont {Y.}~\bibnamefont {Nakajima}}\ and\ \bibinfo {author} {\bibfnamefont {N.}~\bibnamefont {Masuda}},\ }\href {\doibase 10.1007/s00285-014-0770-2} {\bibfield  {journal} {\bibinfo  {journal} {Journal of Mathematical Biology}\ }\textbf {\bibinfo {volume} {70}},\ \bibinfo {pages} {465} (\bibinfo {year} {2015})}\BibitemShut {NoStop}%
\bibitem [{\citenamefont {Khalil}, \citenamefont {San~Miguel},\ and\ \citenamefont {Toral}(2018)}]{khalil2018zealots}%
  \BibitemOpen
  \bibfield  {author} {\bibinfo {author} {\bibfnamefont {N.}~\bibnamefont {Khalil}}, \bibinfo {author} {\bibfnamefont {M.}~\bibnamefont {San~Miguel}}, \ and\ \bibinfo {author} {\bibfnamefont {R.}~\bibnamefont {Toral}},\ }\href {\doibase 10.1103/PhysRevE.97.012310} {\bibfield  {journal} {\bibinfo  {journal} {Physical Review E}\ }\textbf {\bibinfo {volume} {97}},\ \bibinfo {pages} {012310} (\bibinfo {year} {2018})}\BibitemShut {NoStop}%
\bibitem [{\citenamefont {Mobilia}(2003)}]{mobilia_2003}%
  \BibitemOpen
  \bibfield  {author} {\bibinfo {author} {\bibfnamefont {M.}~\bibnamefont {Mobilia}},\ }\href {\doibase 10.1103/PhysRevLett.91.028701} {\bibfield  {journal} {\bibinfo  {journal} {Phys. Rev. Lett.}\ }\textbf {\bibinfo {volume} {91}},\ \bibinfo {pages} {028701} (\bibinfo {year} {2003})}\BibitemShut {NoStop}%
\bibitem [{\citenamefont {Kearns}\ \emph {et~al.}(2009)\citenamefont {Kearns}, \citenamefont {Judd}, \citenamefont {Tan},\ and\ \citenamefont {Wortman}}]{kearns2009behavioural}%
  \BibitemOpen
  \bibfield  {author} {\bibinfo {author} {\bibfnamefont {M.}~\bibnamefont {Kearns}}, \bibinfo {author} {\bibfnamefont {S.}~\bibnamefont {Judd}}, \bibinfo {author} {\bibfnamefont {J.}~\bibnamefont {Tan}}, \ and\ \bibinfo {author} {\bibfnamefont {J.}~\bibnamefont {Wortman}},\ }\href {\doibase 10.1073/pnas.0808147106} {\bibfield  {journal} {\bibinfo  {journal} {Proceedings of the National Academy of Sciences}\ }\textbf {\bibinfo {volume} {106}},\ \bibinfo {pages} {1347} (\bibinfo {year} {2009})}\BibitemShut {NoStop}%
\bibitem [{\citenamefont {Xie}\ \emph {et~al.}(2011)\citenamefont {Xie}, \citenamefont {Sreenivasan}, \citenamefont {Korniss}, \citenamefont {Zhang}, \citenamefont {Lim},\ and\ \citenamefont {Szymanski}}]{xie2011social}%
  \BibitemOpen
  \bibfield  {author} {\bibinfo {author} {\bibfnamefont {J.}~\bibnamefont {Xie}}, \bibinfo {author} {\bibfnamefont {S.}~\bibnamefont {Sreenivasan}}, \bibinfo {author} {\bibfnamefont {G.}~\bibnamefont {Korniss}}, \bibinfo {author} {\bibfnamefont {W.}~\bibnamefont {Zhang}}, \bibinfo {author} {\bibfnamefont {C.}~\bibnamefont {Lim}}, \ and\ \bibinfo {author} {\bibfnamefont {B.~K.}\ \bibnamefont {Szymanski}},\ }\href {\doibase 10.1103/PhysRevE.84.011130} {\bibfield  {journal} {\bibinfo  {journal} {Physical Review E}\ }\textbf {\bibinfo {volume} {84}},\ \bibinfo {pages} {011130} (\bibinfo {year} {2011})}\BibitemShut {NoStop}%
\bibitem [{\citenamefont {Mobilia}, \citenamefont {Petersen},\ and\ \citenamefont {Redner}(2007)}]{mobilia2007role}%
  \BibitemOpen
  \bibfield  {author} {\bibinfo {author} {\bibfnamefont {M.}~\bibnamefont {Mobilia}}, \bibinfo {author} {\bibfnamefont {A.}~\bibnamefont {Petersen}}, \ and\ \bibinfo {author} {\bibfnamefont {S.}~\bibnamefont {Redner}},\ }\href {\doibase 10.1088/1742-5468/2007/08/P08029} {\bibfield  {journal} {\bibinfo  {journal} {Journal of Statistical Mechanics: Theory and Experiment}\ }\textbf {\bibinfo {volume} {2007}},\ \bibinfo {pages} {P08029} (\bibinfo {year} {2007})}\BibitemShut {NoStop}%
\bibitem [{\citenamefont {Yildiz}\ \emph {et~al.}(2013)\citenamefont {Yildiz}, \citenamefont {Ozdaglar}, \citenamefont {Acemoglu}, \citenamefont {Saberi},\ and\ \citenamefont {Scaglione}}]{yildiz2013binary}%
  \BibitemOpen
  \bibfield  {author} {\bibinfo {author} {\bibfnamefont {E.}~\bibnamefont {Yildiz}}, \bibinfo {author} {\bibfnamefont {A.}~\bibnamefont {Ozdaglar}}, \bibinfo {author} {\bibfnamefont {D.}~\bibnamefont {Acemoglu}}, \bibinfo {author} {\bibfnamefont {A.}~\bibnamefont {Saberi}}, \ and\ \bibinfo {author} {\bibfnamefont {A.}~\bibnamefont {Scaglione}},\ }\href {\doibase 10.1145/2538508} {\bibfield  {journal} {\bibinfo  {journal} {ACM Transactions on Economics and Computation}\ }\textbf {\bibinfo {volume} {1}} (\bibinfo {year} {2013}),\ 10.1145/2538508}\BibitemShut {NoStop}%
\bibitem [{\citenamefont {Acemo\u{g}lu}\ \emph {et~al.}(2012)\citenamefont {Acemo\u{g}lu}, \citenamefont {Como}, \citenamefont {Fagnani},\ and\ \citenamefont {Ozdaglar}}]{acemouglu2012opinion}%
  \BibitemOpen
  \bibfield  {author} {\bibinfo {author} {\bibfnamefont {D.}~\bibnamefont {Acemo\u{g}lu}}, \bibinfo {author} {\bibfnamefont {G.}~\bibnamefont {Como}}, \bibinfo {author} {\bibfnamefont {F.}~\bibnamefont {Fagnani}}, \ and\ \bibinfo {author} {\bibfnamefont {A.}~\bibnamefont {Ozdaglar}},\ }\href {\doibase 10.1287/moor.1120.0570} {\bibfield  {journal} {\bibinfo  {journal} {Mathematics of Operations Research}\ }\textbf {\bibinfo {volume} {28}},\ \bibinfo {pages} {1} (\bibinfo {year} {2012})}\BibitemShut {NoStop}%
\bibitem [{\citenamefont {Cao}\ \emph {et~al.}(2024)\citenamefont {Cao}, \citenamefont {Zhang}, \citenamefont {Kou},\ and\ \citenamefont {Zhang}}]{cao2024discrete}%
  \BibitemOpen
  \bibfield  {author} {\bibinfo {author} {\bibfnamefont {W.}~\bibnamefont {Cao}}, \bibinfo {author} {\bibfnamefont {H.}~\bibnamefont {Zhang}}, \bibinfo {author} {\bibfnamefont {G.}~\bibnamefont {Kou}}, \ and\ \bibinfo {author} {\bibfnamefont {B.}~\bibnamefont {Zhang}},\ }\href {\doibase 10.1016/j.inffus.2024.102410} {\bibfield  {journal} {\bibinfo  {journal} {Information Fusion}\ }\textbf {\bibinfo {volume} {109}},\ \bibinfo {pages} {102410} (\bibinfo {year} {2024})}\BibitemShut {NoStop}%
\bibitem [{\citenamefont {Castellano}, \citenamefont {Mu{\~n}oz},\ and\ \citenamefont {Pastor-Satorras}(2009)}]{castellano2009nonlinear}%
  \BibitemOpen
  \bibfield  {author} {\bibinfo {author} {\bibfnamefont {C.}~\bibnamefont {Castellano}}, \bibinfo {author} {\bibfnamefont {M.~A.}\ \bibnamefont {Mu{\~n}oz}}, \ and\ \bibinfo {author} {\bibfnamefont {R.}~\bibnamefont {Pastor-Satorras}},\ }\href {\doibase 10.1103/PhysRevE.80.041129} {\bibfield  {journal} {\bibinfo  {journal} {Physical Review E}\ }\textbf {\bibinfo {volume} {80}},\ \bibinfo {pages} {041129} (\bibinfo {year} {2009})}\BibitemShut {NoStop}%
\bibitem [{\citenamefont {Ramirez}\ \emph {et~al.}(2024)\citenamefont {Ramirez}, \citenamefont {Vazquez}, \citenamefont {San~Miguel},\ and\ \citenamefont {Galla}}]{ramirez2023ordering}%
  \BibitemOpen
  \bibfield  {author} {\bibinfo {author} {\bibfnamefont {L.~S.}\ \bibnamefont {Ramirez}}, \bibinfo {author} {\bibfnamefont {F.}~\bibnamefont {Vazquez}}, \bibinfo {author} {\bibfnamefont {M.}~\bibnamefont {San~Miguel}}, \ and\ \bibinfo {author} {\bibfnamefont {T.}~\bibnamefont {Galla}},\ }\href {\doibase 10.1103/PhysRevE.109.034307} {\bibfield  {journal} {\bibinfo  {journal} {Physical Review E}\ }\textbf {\bibinfo {volume} {109}},\ \bibinfo {pages} {034307} (\bibinfo {year} {2024})}\BibitemShut {NoStop}%
\bibitem [{\citenamefont {Schweitzer}\ and\ \citenamefont {Behera}(2009)}]{schweitzer2009nonlinear}%
  \BibitemOpen
  \bibfield  {author} {\bibinfo {author} {\bibfnamefont {F.}~\bibnamefont {Schweitzer}}\ and\ \bibinfo {author} {\bibfnamefont {L.}~\bibnamefont {Behera}},\ }\href {\doibase 10.1140/epjb/e2009-00001-3} {\bibfield  {journal} {\bibinfo  {journal} {European Physical Journal B}\ }\textbf {\bibinfo {volume} {67}},\ \bibinfo {pages} {301} (\bibinfo {year} {2009})}\BibitemShut {NoStop}%
\bibitem [{\citenamefont {Min}\ and\ \citenamefont {San~Miguel}(2017)}]{min2017fragmentation}%
  \BibitemOpen
  \bibfield  {author} {\bibinfo {author} {\bibfnamefont {B.}~\bibnamefont {Min}}\ and\ \bibinfo {author} {\bibfnamefont {M.}~\bibnamefont {San~Miguel}},\ }\href {\doibase 10.1038/s41598-017-13047-2} {\bibfield  {journal} {\bibinfo  {journal} {Scientific Reports}\ }\textbf {\bibinfo {volume} {7}},\ \bibinfo {pages} {12864} (\bibinfo {year} {2017})}\BibitemShut {NoStop}%
\bibitem [{\citenamefont {Traulsen}\ and\ \citenamefont {Hauert}(2009)}]{traulsen2009stochastic}%
  \BibitemOpen
  \bibfield  {author} {\bibinfo {author} {\bibfnamefont {A.}~\bibnamefont {Traulsen}}\ and\ \bibinfo {author} {\bibfnamefont {C.}~\bibnamefont {Hauert}},\ }\enquote {\bibinfo {title} {Stochastic evolutionary game dynamics},}\ in\ \href {\doibase 10.1002/9783527628001.ch2} {\emph {\bibinfo {booktitle} {Reviews of Nonlinear Dynamics and Complexity}}}\ (\bibinfo  {publisher} {John Wiley \& Sons, Ltd},\ \bibinfo {address} {Hoboken, NJ, USA},\ \bibinfo {year} {2009})\ Chap.~\bibinfo {chapter} {2}, pp.\ \bibinfo {pages} {25--61}\BibitemShut {NoStop}%
\bibitem [{\citenamefont {Masuda}, \citenamefont {Gibert},\ and\ \citenamefont {Redner}(2010)}]{masuda2010heterogeneous}%
  \BibitemOpen
  \bibfield  {author} {\bibinfo {author} {\bibfnamefont {N.}~\bibnamefont {Masuda}}, \bibinfo {author} {\bibfnamefont {N.}~\bibnamefont {Gibert}}, \ and\ \bibinfo {author} {\bibfnamefont {S.}~\bibnamefont {Redner}},\ }\href {\doibase 10.1103/PhysRevE.82.010103} {\bibfield  {journal} {\bibinfo  {journal} {Physical Review E}\ }\textbf {\bibinfo {volume} {82}},\ \bibinfo {pages} {010103} (\bibinfo {year} {2010})}\BibitemShut {NoStop}%
\bibitem [{\citenamefont {Masuda}\ and\ \citenamefont {Redner}(2011)}]{masuda2011can}%
  \BibitemOpen
  \bibfield  {author} {\bibinfo {author} {\bibfnamefont {N.}~\bibnamefont {Masuda}}\ and\ \bibinfo {author} {\bibfnamefont {S.}~\bibnamefont {Redner}},\ }\href {\doibase 10.1088/1742-5468/2011/02/L02002} {\bibfield  {journal} {\bibinfo  {journal} {Journal of Statistical Mechanics: Theory and Experiment}\ }\textbf {\bibinfo {volume} {2011}},\ \bibinfo {pages} {L02002} (\bibinfo {year} {2011})}\BibitemShut {NoStop}%
\bibitem [{\citenamefont {Llabr{\'e}s}, \citenamefont {San~Miguel},\ and\ \citenamefont {Toral}(2023)}]{llabres2023partisan}%
  \BibitemOpen
  \bibfield  {author} {\bibinfo {author} {\bibfnamefont {J.}~\bibnamefont {Llabr{\'e}s}}, \bibinfo {author} {\bibfnamefont {M.}~\bibnamefont {San~Miguel}}, \ and\ \bibinfo {author} {\bibfnamefont {R.}~\bibnamefont {Toral}},\ }\href {\doibase 10.1103/PhysRevE.108.054106} {\bibfield  {journal} {\bibinfo  {journal} {Physical Review E}\ }\textbf {\bibinfo {volume} {108}},\ \bibinfo {pages} {054106} (\bibinfo {year} {2023})}\BibitemShut {NoStop}%
\bibitem [{\citenamefont {Latan{\'e}}(1981)}]{latane}%
  \BibitemOpen
  \bibfield  {author} {\bibinfo {author} {\bibfnamefont {B.}~\bibnamefont {Latan{\'e}}},\ }\href {\doibase 10.1037/0003-066X.36.4.343} {\bibfield  {journal} {\bibinfo  {journal} {American Psychologist}\ }\textbf {\bibinfo {volume} {36}},\ \bibinfo {pages} {343} (\bibinfo {year} {1981})}\BibitemShut {NoStop}%
\bibitem [{\citenamefont {Mobilia}\ and\ \citenamefont {Georgiev}(2005)}]{mobilia2005voting}%
  \BibitemOpen
  \bibfield  {author} {\bibinfo {author} {\bibfnamefont {M.}~\bibnamefont {Mobilia}}\ and\ \bibinfo {author} {\bibfnamefont {I.~T.}\ \bibnamefont {Georgiev}},\ }\href {\doibase 10.1103/PhysRevE.71.046102} {\bibfield  {journal} {\bibinfo  {journal} {Physical Review E}\ }\textbf {\bibinfo {volume} {71}},\ \bibinfo {pages} {046102} (\bibinfo {year} {2005})}\BibitemShut {NoStop}%
\bibitem [{\citenamefont {Holley}\ and\ \citenamefont {Liggett}(1975)}]{holley1975ergodic}%
  \BibitemOpen
  \bibfield  {author} {\bibinfo {author} {\bibfnamefont {R.~A.}\ \bibnamefont {Holley}}\ and\ \bibinfo {author} {\bibfnamefont {T.~M.}\ \bibnamefont {Liggett}},\ }\href {https://www.jstor.org/stable/2959329} {\bibfield  {journal} {\bibinfo  {journal} {The Annals of Probability}\ }\textbf {\bibinfo {volume} {3}},\ \bibinfo {pages} {643} (\bibinfo {year} {1975})}\BibitemShut {NoStop}%
\bibitem [{\citenamefont {Suchecki}, \citenamefont {Egu\'{\i}luz},\ and\ \citenamefont {San~Miguel}(2005)}]{suchecki}%
  \BibitemOpen
  \bibfield  {author} {\bibinfo {author} {\bibfnamefont {K.}~\bibnamefont {Suchecki}}, \bibinfo {author} {\bibfnamefont {V.~M.}\ \bibnamefont {Egu\'{\i}luz}}, \ and\ \bibinfo {author} {\bibfnamefont {M.}~\bibnamefont {San~Miguel}},\ }\href {\doibase 10.1103/PhysRevE.72.036132} {\bibfield  {journal} {\bibinfo  {journal} {Physical Review E}\ }\textbf {\bibinfo {volume} {72}},\ \bibinfo {pages} {036132} (\bibinfo {year} {2005})}\BibitemShut {NoStop}%
\bibitem [{\citenamefont {Taylor}\ \emph {et~al.}(2004)\citenamefont {Taylor}, \citenamefont {Fudenberg}, \citenamefont {Sasaki},\ and\ \citenamefont {Nowak}}]{taylor}%
  \BibitemOpen
  \bibfield  {author} {\bibinfo {author} {\bibfnamefont {C.}~\bibnamefont {Taylor}}, \bibinfo {author} {\bibfnamefont {D.}~\bibnamefont {Fudenberg}}, \bibinfo {author} {\bibfnamefont {A.}~\bibnamefont {Sasaki}}, \ and\ \bibinfo {author} {\bibfnamefont {M.~A.}\ \bibnamefont {Nowak}},\ }\href {\doibase 10.1016/j.bulm.2004.03.004} {\bibfield  {journal} {\bibinfo  {journal} {Bulletin of Mathematical Biology}\ }\textbf {\bibinfo {volume} {66}},\ \bibinfo {pages} {1621} (\bibinfo {year} {2004})}\BibitemShut {NoStop}%
\bibitem [{\citenamefont {Traulsen}, \citenamefont {Claussen},\ and\ \citenamefont {Hauert}(2005)}]{traulsen2005coevolutionary}%
  \BibitemOpen
  \bibfield  {author} {\bibinfo {author} {\bibfnamefont {A.}~\bibnamefont {Traulsen}}, \bibinfo {author} {\bibfnamefont {J.~C.}\ \bibnamefont {Claussen}}, \ and\ \bibinfo {author} {\bibfnamefont {C.}~\bibnamefont {Hauert}},\ }\href {\doibase 10.1103/PhysRevLett.95.238701} {\bibfield  {journal} {\bibinfo  {journal} {Physical Review Letters}\ }\textbf {\bibinfo {volume} {95}},\ \bibinfo {pages} {238701} (\bibinfo {year} {2005})}\BibitemShut {NoStop}%
\bibitem [{\citenamefont {Kitching}\ and\ \citenamefont {Galla}(2024)}]{kitching2024qdeformed}%
  \BibitemOpen
  \bibfield  {author} {\bibinfo {author} {\bibfnamefont {C.~R.}\ \bibnamefont {Kitching}}\ and\ \bibinfo {author} {\bibfnamefont {T.}~\bibnamefont {Galla}},\ }\href {\doibase 10.1103/PhysRevE.110.064319} {\bibfield  {journal} {\bibinfo  {journal} {Phys. Rev. E}\ }\textbf {\bibinfo {volume} {110}},\ \bibinfo {pages} {064319} (\bibinfo {year} {2024})}\BibitemShut {NoStop}%
\bibitem [{\citenamefont {Traulsen}, \citenamefont {Pacheco},\ and\ \citenamefont {Nowak}(2007)}]{traulsen2007pairwise}%
  \BibitemOpen
  \bibfield  {author} {\bibinfo {author} {\bibfnamefont {A.}~\bibnamefont {Traulsen}}, \bibinfo {author} {\bibfnamefont {J.~M.}\ \bibnamefont {Pacheco}}, \ and\ \bibinfo {author} {\bibfnamefont {M.~A.}\ \bibnamefont {Nowak}},\ }\href {\doibase 10.1016/j.jtbi.2007.01.002} {\bibfield  {journal} {\bibinfo  {journal} {Journal of Theoretical Biology}\ }\textbf {\bibinfo {volume} {246}},\ \bibinfo {pages} {522} (\bibinfo {year} {2007})}\BibitemShut {NoStop}%
\bibitem [{\citenamefont {Gillespie}(1976)}]{gillespie1976general}%
  \BibitemOpen
  \bibfield  {author} {\bibinfo {author} {\bibfnamefont {D.~T.}\ \bibnamefont {Gillespie}},\ }\href {\doibase 10.1016/0021-9991(76)90041-3} {\bibfield  {journal} {\bibinfo  {journal} {Journal of Computational Physics}\ }\textbf {\bibinfo {volume} {22}},\ \bibinfo {pages} {403} (\bibinfo {year} {1976})}\BibitemShut {NoStop}%
\bibitem [{\citenamefont {Vazquez}, \citenamefont {Castell\'o},\ and\ \citenamefont {San~Miguel}(2010)}]{castello}%
  \BibitemOpen
  \bibfield  {author} {\bibinfo {author} {\bibfnamefont {F.}~\bibnamefont {Vazquez}}, \bibinfo {author} {\bibfnamefont {X.}~\bibnamefont {Castell\'o}}, \ and\ \bibinfo {author} {\bibfnamefont {M.}~\bibnamefont {San~Miguel}},\ }\href {\doibase 10.1088/1742-5468/2010/04/P04007} {\bibfield  {journal} {\bibinfo  {journal} {Journal of Statistical Mechanics: Theory and Experiment}\ }\textbf {\bibinfo {volume} {2010}},\ \bibinfo {pages} {P04007} (\bibinfo {year} {2010})}\BibitemShut {NoStop}%
\bibitem [{\citenamefont {Abrams}\ and\ \citenamefont {Strogatz}(2003)}]{abrams2003modelling}%
  \BibitemOpen
  \bibfield  {author} {\bibinfo {author} {\bibfnamefont {D.~M.}\ \bibnamefont {Abrams}}\ and\ \bibinfo {author} {\bibfnamefont {S.~H.}\ \bibnamefont {Strogatz}},\ }\href {\doibase 10.1038/424900a} {\bibfield  {journal} {\bibinfo  {journal} {Nature}\ }\textbf {\bibinfo {volume} {424}},\ \bibinfo {pages} {900} (\bibinfo {year} {2003})}\BibitemShut {NoStop}%
\bibitem [{\citenamefont {Mellor}, \citenamefont {Mobilia},\ and\ \citenamefont {Zia}(2016)}]{Mellor2016characterization}%
  \BibitemOpen
  \bibfield  {author} {\bibinfo {author} {\bibfnamefont {A.}~\bibnamefont {Mellor}}, \bibinfo {author} {\bibfnamefont {M.}~\bibnamefont {Mobilia}}, \ and\ \bibinfo {author} {\bibfnamefont {R.~K.~P.}\ \bibnamefont {Zia}},\ }\href {\doibase 10.1209/0295-5075/113/48001} {\bibfield  {journal} {\bibinfo  {journal} {Europhysics Letters}\ }\textbf {\bibinfo {volume} {113}},\ \bibinfo {pages} {48001} (\bibinfo {year} {2016})}\BibitemShut {NoStop}%
\bibitem [{\citenamefont {Mellor}, \citenamefont {Mobilia},\ and\ \citenamefont {Zia}(2017)}]{mellor2017heterogeneous}%
  \BibitemOpen
  \bibfield  {author} {\bibinfo {author} {\bibfnamefont {A.}~\bibnamefont {Mellor}}, \bibinfo {author} {\bibfnamefont {M.}~\bibnamefont {Mobilia}}, \ and\ \bibinfo {author} {\bibfnamefont {R.~K.~P.}\ \bibnamefont {Zia}},\ }\href {\doibase 10.1103/PhysRevE.95.012104} {\bibfield  {journal} {\bibinfo  {journal} {Physical Review E}\ }\textbf {\bibinfo {volume} {95}},\ \bibinfo {pages} {012104} (\bibinfo {year} {2017})}\BibitemShut {NoStop}%
\bibitem [{\citenamefont {Raducha}\ and\ \citenamefont {San~Miguel}(2022)}]{raducha2022}%
  \BibitemOpen
  \bibfield  {author} {\bibinfo {author} {\bibfnamefont {T.}~\bibnamefont {Raducha}}\ and\ \bibinfo {author} {\bibfnamefont {M.}~\bibnamefont {San~Miguel}},\ }\href {\doibase 10.1038/s41598-022-07195-3} {\bibfield  {journal} {\bibinfo  {journal} {Scientific Reports}\ }\textbf {\bibinfo {volume} {12}},\ \bibinfo {pages} {3373} (\bibinfo {year} {2022})}\BibitemShut {NoStop}%
\bibitem [{\citenamefont {Mobilia}(2012)}]{mobilia2012stochastic}%
  \BibitemOpen
  \bibfield  {author} {\bibinfo {author} {\bibfnamefont {M.}~\bibnamefont {Mobilia}},\ }\href {\doibase 10.1103/PhysRevE.86.011134} {\bibfield  {journal} {\bibinfo  {journal} {Physical Review E}\ }\textbf {\bibinfo {volume} {86}},\ \bibinfo {pages} {011134} (\bibinfo {year} {2012})}\BibitemShut {NoStop}%
\bibitem [{\citenamefont {Mobilia}(2013)}]{mobilia2013evolutionary}%
  \BibitemOpen
  \bibfield  {author} {\bibinfo {author} {\bibfnamefont {M.}~\bibnamefont {Mobilia}},\ }\href {\doibase https://doi.org/10.1016/j.chaos.2013.07.011} {\bibfield  {journal} {\bibinfo  {journal} {Chaos, Solitons \& Fractals}\ }\textbf {\bibinfo {volume} {56}},\ \bibinfo {pages} {113} (\bibinfo {year} {2013})}\BibitemShut {NoStop}%
\bibitem [{\citenamefont {Szolnoki}, \citenamefont {Perc},\ and\ \citenamefont {Mobilia}(2014)}]{szolnoki2014facilitators}%
  \BibitemOpen
  \bibfield  {author} {\bibinfo {author} {\bibfnamefont {A.}~\bibnamefont {Szolnoki}}, \bibinfo {author} {\bibfnamefont {M.}~\bibnamefont {Perc}}, \ and\ \bibinfo {author} {\bibfnamefont {M.}~\bibnamefont {Mobilia}},\ }\href {\doibase 10.1103/PhysRevE.89.042802} {\bibfield  {journal} {\bibinfo  {journal} {Physical Review E}\ }\textbf {\bibinfo {volume} {89}},\ \bibinfo {pages} {042802} (\bibinfo {year} {2014})}\BibitemShut {NoStop}%
\bibitem [{\citenamefont {Ohtsuki}\ and\ \citenamefont {Nowak}(2006)}]{ohtsuki2006replicator}%
  \BibitemOpen
  \bibfield  {author} {\bibinfo {author} {\bibfnamefont {H.}~\bibnamefont {Ohtsuki}}\ and\ \bibinfo {author} {\bibfnamefont {M.~A.}\ \bibnamefont {Nowak}},\ }\href {\doibase 10.1016/j.jtbi.2006.06.004} {\bibfield  {journal} {\bibinfo  {journal} {Journal of Theoretical Biology}\ }\textbf {\bibinfo {volume} {243}},\ \bibinfo {pages} {86} (\bibinfo {year} {2006})}\BibitemShut {NoStop}%
\bibitem [{\citenamefont {Smith}(1982)}]{smith1982evolution}%
  \BibitemOpen
  \bibfield  {author} {\bibinfo {author} {\bibfnamefont {J.~M.}\ \bibnamefont {Smith}},\ }\href {\doibase 10.1017/CBO9780511806292} {\emph {\bibinfo {title} {Evolution and the Theory of Games}}}\ (\bibinfo  {publisher} {Cambridge University Press},\ \bibinfo {address} {Cambridge, UK},\ \bibinfo {year} {1982})\BibitemShut {NoStop}%
\bibitem [{\citenamefont {Hofbaur}\ and\ \citenamefont {Schalg}(2000)}]{hofbaur2000sophisticated}%
  \BibitemOpen
  \bibfield  {author} {\bibinfo {author} {\bibfnamefont {J.}~\bibnamefont {Hofbaur}}\ and\ \bibinfo {author} {\bibfnamefont {K.~H.}\ \bibnamefont {Schalg}},\ }\href {\doibase 10.1007/s001910000049} {\bibfield  {journal} {\bibinfo  {journal} {Journal of Evolutionary Economics}\ }\textbf {\bibinfo {volume} {10}},\ \bibinfo {pages} {523} (\bibinfo {year} {2000})}\BibitemShut {NoStop}%
\bibitem [{\citenamefont {Traulsen}, \citenamefont {Nowak},\ and\ \citenamefont {Pacheco}(2006)}]{traulsen2006stochastic}%
  \BibitemOpen
  \bibfield  {author} {\bibinfo {author} {\bibfnamefont {A.}~\bibnamefont {Traulsen}}, \bibinfo {author} {\bibfnamefont {M.~A.}\ \bibnamefont {Nowak}}, \ and\ \bibinfo {author} {\bibfnamefont {J.~M.}\ \bibnamefont {Pacheco}},\ }\href {\doibase 10.1103/PhysRevE.74.011909} {\bibfield  {journal} {\bibinfo  {journal} {Physical Review E}\ }\textbf {\bibinfo {volume} {74}},\ \bibinfo {pages} {011909} (\bibinfo {year} {2006})}\BibitemShut {NoStop}%
\bibitem [{\citenamefont {Mukhopadhyay}\ and\ \citenamefont {Chakraborty}(2021)}]{muk2021replicator}%
  \BibitemOpen
  \bibfield  {author} {\bibinfo {author} {\bibfnamefont {A.}~\bibnamefont {Mukhopadhyay}}\ and\ \bibinfo {author} {\bibfnamefont {S.}~\bibnamefont {Chakraborty}},\ }\href {\doibase 10.1063/5.0032311} {\bibfield  {journal} {\bibinfo  {journal} {Chaos: An Interdisciplinary Journal of Nonlinear Science}\ }\textbf {\bibinfo {volume} {31}},\ \bibinfo {pages} {023123} (\bibinfo {year} {2021})}\BibitemShut {NoStop}%
\bibitem [{\citenamefont {Kitching}(2025)}]{github_zealots}%
  \BibitemOpen
  \bibfield  {author} {\bibinfo {author} {\bibfnamefont {C.~R.}\ \bibnamefont {Kitching}},\ }\href {https://github.com/C-Kitching/breaking-coexistence-zealotry-vs-nonlinear-social-impact} {\enquote {\bibinfo {title} {{GitHub: Breaking co-existence zealotry vs nonlinear social impact}},}\ } (\bibinfo {year} {2025})\BibitemShut {NoStop}%
\bibitem [{\citenamefont {Lewenstein}, \citenamefont {Nowak},\ and\ \citenamefont {Latan\'e}(1992)}]{lewenstein}%
  \BibitemOpen
  \bibfield  {author} {\bibinfo {author} {\bibfnamefont {M.}~\bibnamefont {Lewenstein}}, \bibinfo {author} {\bibfnamefont {A.}~\bibnamefont {Nowak}}, \ and\ \bibinfo {author} {\bibfnamefont {B.}~\bibnamefont {Latan\'e}},\ }\href {\doibase 10.1103/PhysRevA.45.763} {\bibfield  {journal} {\bibinfo  {journal} {Physical Review A}\ }\textbf {\bibinfo {volume} {45}},\ \bibinfo {pages} {763} (\bibinfo {year} {1992})}\BibitemShut {NoStop}%
\bibitem [{\citenamefont {Llabr\'es}, \citenamefont {San~Miguel},\ and\ \citenamefont {Toral}(2025)}]{jaume_future}%
  \BibitemOpen
  \bibfield  {author} {\bibinfo {author} {\bibfnamefont {J.}~\bibnamefont {Llabr\'es}}, \bibinfo {author} {\bibfnamefont {M.}~\bibnamefont {San~Miguel}}, \ and\ \bibinfo {author} {\bibfnamefont {R.}~\bibnamefont {Toral}},\ }\href {https://arxiv.org/abs/2505.11358} {\enquote {\bibinfo {title} {preprint arxiv:2505.11358},}\ } (\bibinfo {year} {2025})\BibitemShut {NoStop}%
\bibitem [{\citenamefont {N{\aa}sell}(1996)}]{naasell1996quasi}%
  \BibitemOpen
  \bibfield  {author} {\bibinfo {author} {\bibfnamefont {I.}~\bibnamefont {N{\aa}sell}},\ }\href {\doibase 10.2307/1428186} {\bibfield  {journal} {\bibinfo  {journal} {Advances in Applied Probability}\ }\textbf {\bibinfo {volume} {28}},\ \bibinfo {pages} {895} (\bibinfo {year} {1996})}\BibitemShut {NoStop}%
\bibitem [{\citenamefont {Collet}, \citenamefont {Mart{\'\i}nez},\ and\ \citenamefont {San~Mart{\'\i}n}(2013)}]{collet2013quasi}%
  \BibitemOpen
  \bibfield  {author} {\bibinfo {author} {\bibfnamefont {P.~K.}\ \bibnamefont {Collet}}, \bibinfo {author} {\bibfnamefont {S.}~\bibnamefont {Mart{\'\i}nez}}, \ and\ \bibinfo {author} {\bibfnamefont {J.}~\bibnamefont {San~Mart{\'\i}n}},\ }\href@noop {} {\emph {\bibinfo {title} {Quasi-stationary distributions}}}\ (\bibinfo  {publisher} {Springer Berlin, Heidelberg},\ \bibinfo {address} {Berlin, Germany},\ \bibinfo {year} {2013})\BibitemShut {NoStop}%
\bibitem [{\citenamefont {Belki\'{c}}(2019)}]{belkic2019trinominal}%
  \BibitemOpen
  \bibfield  {author} {\bibinfo {author} {\bibfnamefont {D.}~\bibnamefont {Belki\'{c}}},\ }\href {\doibase 10.1007/s10910-018-0985-3} {\bibfield  {journal} {\bibinfo  {journal} {Journal of Mathematical Chemistry}\ }\textbf {\bibinfo {volume} {59}},\ \bibinfo {pages} {59} (\bibinfo {year} {2019})}\BibitemShut {NoStop}%
\end{thebibliography}
\end{document}